\documentclass[french,letterpaper,11pt]{book}

\def\mystretch{1.23}
\renewcommand{\baselinestretch}{\mystretch}

\usepackage{amsmath,amsfonts,amssymb,epsfig,graphicx,fancyheadings,caption,rotating}
\usepackage{feynmf}
\graphicspath{{fig/}}
\DeclareGraphicsExtensions{.eps,.eps.gz}
\usepackage[french]{babel}
\usepackage{lscape}

\makeatletter

\renewcommand{\@makefntext}[1]{%
  \renewcommand{\baselinestretch}{1}%
  \footnotesize%
  \noindent\makebox[1.8em][r]{\@thefnmark.\ }#1%
  \renewcommand{\baselinestretch}{\mystretch}%
}

\def\clearoddpage{\clearpage\ifodd\c@page\else
    \hbox{}\newpage\if@twocolumn\hbox{}\newpage\fi\fi}

\makeatother

\usepackage[french]{minitoc}

\begin{document}

\bibliographystyle{apalike}

\pagenumbering{arabic}

\def\gev{{\mathrm \, GeV}}

\thispagestyle{empty}
\include{titre}

\renewcommand{\baselinestretch}{1}
\normalsize

\clearoddpage


\pagestyle{empty}

\vspace{15cm}

\begin{center}

\huge{ \bf Review of double beta experiments  } \\

\vspace{0.3cm}

\Large{\it (May 2012)  } \\

\vspace{1.5cm}

\Large{Xavier Sarazin}

\vspace{0.5cm}

\Large{\it LAL, Univ Paris-Sud, CNRS/IN2P3, Orsay, France }

\vspace{1.5cm}

\LARGE{\it Abstract  }

\end{center}

\normalsize{This paper is the first part of the manuscript written in April 2012 for my academic {\it Accreditation to supervise research}.
It offers a review of the double beta experimental techniques. My purpose is to detail, for each technique, the different origins of background, how they can be identified, and how they can be reduced. Advantages and limitations are discussed.
This review is organized as follows. First, the question of the possible Majorana nature for the neutrino is presented and the physic of neutrinoless double beta decay is summarized.  Then I begin by presenting the tracko-calo NEMO-3 and SuperNEMO experiments. I've worked on these two experiments since 15 years. So it was natural to start with them with a relatively more exhaustive description. I will then present the germanium technique. I will then review the bolometer technique. I will describe in detail the recent progress in scintillating bolometers because I think that it is one of the most promising techniques. Finally I will review the large liquid scintillator detectors and Xenon TPC.
The last chapter offers a summary of the different techniques and projects.}

\cleardoublepage

\clearoddpage



\pagestyle{fancy}


\tableofcontents
\clearoddpage





\chapter{The majorana neutrino and the neutrinoless double beta decay}


The neutrino is one of the most puzzling elementary particle with very unique properties. It has no electrical charge\footnote{The limit on $\bar{\nu}_e$ magnetic moment gives $q/e <3.7 \ 10^{-12}$ and astrophysical considerations give $q/e < 2 \ 10^{-14}$~\cite{nu-charge}}, it is thus only sensitive to weak interaction and its mass is very light. The absence of electrical charge could be considered as a minor characteristic. But it is not. Ettore Majorana showed that a neutral elementary particle which does not contain any discrete quantum number (as the neutrino), can be described by a so-called Majorana field, in which the distinction between matter and antimatter vanishes~\cite{majorana}. In other words, a neutrino might be identical to its own anti-particle. 



If the neutrino is a Majorana neutrino, an important consequence is that Lepton Number Violation (LNV) must occur~\cite{giunti}\cite{kayser}. LNV is a required condition for Grand Unified Theories (GUT), in which quarks and leptons are components of the same multiplet, and hence both lepton and baryon numbers are not expected to be conserved quantities. Leptogenesis is an example of model, which uses the LNV from the decay of heavy Majorana neutrinos to produce the observed asymetry of matter and antimatter in the Universe. 
Another motivation for the Majorana neutrino is the see-saw mechanism~\cite{mohapatra}\cite{giunti}, which splits the Majorana mass term of neutrinos in light and heavy Majorana neutrinos and thus could explain the very small mass of the observed neutrinos, with the condition that the mass of the heavy one is at the GUT energy scale of about 10$^{15}$~GeV.

The most sensitive method to answer the nature of the neutrino is the search of the neutrinoless double beta decay ($\beta \beta 0 \nu$).

\section{The neutrinoless double beta decay}

\begin{figure}[!h]
  \centering
  \includegraphics[scale=0.33]{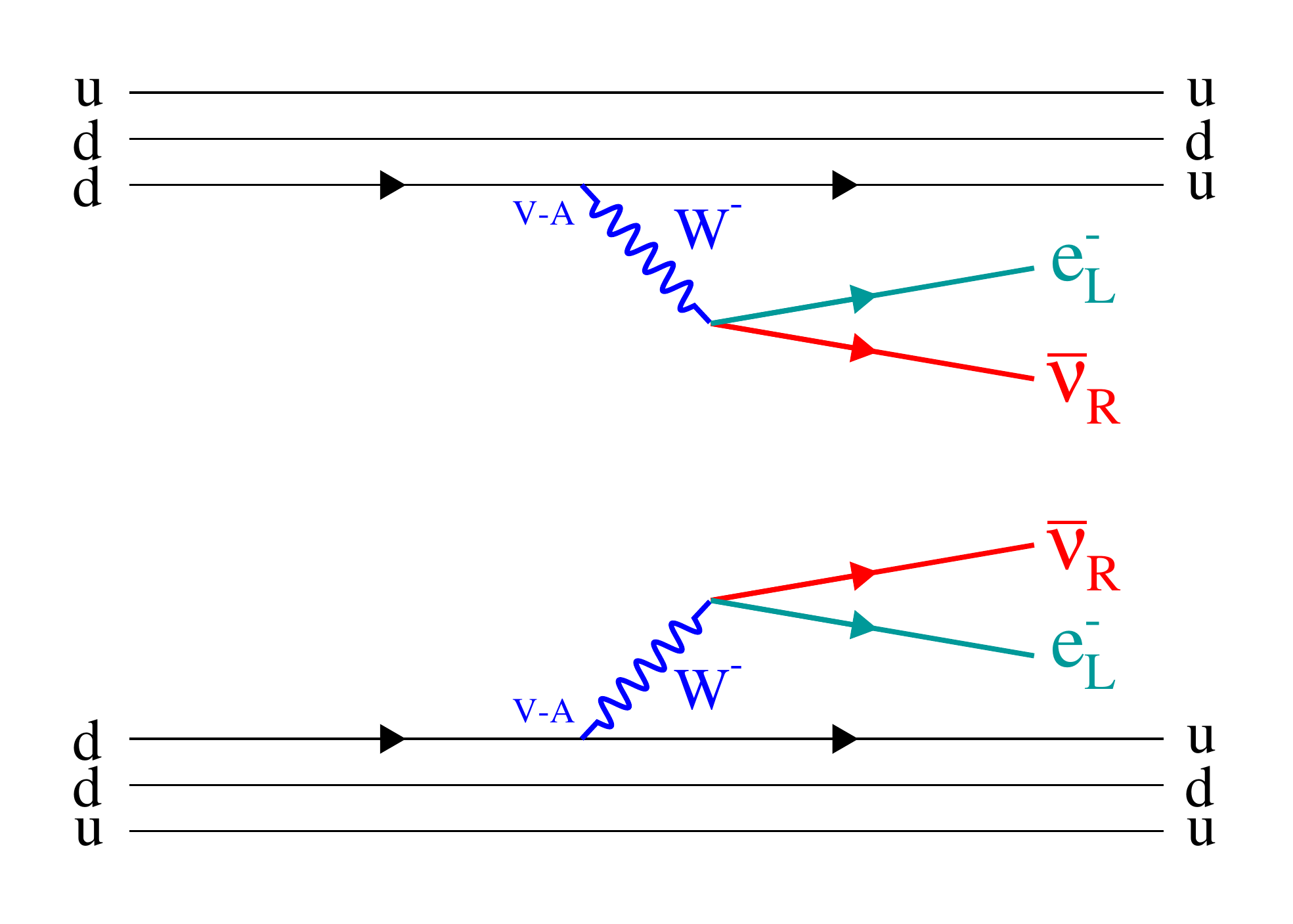}
  \includegraphics[scale=0.33]{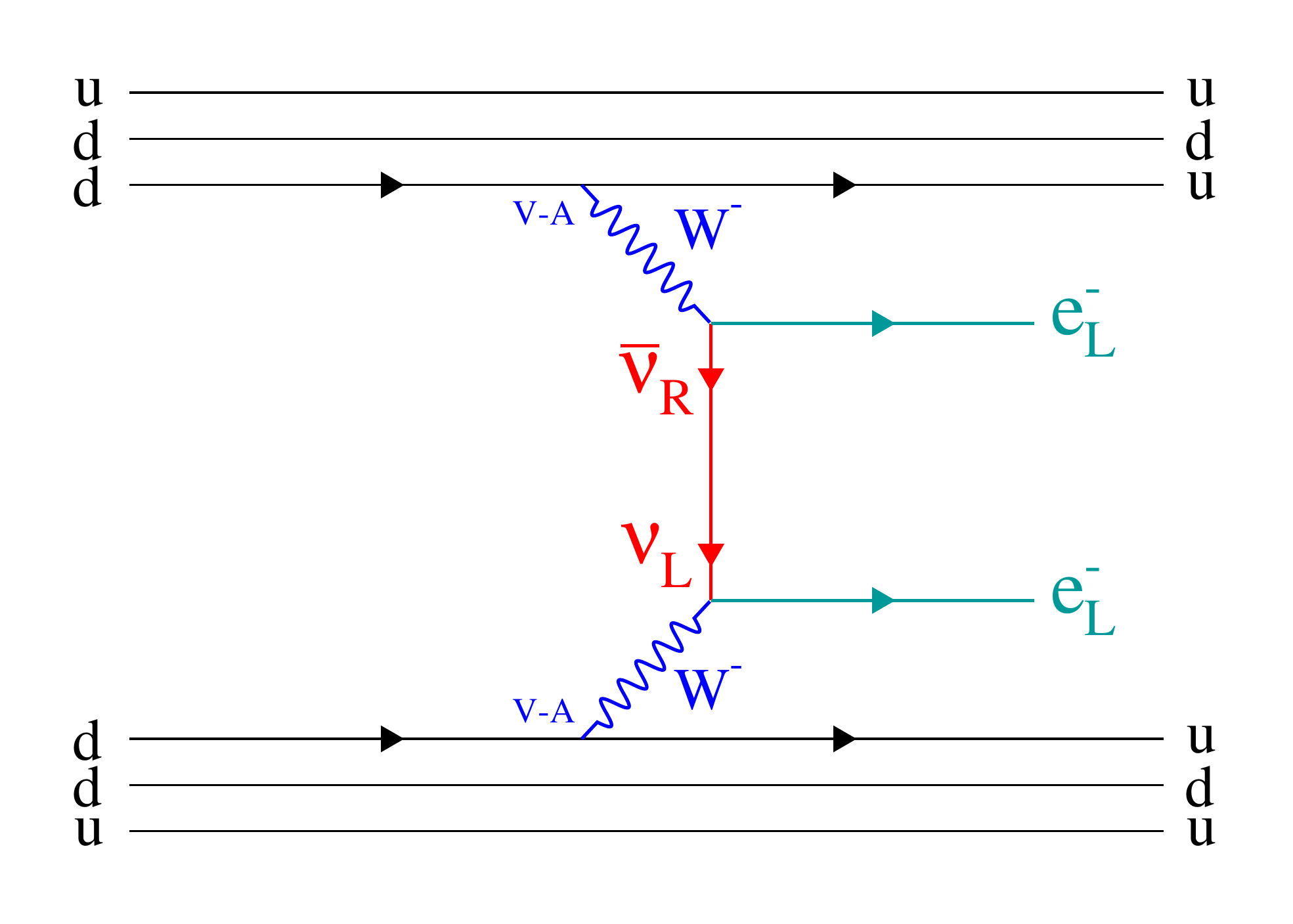}
  \includegraphics[scale=0.33]{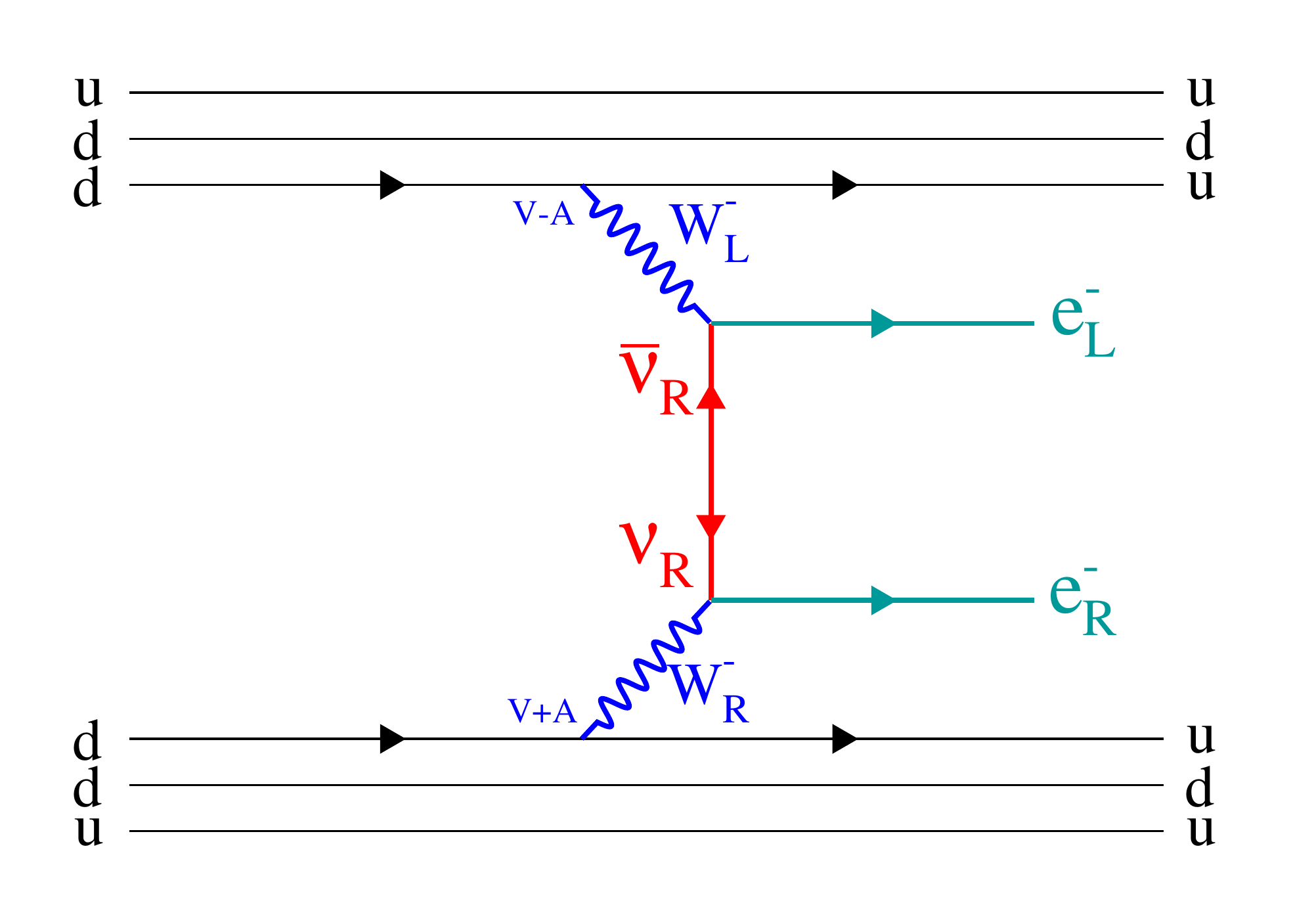}
  \includegraphics[scale=0.33]{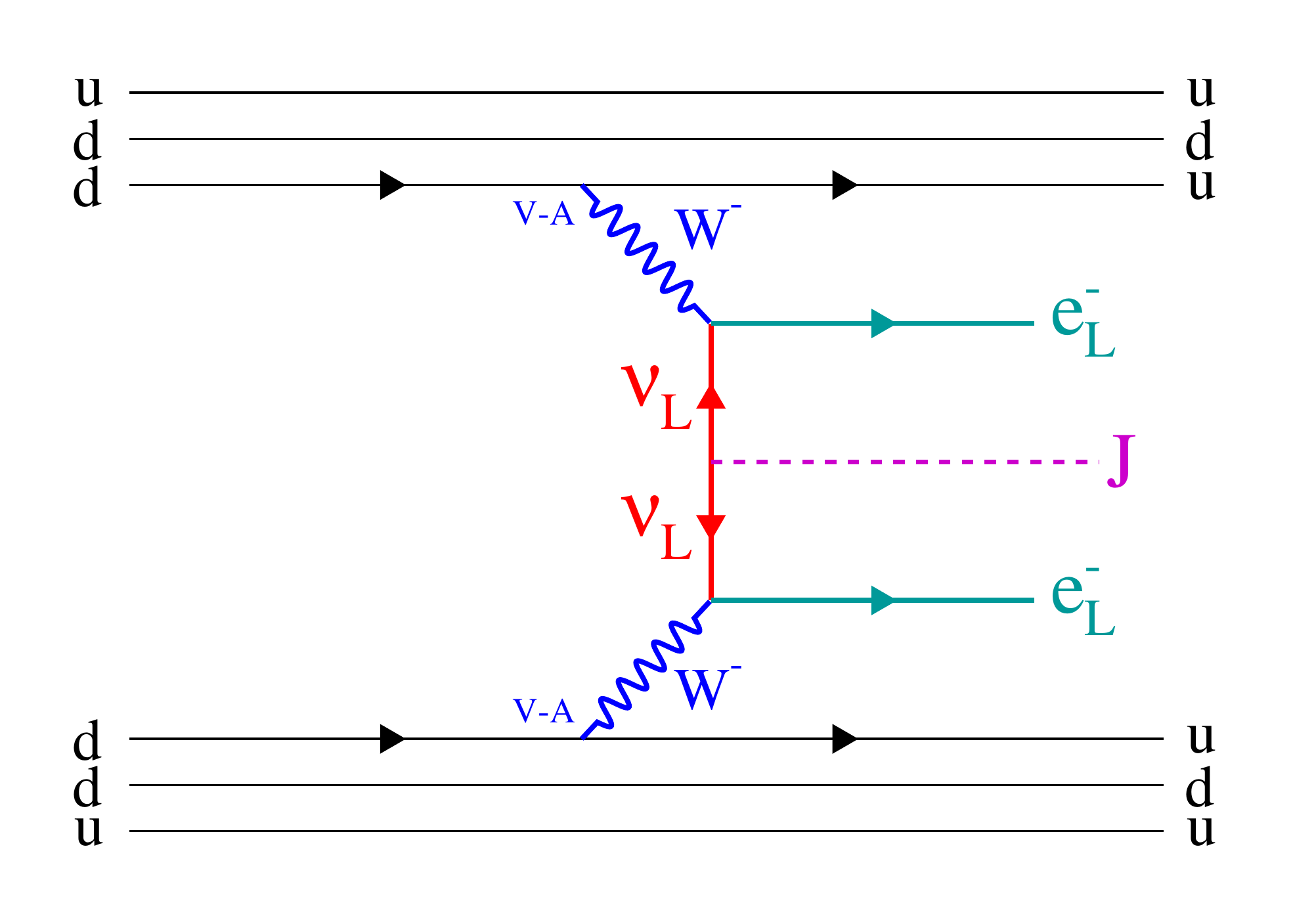}
  \caption{Feynman diagram of the double beta processes. Upper left: standard $\beta \beta$-decay with emission of two neutrinos; Upper right: neutrinoless $\beta \beta 0 \nu$-decay in the case of the exchange of a virtual Majorana neutrino; Lower left: $\beta \beta 0 \nu$-decay in the case of right handed weak coupling, Lower right: $\beta \beta 0 \nu$-decay in the case of a Majoron emission.}
  \label{fig:graph-bb}
\end{figure}

The standard double beta decay with emission of two electrons and two neutrinos ($\beta \beta 2 \nu$) is a second order process of $\beta$-decay, which is produced by isotopes whose $\beta$-decay is forbiden (due to a higher energy level of the daughter nuclei or due to angular momentum conservation). 
$$  (A,Z)  \rightarrow (A,Z-2) + 2e^- + 2\nu_e $$
The Feynman diagram is shown in Figure~\ref{fig:graph-bb}-a. This standard process is very rare but it has been already observed for 7 isotopes with a hal-life varying from about $7 \ 10^{18}$~years for $^{100}$Mo and $^{150}$Nd, to about $10^{21}$~years for $^{76}$Ge and $^{136}$Xe, and about $2 \ 10^{24}$~years for $^{128}$Te. Table~\ref{tab:current-limits} lists the double beta isotopes used in the search of the $\beta \beta 0 \nu$-decay and their measured $\beta \beta 2 \nu$ half-life. 

If we now consider that the neutrino is a Majorana particle, identical to its own antiparticle, then it becomes possible that the neutrino emitted at the first vertex of the $W$ boson decay is absorbed by the second $W$ vertex and thus only two electrons are emitted by the nucleus (the exchange neutrino is virtual). It corresponds to the neutrinoless double beta decay ($\beta \beta 0 \nu$) where two neutrons decay into two protons emitting only two electrons: 
$$  (A,Z)  \rightarrow (A,Z-2) + 2e^- $$
This process violates the lepton number by two units ($\Delta L = 2$), and is thus forbidden by the Standard Model. 

Experimentaly, the two decay modes, $\beta \beta 2 \nu$ and $\beta \beta 0 \nu$  are distinguished by the fact that the sum of the two electron energies in $\beta \beta 0 \nu$-decay is constant and equal to the transition
energy $Q_{\beta\beta}$ while it varies continuously in $\beta \beta 2 \nu$-decay up to the same energy as its limit (with a maximum around 1/3~$Q_{\beta\beta}$).

The $\beta\beta 0\nu$-decay was first proposed by Furry in 1939~\cite{furry}. We note that Furry expected a larger $\beta\beta 0\nu$-decay rate since only two particles are emitted in the $\beta\beta 0\nu$-decay, corresponding to a larger phase space factor. However, because of the absence of right-handed current in the weak interaction (not yet known at the time of Furry's calculation), the helicity of the neutrino emitted at the first vertex  must be flipped in order to be absorbed by the second vertex. It is possible since the neutrino is a massive particle, and therefore there is a tiny admixtures of opposite helicity in neutrino of order of $m_{\nu}/E$ where $m_{\nu}$ is the neutrino mass and $E$ its energy.
The $\beta\beta 0\nu$-decay rate is then strongly suppressed and is proportional to $m_{\nu}/E$.

\begin{figure}[!h]
  \centering
  \includegraphics[scale=0.4]{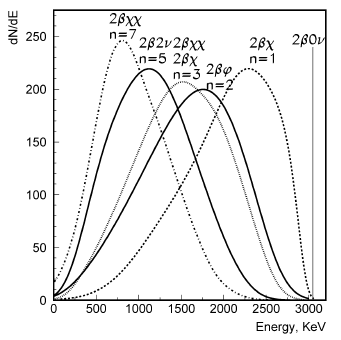}
  \caption{Energy spectra of different modes of $\beta\beta$-decay of $^{100}$Mo with the emission of Majoron.}
  \label{fig:majoron-spectra}
\end{figure}

The mechanism which as been described is only one possible process (considered as the standard process) but other mechanisms are possible and any source of lepton number violation can induce $\beta \beta 0 \nu$ decay and contribute to its amplitude.
For instance if there is a right-handed component in the weak interaction (V+A current), the right-handed neutrino emitted at the first vertex is directly absorbed via V+A current by the second $W$ boson without requirement of flip helicity (see Figure~\ref{fig:graph-bb}). 
In this case the expected angular distribution between the two emitted electrons is different from the distribution in the case of the standard mechanism. As it will be discussed later, only tracko-calo experiments like NEMO-3 and SuperNEMO can measure the angular distribution and therefore can distinguish  the two processes in case of a positive signal.
Another possible mechanism is the exchange of a supersymetric particle $\chi$, instead of a Majorana neutrino, with short-range or long-range $R$-parity violating SUSY contributions. 
See the complete review of possible mechanisms in~\cite{rodejohann}.
Finally another alternative mechanism is the emission of a Majoron, a goldstone (massless) boson related to the $L-B$ symetry breaking (see Figure~\ref{fig:graph-bb}). 
In this case, the $\beta\beta$ energy sum spectrum of the two electrons is expected to be distorded. The deformation depends on the spectral index (or number of emitted Majoron) as illustrated in Figure~\ref{fig:majoron-spectra}. 

\begin{figure}[!h]
  \centering
  \includegraphics[scale=0.4]{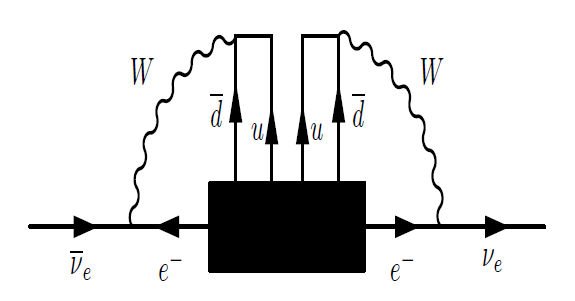}
  \caption{Black-box diagram equivalent to the $\beta\beta 0\nu$ diagram.}
  \label{fig:black-box}
\end{figure}

Even if many mechanisms are possible to produce $\beta \beta 0 \nu$-decay, any observation of  $\beta \beta 0 \nu$-decay would prove that the neutrino is a Majorana particle. Indeed all realizations of the $\beta \beta 0 \nu$-decay are connected to a Majorana neutrino mass via the black-box diagramm illustrated in Figure~\ref{fig:black-box}, or the Schechter-Valle theorem~\cite{schechter-valle}\cite{kayser}. The result is a $\bar{\nu}_e \rightarrow \nu_e$ transition, which is nothing but a Majorana mass term. However this is a tiny mass generated, at the 4-loop level and of the order of 10$^{-23}$~eV~\cite{duerr}, too small to explain the lower limit on the neutrino masses from the oscillation experiments. The neutrino mass obviously is generated via another mechanism. 

We should also note that there are similar processes called neutrinoless double beta$^+$ decay $(0\nu \beta^+ \beta^+)$, or $\beta^+$-decay electron capture $(0\nu \beta^+ EC)$, or double electron capture $(0\nu EC \ EC)$ of bound state electrons $e_b^-$, which can also be searched for:
$$  (A,Z)  \rightarrow (A,Z-2) + 2e^+  \ \ \  (0\nu \beta^+ \beta^+) $$
$$  e_b^- + (A,Z)  \rightarrow (A,Z-2) + e^+  \ \ \  (0\nu \beta^+ EC) $$
$$  2e_b^- + (A,Z)  \rightarrow (A,Z-2)^{\ast} \ \ \  (0\nu EC \ EC)   $$
Observation of one of those processes would also imply the non-conservation of the lepton number. However the creation of one or two positrons reduces the phase space factor. Therefore the $(0\nu \beta^+ \beta^+)$ or $(0\nu \beta^+ EC)$ decay rates are strongly reduced. It is also the case for $(0\nu EC \ EC)$ although the $^{152}$Gd-$^{152}$Sm transition has recently been identified as a possible interesting candidate~\cite{ec-ec}. 
We will focus in this review on the $\beta \beta 0 \nu$-decay. 

\section{Constraints from neutrino oscillations}

If we assume that the dominant lepton number violation mechanism at low energies is the light Majorana neutrino exchange, the half-life of $\beta \beta 0 \nu$-decay can be written as:
$$ \left( T_{1/2}^{0\nu} \right)^{-1} = G_{0\nu}(Q_{\beta\beta},Z) \left| M_{0\nu} \right|^2 \frac{\langle m_{ee} \rangle^2}{m_e^2}  $$
where $G_{0\nu}(Q_{\beta\beta},Z)$ is the phase space factor. It contains the kinematic information about the final state particles, and is exactly calculable to the precision of the input parameters 
(see Table~\ref{tab:g0nu-nme}). $\left| M_{0\nu} \right|$ is the nuclear matrix element, $m_e$ is  the mass of the electron, and $\langle m_{ee} \rangle$ is the effective Majorana  mass of the electron neutrino. It is defined as:
$$ \langle m_{ee} \rangle = \left| \sum_i U_{ei}^2 m_i \right| $$
where $m_i$ are the neutrino mass eigenstates and $U_{ei}$ are the elements of the neutrino mixing Pontecorvo-Maki-Nakagawa-Sakata (PMNS) matrix $U$. 

The PMNS matrix $U$ has been introduced to explain the observed neutrinos oscillations. Its formalism is that in the charged current term of electroweak interactions, the neutrino flavour states $\nu_e$, $\nu_{\mu}$ and $\nu_{\tau}$ are superpositions of neutrino mass states $\nu_1$, $\nu_2$ and $\nu_3$:
$$ \nu_{\alpha} =  U_{\alpha i}^{\ast} \nu_i $$ 
where $\alpha=e,\mu,\tau$ and $i=1,2,3$. The PMNS mixing matrix $U$ is unitary and can be written in its standar parametrization (see~\cite{rodejohann} or \cite{giunti} for instance) as
$$ U = \left(
\begin{array}{ccc}
c_{12}c_{13}                                 & s_{12}c_{13}                                & s_{13}e^{-i\delta}\\
-s_{12}c_{23} -c_{12}s_{23}s_{13}e^{i\delta} & c_{12}c_{23} -s_{12}s_{23}s_{13}e^{i\delta} & s_{23}c_{13}   \\
s_{12}s_{23} -c_{12}c_{23}s_{13}e^{i\delta}  & -c_{12}s_{23} -s_{12}c_{23}s_{13}e^{i\delta}& c_{23}c_{13}
\end{array} \right) P $$
where $c_{ij}=\cos \theta_{ij}$, $s_{ij}=\sin \theta_{ij}$ and $\delta$ is the possible {\it Dirac phase} which could produce a CP violation in neutrino oscillations. A diagonal phase matrix $P$ has been introduced, containing the two {\it Majorana phases} $\alpha$ and $\beta$ (physical if neutrinos are Majorana):
$$ P = diag(1,e^{i\alpha},e^{i(\beta+\delta)}) $$

\begin{figure}[!h]
  \centering
  \includegraphics[scale=0.4]{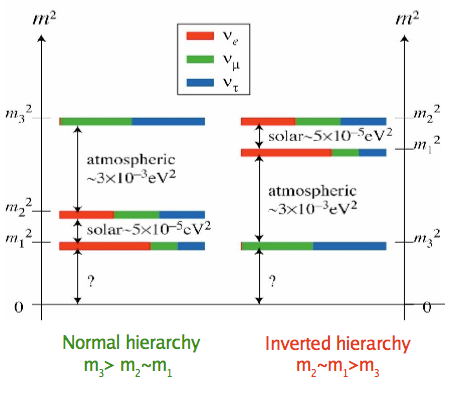}
  \caption{Patterns in the case of normal or inverted hierarchy of the eigenstate neutrino masses. Colors show the mixing of the flavour neutrino eigenstates inside the mass neutrino eigenstates.}
  \label{fig:neutrino-pattern}
\end{figure}

\begin{figure}[!h]
  \centering
  \includegraphics[scale=0.4]{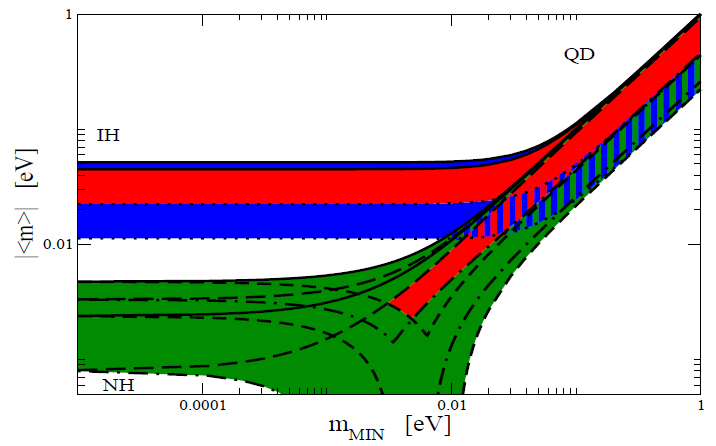}
  \caption{The predicted value of $\langle m_{ee} \rangle$ as a function of $m_{min}$, obtained using the $2\sigma$ allowed ranges of $\Delta m_{atm}^2$, $ \Delta m_{\odot}^2$, $\sin^2 \theta_{\odot}^2$ and $\sin^2 \theta_{13}^2$. The regions shown in red correspond to violation of CP-symmetry while the rest of the regions correspond to the four different sets of CP-conserving values of the two Majorana phases with the green region for normal hierarchy and blue region for inverted hierarchy.}
  \label{fig:petcov-2008}
\end{figure}

The effective Majorana mass of the electron neutrino can be written in term of the neutrino oscillation parameters:
$$ \langle m_{ee} \rangle = \left| \sum_i U_{ei}^2 m_i \right| = \left| m_1 c_{12}^2 c_{13}^2 + (m_2 s_{12}^2 c_{13}^2)e^{2i\alpha} + (m_3 s_{13}^2)e^{2i\beta} \right| $$
Results of the neutrino oscillation measurements provide constraints on the variation of the effective Majorana mass $\langle m_{ee} \rangle$ as a function of the neutrino mass scale.
However, the constraints depend on the neutrino mass pattern, which is unfortunatly not known. 
Indeed, neutrino oscillations are only sensitive to the absolute value of the difference of the square of the neutrino mass $\left| \Delta m^2 \right|$. Since we cannot measure the sign of $\Delta m_{atm}^2$ in the atmospheric neutrino oscillation, there are two possible patterns of the eigenstate neutrino masses, as illustrated in Figure~\ref{fig:neutrino-pattern}: 
\begin{itemize}
\item In the normal hierarchy (NH) case, the gap between the two lightest mass eigenstates corresponds to the small mass difference, measured by solar experiments. 
\item In the inverted hierarchy (IH) case, the gap between the two lightest states corresponds to the large mass difference, measured by atmospheric experiments. 
\item Finally, the particular case in which the neutrino mass differences are very small compared with their absolute scale dorresponds to quasi-degenerate (QD) pattern.
\end{itemize}
Figure~\ref{fig:petcov-2008} shows the predicted value of $\langle m_{ee} \rangle$ as a function of the smallest mass of neutrinos $m_{min}$~\cite{pascoli-petcov}, with the three possible neutrino mass patterns:
\begin{itemize}
\item In the case of normal hierarchy, $\langle m_{ee} \rangle$ can be as low as 1~meV. In the ``catastrophe scenario'',  $\langle m_{ee} \rangle$ can vanish due to cancelation of the coefficients by appropriate Majorana phases. 
\item In the case of inverted hierarchy, the expected range is $~10 \mathrm{meV} < \langle m_{ee} \rangle < 50 \mathrm{meV}$~\cite{dueck-rodejohann-zuber}. The lower limit depends upon the exact value of $\sin^2 \theta_{12}$ and on the Majorana phase pattern. In case of low $\sin^2 \theta_{12}$ (0.27) or in case of violation of CP-symetry, the lower limit can be 20~meV. 
\item In the case of quasi-degenerated neutrino mass pattern, larger values for  $\langle m_{ee} \rangle$ can be obtained, , approximately above 50 meV.
\end{itemize}
I remind that these constraints are valid only for the standard $\beta\beta 0\nu$ mechanism with the exchange of a virtual light Majorana neutrino. 
As previously said, other mechanisms can also contribute and thus can increase (positive interferences) or decrease (negative interferences) the $\beta\beta 0\nu$-decay rate.

In conclusion, let's try first to detect a $\beta\beta 0\nu$-decay, whatever the expectations !

\section{The nuclear matrix elements}

Theoretical uncertainties for the nuclear matrix element (NME) calculations are the main limitation in making interpretation of $\beta\beta 0\nu$-decay or in comparing the sensitivity (or hopefuly signal amplitude !)  between different experiments measuring different isotopes. 
We can list 5 methods to calculate the NME: the Quasi-particle Random Phase Approximation (QRPA, including its many variants and evolution steps)~\cite{qrpa-tubingen}\cite{qrpa-suhonen}, the Nuclear Shell Model (NSM)~\cite{shell-model}, the Interacting Boson Model (IBM)~\cite{ibm}, the Generating Coordinate Method (GCM)~\cite{gcm}, and the Projected Hartree-Fock-Bogoliubov model (PHFB)~\cite{phfb}.
Basicaly, QRPA uses a large number of active nucleons in a large space but with a specific type of correlation suited for collective motion, whereas NSM uses a small number of nucleons within a limited space.  

I would like to underline that the NME calculations for the standard $\beta \beta 2 \nu$-decay  and the $\beta \beta 0 \nu$-decay are very different. The $\beta \beta 2 \nu$ process involves only Gamow-Teller transitions through intermediate 1$^+$ states only because of low momentum transfer. In contrast, for $\beta \beta 0 \nu$-decay, the exchange Majorana neutrino has a relatively high momentum of about $q \approx 100$~MeV (corresponding to the average distance $r \approx 1/q \approx 1$~fm between the two decaying neutrons). Therefore the $\beta \beta 0 \nu$ process involves all the J$^+$ intermediate states and it is evaluated at two pointlike Fermi vertices containing a Fermi and a Gamow-Teller part. 

Important studies of various nuclear effects and impact on the NME calculation have been performed in the last years both in QRPA and in NSM. It has been sumarized in~\cite{elliot-2011}. I mention few examples:
\begin{itemize}
\item[\textbullet] The main contribution to NME comes from short internucleon distances ($r<2-3$~fm)~\cite{qrpa-short-range}, and the nucleons tend to overlap. Short-range correlations (SRC) now take the hard core repulsion into account. There are different methods to treat SRC: Jastrow-like function, Unitary Correlation Operator Method (UCOM) and Coupled Cluster Method (CCM). The Jastrow method leads to a reduction of the NME of $\approx 20\%$ while UCOM and CCM, which are favored models, reduce NME by $\approx 5\%$~\cite{qrpa-short-range}\cite{qrpa-tubingen}.
\item[\textbullet] The difference in the nucleon configuration of the initial and final nuclei is an important input to NME. Reactions such as (d,p),  ($\alpha$,$^3$He) and (d,$^3$He) were used to study the occupation of valence neutron or proton orbits in $^{76}$Ge and $^{76}$Se~\cite{ge76-n}\cite{ge76-p}. These data were then used to constrain calculations of NME by both the shell model~\cite{sm-ge76-np} and QRPA~\cite{qrpa-tubingen}\cite{qrpa-suho-ge76} methods. For the Shell Model, the value of the NME is enhanced by about 15$\%$ compared to previous calculations, whereas in the QRPA the NMEs are reduced by 20$\%$~\cite{sm-ge76-np}. This diminishes the discrepancies between both approaches. This points out the importance of spectroscopic information in order to test the validity of the NME calculations. 
\item[\textbullet] It is very difficult to calculate the relative strengths of the virtual J$^+$ states of the intermediate nuclei involved in the $\beta\beta 0 \nu$ decay. Muon capture on the final nucleus also excites all these states~\cite{muon-capture}, and therefore provides additional experimental data for the theoretical calculations.
\item[\textbullet] Accurate measurement of the $\beta\beta 2 \nu$ half-lives along with electron capture of the intermediate nuclei can help determine the $g_{pp}$ parameter used in QRPA (particle-particle interaction strength). 
\item[\textbullet] Finally, being able to calculate the $\beta\beta 2\nu$ half-life can be also considered a necessary, but not a sufficient condition for demonstrating a correct $\beta\beta 0\nu$ NME calculation.
\end{itemize}

A complete compilation of the most recent NME calculations (with the 5 different techniques) have been recently proposed in~\cite{dueck-rodejohann-zuber}.  It has been scaled to a radius $r=1.2$~fm and to a free nucleon axial-vector coupling constant $g_A=1.25$ (QRPA NME are reduced by about 15$\%$ if one use a quenched constant $g_A=1$~\cite{qrpa-tubingen}). 
In the following I will use this compilation which is summarized in table~\ref{tab:g0nu-nme}.

We see that the NME's can vary by a factor 2 to 3. In most cases the results of the Shell Model calculations are the smallest ones, while the largest ones may come from the IBM, QRPA or GCM. It results to a factor of uncertainty of about 4 to 10 on the required sensitivity to $T_{1/2}^{0\nu}$ or on the required mass of isotope, when we compare different experiments using different isotopes. 

Finally, all these calculations discussed here have been performed for the standard mechanism with the exchange of a virtual light Majorana neutrino. Rare and old calculations have been done in the case of V+A mechanism or in the case of exchange of SUSY particles with R-parity violation.

\section{Current limits on the effective mass and required sensitivities for the future}

We list in table~\ref{tab:current-limits} the current limits on the $\beta\beta 0\nu$ half-life $T_{1/2}^{0\nu}$, obtained with various experiments. 
Except KAMLAND-Zen, all the other experiments are finished.
4 experiments have reached the ``10$^{24}$ club'', it means they have set a limit on $T_{1/2}^{0\nu}$ higher than 10$^{24}$~years: Heidelberg-Moscow/IGEX with about 10~kg of $^{76}$Ge, NEMO-3 with about 7~kg of $^{100}$Mo, CUORICINO with about 10~kg of $^{130}$Te and recently Kamland-Zen with about 300~kg of $^{136}$Xe. 
The corresponding lower and upper limits on the effective Majorana mass are in the same range, about 0.2 and 0.7~eV, respectively. 
We note that, although the limit obtained on $T_{1/2}^{0\nu}$ with $^{76}$Ge is a factor 10 higher than for $^{100}$Mo or $^{130}$Te, the limit on the effective neutrino mass is in the same range. It is mostly due to a lower phase space factor for  $^{76}$Ge. 

The observation of a $\beta\beta 0\nu$ signal in Ge, claimed by a part of the Heidelberg-Moscow collaboration, will be discussed in Chapter~\ref{chap:germanium} (germanium experiments).
The reader can also refer to V.I. Tretyak's paper~\cite{false-discoveries}, which summarizes all the ``discoveries'' of $\beta\beta$ decays (including $\beta\beta 0\nu$) which were disproved in the subsequent investigations.

\begin{table}[!h]
\centering
\begin{tabular}{ccccccc}
\hline
\hline
Isotope & $T_{1/2}^{2\nu}$ (yr) & Experiment & $T_{1/2}^{0\nu}$ (yr)   & Experiment  & \multicolumn{2}{c}{$\langle m_{ee} \rangle$ (eV)} \\
           &                    &            & (90$\%$ C.L.)           &             & Min.     &  Max.   \\
           &                    &            &                         &             &          &         \\
\hline
$^{48}$Ca  & $4.2_{-1.0}^{+2.1}$ 10$^{19}$ & NEMO-3 & 5.8 10$^{22}$ &  CANDLES~\cite{candle}    & 3.55      &  9.91  \\
$^{76}$Ge  & $1.5 \pm 0.1 \ 10^{21}$ & HDM & {\bf 1.9 10$^{25}$} &  HDM~\cite{klapdor-2001}        & {\bf 0.21} &  {\bf 0.53}  \\
$^{82}$Se  & $9.0 \pm 0.7 \ 10^{19}$ & NEMO-3     & 3.2 10$^{23}$ &  NEMO-3~\cite{nemo3-taup11}     & 0.85      &  2.08  \\
$^{96}$Zr  & $2.0 \pm 0.3 \ 10^{19}$ & NEMO-3     & 9.2 10$^{21}$ &  NEMO-3~\cite{nemo3-zr}     & 3.97      &  14.39 \\
$^{100}$Mo & $7.1 \pm 0.4 \ 10^{18}$ & NEMO-3     & {\bf 1.0 10$^{24}$} &  NEMO-3~\cite{nemo3-taup11}     & {\bf 0.31} &  {\bf 0.79}  \\
$^{116}$Cd & $3.0 \pm 0.2 \ 10^{19}$ & NEMO-3     & 1.7 10$^{23}$ & SOLOTVINO~\cite{danevich-2003}   & 1.22      &  2.30    \\
$^{130}$Te & $0.7 \pm 0.1 \ 10^{21}$ & NEMO-3     & {\bf 2.8 10$^{24}$} & CUORICINO~\cite{cuoricino-final-result}   & {\bf 0.27} &  {\bf 0.57}  \\
$^{136}$Xe & $2.38 \pm 0.14 \ 10^{21}$ & Kamland & {\bf 5.7 10$^{24}$} & Kamland-Zen~\cite{kamland-zen} & {\bf 0.25} & {\bf 0.6} \\
$^{150}$Nd & $7.8 \pm 0.7 \ 10^{18}$ & NEMO-3     & 1.8 10$^{22}$ & NEMO-3~\cite{nemo3-nd}      & 2.35      &   8.65   \\
\hline
\hline
\end{tabular}
\caption{$\beta\beta 2 \nu$ half-lives and $\beta\beta 0 \nu$ half-life limits measured in a variety of experiments. Last two rows show the minimal and maximal upper limits on the effective majorana neutrino mass $\langle m_{ee} \rangle$, using NME's from table~\ref{tab:g0nu-nme}.}
\label{tab:current-limits}
\end{table}

\begin{figure}[!h]
  \centering
  \includegraphics[scale=0.3]{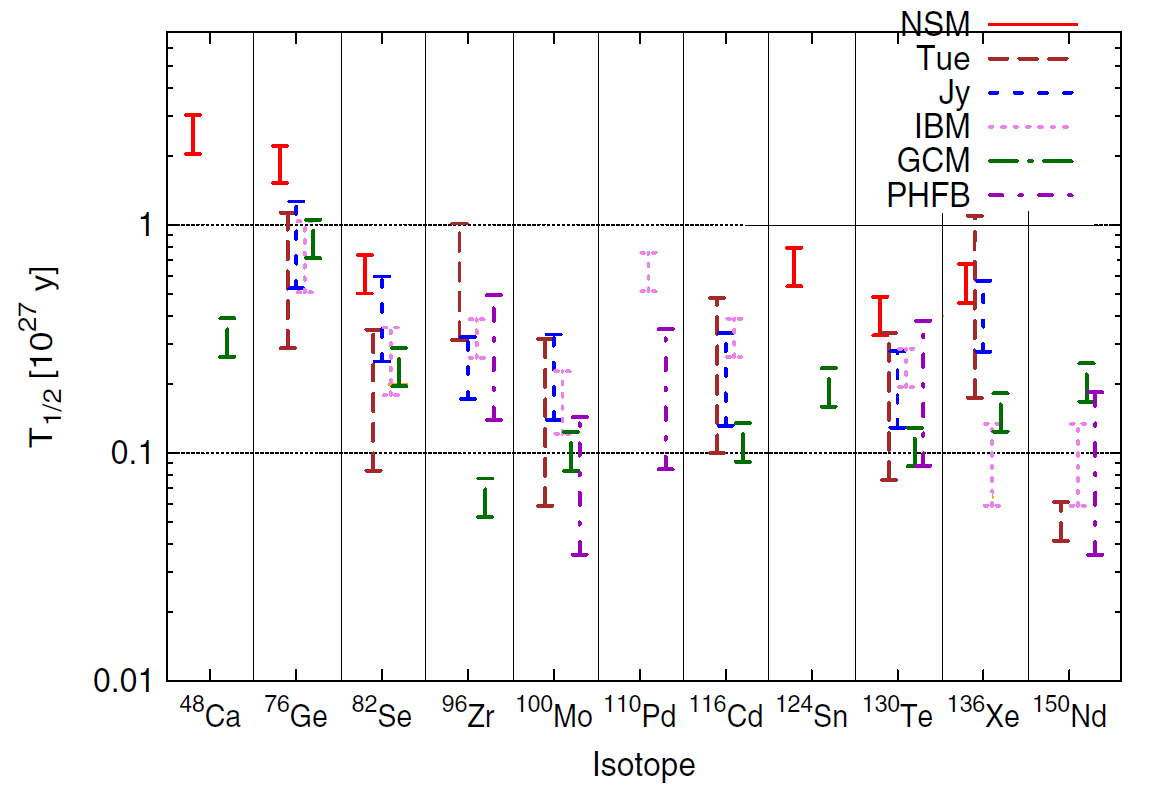}
  \caption{Required half-life sensitivities in order to start exploring the inverted hierarchy, corresponding to an effective Majorana neutrino mass of about 50~meV (from~\cite{dueck-rodejohann-zuber}).}
  \label{fig:sensitivity-50mev}
\end{figure}

Figure~\ref{fig:sensitivity-50mev} shows the required sensitivity on $T_{1/2}^{0\nu}$ in order to start exploring the inverted hierarchy region, corresponding to an upper limit of $\langle m_{ee} \rangle = 50$~meV. For a given isotope, there is an uncertainty of $\approx 5$ to 10 due to the NME's, as discussed above. But roughly speaking, we see that a sensitivity of $\approx 10^{26}$~years for $^{100}$Mo and up to $\approx 10^{27}$~years for $^{76}$Ge is needed.


\section{Experimental aspects}


In case of no signal, the half-life sensitivity for a $\beta\beta 0\nu$ experiment is given by:
$$ T_{1/2}^{0\nu} >  \ln 2 \frac{\mathcal{N} M \varepsilon}{A}\frac{T_{obs}}{N_{excl}} $$
where $M$ is the mass of enriched isotope, $A$ its atomic mass, $\mathcal{N}$ the Avogadro, $\varepsilon$ the $\beta\beta 0 \nu$ efficiency of the detector, $T_{obs}$ is the duration of the measurement and $N_{excl}$ is the number of excluded $\beta\beta 0 \nu$ events.

From this relation, and due to the low rates of the $\beta\beta 0 \nu$ process, it is clear that the essential requirement of the double beta experiments is to achieve an extremely low radioactive background and large mass of isotopes. 

It is also important to note that in the case of no background (a dream !), the half-life sensitivity increases as the duration of observation $T_{obs}$, while in the case of relatively larger background, it increases only as $\sqrt{T_{obs}}$. 

A large variety of experimental techniques have been developped for the search of $\beta\beta 0 \nu$ decay. 
There are basicaly two types of approaches: the calorimetric and the tracko-calo methods.

In the calorimetric technique, the source is embedded in the detector itself which provides a high detection efficiency. 
With a proper choice of detector, a very high energy resolution up to FWHM=0.1$\%$ at $Q_{\beta\beta}$ can be achieved as in Germanium or bolometer detectors. 
However it is difficult to reconstruct the event topology and to identify the background components (with the exception of Xenon TPC but at the price of a lower energy resolution). The capacity to prove that the observation is indeed a $\beta\beta 0 \nu$-decay, and not an unidentified background, is then limited. 
Recently large existing liquid scintillator detectors, initialy developped for neutrino oscillation measurements (Kamland, SNO), have been reused as $\beta\beta$ detectors by adding isotope inside the liquide scintillator. It allows to reach quickly a large amount of isotope ($\approx 100$~kg) but with a limited energy resolution and thus a non negligible background.

The tracko-calo method separates the detector from the source and the detector combines a calorimeter and a tracking detector. It allows to reconstruct directly the track of each of the two emitted electrons from the source foil and also to identify and measure each background component. However the price is a lower $\beta\beta 0 \nu$ efficiency and a lower energy resolution.
We mention that it is also the most sensitive technique for the search of the $\beta \beta 0 \nu$ with V+A right-handed weak current, since it provides the angular distribution between the two emitted electrons.

We emphasize that the search of $\beta\beta 0 \nu$-decay requires several experimental techniques and more than one isotope.
This is because there could be unknown background and gamma transitions, and a line observed at the end point in one isotope does not necessarily imply that $\beta\beta 0\nu$ decay was discovered. Nuclear matrix elements are also not very well known. 

\vspace{1cm}

In the next chapters, I will review the different $\beta\beta$ experimental approachs. My purpose is to detail, for each technique, the different origins of background, how they can be identified, and how they can be reduced. Advantages and limitations will be discussed.

I will first present the tracko-calo NEMO-3 and SuperNEMO experiments. I work on these two experiments since 15 years. So it was natural to start with them with a relatively more exhaustive description. I will then present the germanium technique. This is today the most sensitive technique on the $\beta\beta 0\nu$ half-life. I will then review the bolometer technique. I will describe in detail the recent progress in scintillating bolometers because I think that it is one of the most promising technique. Finally I will review the large liquid scintillator detectors and Xenon TPC. 

The reader who has not the courage to read this entire review, is allowed to go directly to the summary given in Chapter~\ref{chap:summary}.

\begin{landscape}

\begin{table}[!h]
\centering
\begin{tabular}{cccccccccc}
\hline
\hline
        &                        &                   &             &  \multicolumn{6}{c}{Nuclear Matrix Elements} \\
Isotope & $G^{0\nu}$             & $Q_{\beta\beta}$  & Nat. ab. &  NSM~\cite{shell-model}\cite{sm-ge76-np} & QRPA-Tu.~\cite{qrpa-tubingen} & QRPA-Jy~\cite{qrpa-suhonen} & IBM~\cite{ibm} & GCM~\cite{gcm} & PHFB~\cite{phfb} \\
        & (10$^{-14}$~y$^{-1}$) & (keV)             & ($\%$)      &  (UCOM) & (CCM) & (UCOM) & (Jastrow) & (UCOM) & (Mixed) \\
\hline
$^{48}$Ca  &  6.35  & 4273.7  & 0.187 & 0.85 & ...       & ...         & ...     & 2.37 & ...        \\
$^{76}$Ge  &  0.623 & 2039.1  & 7.8   & 3.26 & 4.44-7.24 & 4.19-5.36 & 4.64-5.47 & 4.6  & ...        \\
$^{82}$Se  &  2.70  & 2995.5  & 9.2   & 2.64 & 3.85-6.46 & 2.94-3.72 & 3.80-4.41 & 4.22 & ...        \\
$^{96}$Zr  &  5.63  & 3347.7  & 2.8   & ...  & 1.56-2.31 & 2.76-3.12 & 2.53      & 5.65 & 2.24-3.46  \\
$^{100}$Mo &  4.36  & 3035.0  & 9.6   & ...  & 3.17-6.07 & 3.10-3.93 & 3.73-4.22 & 5.08 & 4.71-7.77  \\
$^{110}$Pd &  1.40  & 2004.0  & 11.8  & ...  & ...       & ...         & 3.62    & ...  & 5.33-8.91  \\
$^{116}$Cd &  4.62  & 2809.1  & 7.6   & ...  & 2.51-4.52 & 3.00-3.94 & 2.78      & 4.72 & ...        \\
$^{124}$Sn &  2.55  & 2287.7  & 5.6   & 2.62 & ...       & ...         & ...     & 4.81 & ...        \\
$^{130}$Te &  4.09  & 2530.3  & 34.5  & 2.65 & 3.19-5.50 & 3.48-4.22 & 3.37-4.06 & 5.13 & 2.99-5.12  \\
$^{136}$Xe &  4.31  & 2461.9  & 8.9   & 2.19 & 1.71-3.53 & 2.38-2.80 & 3.35      & 4.2  & ...        \\
$^{150}$Nd &  19.2  & 3367.3  & 5.6   & ...  & 3.45      & ...       & 2.32-2.89 & 1.71 & 1.98-3.7   \\
\hline
\hline
\end{tabular}
\caption{A  compilation of the most recent NME calculations, from~\cite{dueck-rodejohann-zuber} (NME have been normalized with $r=1.2$~fm).}
\label{tab:g0nu-nme}
\end{table}

\end{landscape}

%
%

\chapter{NEMO Tracko-Calo Experiments}
\label{chap:nemo}

At the end of the 80's, it was decided in France to start developping a serie of tracko-calo detectors, so-called NEMO detectors. The $\beta\beta$ sources are in the form of very thin and large foils and are separated from the detector. The combination of a tracking detector and a calorimeter provides both the measurement of the $\beta\beta$ energy spectrum and the direct reconstruction of the tracks of the two emitted electrons from the source foil. The efficiency to reject the background is therefore very high. However the energy resolution and the efficiency to detect a possible $\beta\beta 0\nu$ signal are relatively low compared to pure calorimetric detectors. Moreover the size of the detector must be relatively large in order to contain a large mass of $\beta\beta$ source foils. 

I will first present the result of the NEMO-3 experiment which took data in the Modane underground laboratory (LSM) from early 2003 up to january 2011 and measured several double beta isotopes for a total mass of $\approx 10$~kg. The two main isotopes for the $\beta\beta 0 \nu$ search were $^{100}$Mo (35~kg.y of exposure) and $^{82}$Se (4.5~kg.y of exposure). 

I will then present the SuperNEMO project which is based on an extension and an improvement of the experimental techniques used in NEMO-3.

%
%

\section{NEMO-3 Experiment}


\subsection{Description of the detector}

\begin{figure}[!h]  
\centering
\includegraphics[scale=0.4]{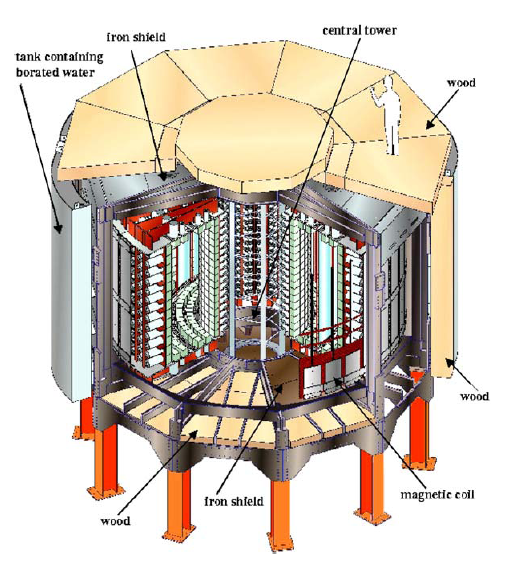}
\caption{An exploded view of the NEMO-3 detector. Note the
coil, iron g-ray shield, and the two different types of neutron
shields, composed of water tanks and wood. The paraffin shield
under the central tower is not shown on the picture.}
\label{fig:nemo3-layout}
\end{figure}

The NEMO-3 detector is cylindrical in design and divided into 20 equal sectors. Figure~\ref{fig:nemo3-layout} shows a schematic view of the detector. 

The source foils are in the form of very thin strips (40 to 60~mg/cm$^2$) and are fixed vertically. It corresponds to a large cylinder of 3.1m in diameter and 2.5m in height ($\approx 20$~m$^2$). 
Different sources of double beta emitters have been installed. 
The two main isotopes are $^{100}$Mo (6914~g, 12 sectors) and $^{82}$Se (932~g, 2.5 sectors) and are devoted for the $\beta \beta 0 \nu$ search.
Other isotopes have been also added in relatively smaller mass for the $\beta \beta 2 \nu$ measurement: 
$^{116}$Cd (405g, 1 sector), $^{130}$Te (454~g, 2 sectors), $^{150}$Nd (37~g), $^{96}$Zr (9~g) and $^{48}$Ca (7~g).
Also 1.5 sectors equipped with natural tellerium (614~g of TeO$_2$) and 1~sector equipped with pure copper (621~g) are used for external background measurements. 

On both sides of the sources, there is a gaseous tracking detector which consists of 6180 open drift cells operating in the Geiger mode allowing three-dimensional track reconstruction.
To minimize the multiple scattering, the gas is a mixture of 95$\%$ helium, 4$\%$ ethyl alcohol, 1$\%$ argon, and
0.1$\%$ water. 
The wire chamber is surrounded by a calorimeter which consists of 1940 plastic scintillator blocks coupled to very low radioactive photomultipliers (PMT's).
The energy resolution (FWHM) of the calorimeter is about 15$\%$ at 1 MeV for the scintillators equipped with the 5'' PMT's on the external wall and 17$\%$ for the 3'' PMT's on the internal wall. 
The resolution of the summed energy of the two electrons in the $\beta\beta 0 \nu$ decay is mainly a convolution of the energy resolution of the calorimeter and the fluctuation in the electron energy loss in the foil source which gives a non-Gaussian tail. 
The FWHM of the expected two-electron energy spectrum of the $\beta\beta 0 \nu$ decay is 350 keV. 
Timing information of the PMT's signals is used to discriminate between external particles crossing the detector and internal particles emitted from the source foils, allowing background studies and rejection. 
The time resolution is around 250~ps (r.m.s.) for 1~MeV electrons. 

A solenoid surrounding the detector produces a 25 gauss magnetic field in order to distinguish electrons from positrons with an efficiency of about 95$\%$ for 1~MeV electrons.

An external shield of 19~cm of low radioactivity iron, a water shield, and a wood shield cover the detector to reduce external $\gamma$'s and neutrons.

We started taking data in February 2003. But we observed a Radon contamination inside the tracking detector, ten times too high. A quick analysis showed that it was due to a diffusion of external Radon present in the lab. Thus, end of 2004, a Radon-tight structure surrounding the detector has been installed and free-Radon air passing through charcoal at low temperature (-50°C) has been flushed inside this buffer volume in order to reduce the observed Radon contamination inside the detector. 
Data taken between February 2003 until september 2004 correspond to Phase 1 with Radon background.
Data taken from January 2005 until January 2011 correspond to Phase 2 with low Radon background.

\begin{figure}[!h]  
\centering
\includegraphics{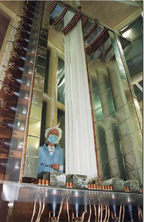}
\includegraphics{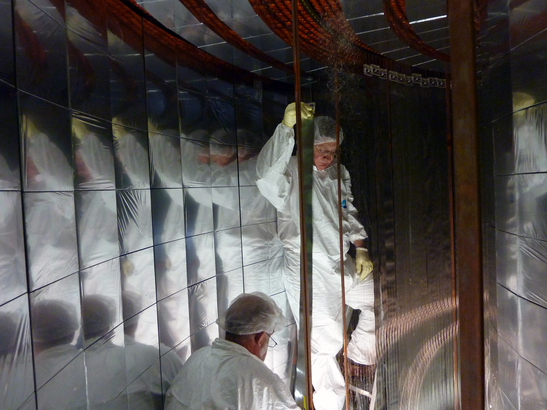}
\caption{(Left) View of a sector inside the LSM clean room during the assembly of the source foil. Note the double beta foil strips (in white) and the copper calibration tube besides, the scintillator blocks wrapped with aluminized mylar, the PMT's in the back surrounding by black tubes and the copper rings on the top and bottom of the tracking Geiger cells (the wires are too thin to be seen here. (Right) view inside of the detector during the dismounting of the sources foils in April 2011.}
\label{fig:nemo3-photo-inside}
\end{figure}

\begin{figure}[!h]  
\centering
\includegraphics[scale=4.]{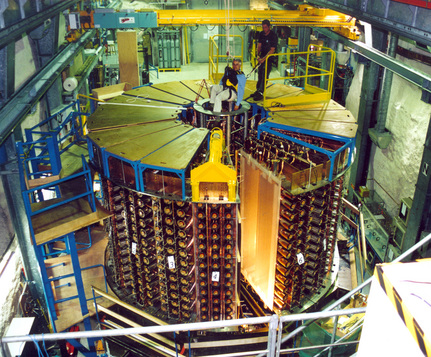}
\includegraphics{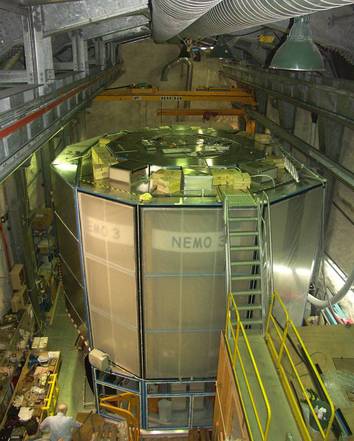}
\caption{(Left) View of the NEMO-3 detector in the LSM before the installation of the last sector in August 2000. (Right) View of the complete detector in LSM with the Radon-tight structure surrounding the detector.}
\label{fig:nemo3-photo}
\end{figure}

Some pictures of the detector are given in Figures~\ref{fig:nemo3-photo-inside} and \ref{fig:nemo3-photo}.
A complete description of the NEMO-3 detector is given in the {\it Technical design and performance} article in reference~\cite{nemo3-tdr}. More detailed informations can also be found in~\cite{nemo3-hdr-coco}.

The trigger condition requires at least 1~PMT with an energy above 150~keV and three Geiger cells fired. With these conditions, the trigger rate was only about 5~Hz. 

\begin{figure}[!h]  
\centering
\includegraphics[scale=0.4]{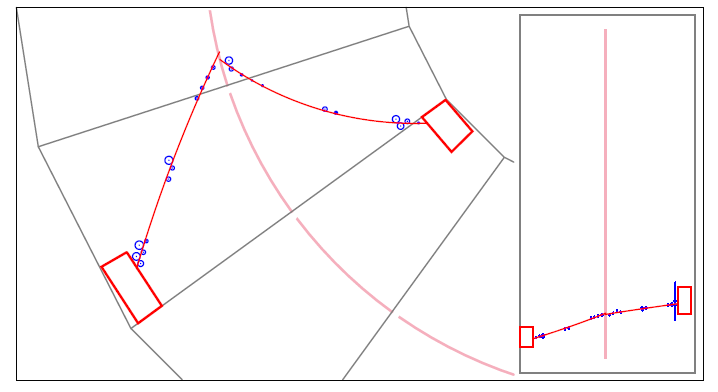}
\caption{Transverse (left) and longitudinal (right) view of a reconstructed $\beta\beta 0\nu$ candidate event selected from the data with a two-electron energy sum of 2812~keV.}
\label{fig:nemo3-event}
\end{figure}

A two-electron (2e$^-$) event (see Figure~\ref{fig:nemo3-event}) candidate for a $\beta\beta$ decay is defined as follows: two tracks come from the same vertex on the source foil, each track must be associated with a fired scintillator, its curvature must correspond to a negative charged particle emitted from the source, and the time of flight must correspond to the two electrons being emitted from the same source position. For each electron an energy threshold of 200~keV is applied.

\subsection{Energy, timing and efficiency calibrations}

\begin{figure}[!h]  
\centering
\includegraphics[scale=0.4]{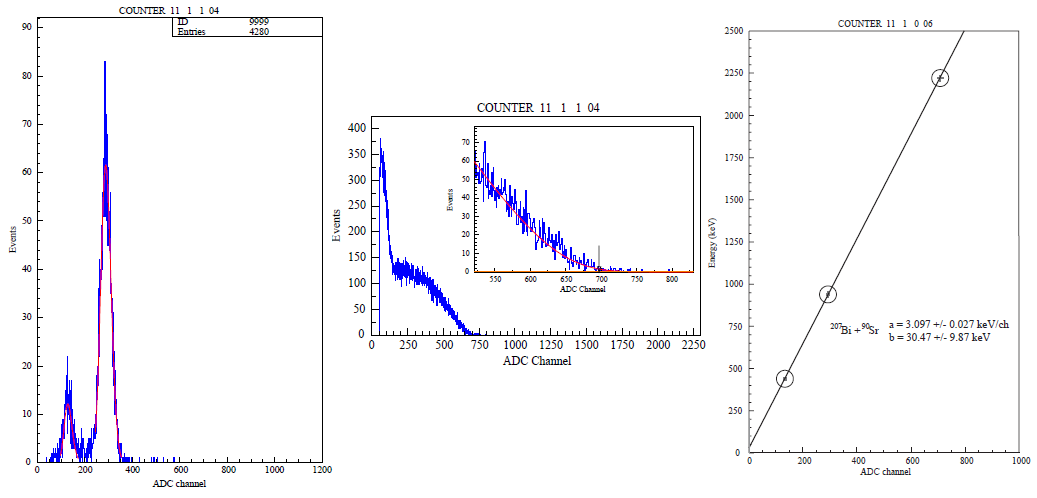}
\caption{(Left) Spectral fit of the 482 and 976~keV $\gamma$-rays coming from $^{207}$Bi decays for one counter ($\approx 3$~keV/channel); (Center) Energy spectrum of the $\beta$-decay of $^{90}$Sr (low energy) and $^{90}$Y with the full spectrum and in zoom the fit to the high-energy tail (end-point) of the spectrum, which is made with a function describing the shape of a single $\beta$ spectrum of $^{90}$Y, convolved with the energy resolution function of the scintillator bloc  and taking into account the mean energy loss of the electrons; (Right) the linear fit of the three calibration points with the energy offset at null charge of $\approx 30$~keV.}
\label{fig:nemo3-calib}
\end{figure}

Absolute energy calibrations were carried out every $\approx 40$ days using $^{207}$Bi sources which provides  conversion electrons of 482 and 976~keV (K lines) with a branching ratio of 1.5 and 7.0$\%$, respectively. A calibration run took about 24~hours to collect enough statistics. 
Figure~\ref{fig:nemo3-calib} shows an example of energy spectrum obtained for one counter after a calibration run.  
A dedicated long run at the begining of the NEMO-3 was also performed using $^{90}$Sr source. The measurement of the end-point of the $\beta$ spectrum of $^{90}$Y (daughter of $^{90}$Sr) at 2.283~MeV gave one additional calibration point to control the linearity. 
We mention also that the integration of all the $^{207}$Bi calibration runs over the 7 years data taking gave enough statistics to observe also the rare 1682~keV conversion electron (branching ratio of only 0.02$\%$) in $^{207}$Bi. It allowed to check a posteriori the energy linearity up to 1.7 MeV of each counter during all the period of NEMO-3 running.

The calibration lines obtained with the two $^{207}$Bi peaks as well as the fit combining $^{207}$Bi and $^{90}$Sr results do not necessarily intersect the origin of the axes. The extrapolated energy offset at ADC equal to zero (electronic pedestal subtracted) is in average 33~keV (after impact correction, see below) with an uncertainty of about 3~keV. This effect was previously observed with data obtained with an electron spectrometer during the test and assembly of the scintillator blocks of NEMO-3. It is due to a non linearity of plastic scintillator at low energy, below about 100~keV (quenching of low energy electrons), already observed and reported in litterature in other experiments. 

\begin{figure}[!h]  
\centering
\includegraphics[scale=0.4]{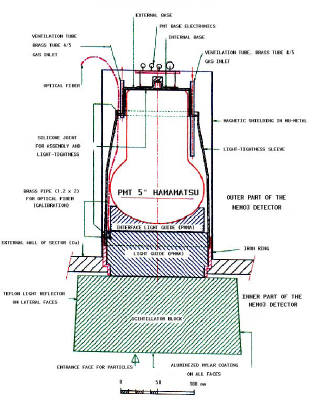}
\caption{Layout of an external scintillator block coupled via to successive optical guides to a 5'' PMT.}
\label{fig:nemo3-block-layout}
\end{figure}

By adding $^{207}$Bi calibration runs, one can also measure the response of each scintillator block to 1~MeV electron as a function of the impact position of the electron track on the entrance surface of the scintillator. This effect was previously observed with data obtained with the electron spectrometer in CENBG Bordeaux during the NEMO-3 calorimeter assembly. The impact corrections are relatively weak for the scintillator blocks equipped with 3'' PMT's, typicaly $1-2\%$ (with $3 \times 3$ corrections points), but is stronger for those equipped with 5'' PMT's, up to $10\%$ (with $5 \times 5$ corrections points). This effect has a non-negligible consequence on the energy resolution and is thus corrected offline by applying the impact correction factors measured for each scintillator block. 
This non uniform response of the block is due to the mechanical design of the coupling of the PMT to the scintillator as shown in Figure~\ref{fig:nemo3-block-layout}.  
The optical guide is cylindrical in design with a diameter equal to the PMT one. However the scintillator block is rectangular in design with a size (side) almost twice larger.
As it will be discussed in the next section (relating to SuperNEMO), the limited energy resolution of NEMO-3 is also strongly correlated to this design. 

\begin{figure}[!h]  
\centering
\includegraphics[scale=0.4]{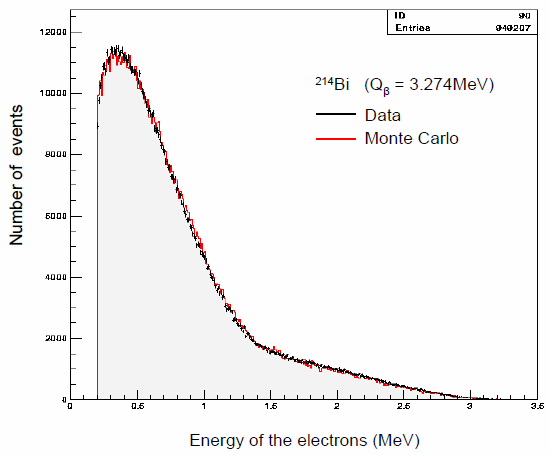}
\caption{Energy spectrum of the $^{214}$Bi $\beta$-decay from the wires measured with the ($e^-$, delay-$\alpha$)~\cite{arnold-note-1}. The black histogram corresponds to the data and the red one to the Monte-Carlo simulations.}
\label{fig:nemo3-bi214-spectrum}
\end{figure}

We mention also an independant test which has been performed in order to verify the energy calibration from 2 to 3.3~MeV~\cite{arnold-note-1}. The $\beta$-decay of the $^{214}$Bi ($Q_{\beta}=3.274$~MeV), which is present on the surface of the wires, has been measured via the $^{214}$Bi-$^{214}$Po cascade by selecting an $e^-$ and a delayed $\alpha$ emitted from the same vertex inside the tracking chamber (see next section). Figure~\ref{fig:nemo3-bi214-spectrum} shows the reconstructed $\beta$ energy spectrum of the $^{214}$Bi, in very good agreement with the simulations above 0.5~MeV and up to 3.3~MeV (the small discrepency at lower energy is due to the energy losses of the electrons on the wires). This test gives a cross-check of the energy calibration between 2.2 and 3.3~MeV.

The energy measurement is meaningfull only if the energy calibration of each counter is stable in time and if any possible drift is well controlled. 
A laser calibration system was used to survey the asbolute energy. Two laser runs were performed every day. The gain peak stability is surveyed with an accuracy of about $3\%$. During a 40 days time interval between two consecutive absolute $^{207}$Bi calibration runs: in average $\approx 95\%$ of PMT's appeared stable within the $3\%$ accuracy and no gain correction was applied; and $\approx 5\%$ of PMT's are detected unstable because of an apparent gain instability (unstable PMT or HV or unstable laser response for this channel after reference correction). These apparent unstable PMT's were rejected for the analysis (also suppressed for the Monte Carlo simulation for the same effective duration).
However laser survey corrections can be applied to $\approx 70\%$ of the available data, and the effective $\beta\beta 0 \nu$ efficiency becomes  $\epsilon (\beta\beta 0 \nu) = 12.5 \%$. 

For timing calibration, first the relative offsets for each channel are determined with a $^{60}$Co source, which emits two coincident $\gamma$-rays with energies of 1332 and 1173~keV. Spectra of arrival time differences are collected to establish time delays between the 1940 channels. Then the two-electrons events from $^{207}$Bi (corresponding to the two conversion electrons of 0.5 and 1~MeV emitted simultaneously) are used for time alignment and time resolution measurement with electrons. 
Possible drift or unstability of the TDC between two consecutive $^{207}$Bi calibration run are also studied with daily laser runs~\cite{mathieu-these}.

The efficiency of the detector to detect two electron events ($\beta \beta$-like events) has been directly measured with dedicated $^{207}$Bi runs using four low active ($\sim$~10~Bq) and calibrated $^{207}$Bi sources at four opposite position inside the detector. The two conversion electrons emitted simultaneously by the $^{207}$Bi sources were selected. Knowing the activity of the Bi source, the two-electrons detection efficiency was measured. It was in agreement with the Monte Carlo efficiency within a systematic error of $5\%$. 

The two electron events emitted by $^{207}$Bi sources were also used to measure the vertex resolution for the two-electron channel used to reconstruct $\beta \beta$ events. If we define the transverse dispersion $\delta R \phi$ and the the longitudinal dispersion $\delta Z$ as the distance between the vertices associated with the two reconstructed tracks, the vertex resolution for $\beta \beta$ events is $\sigma(\delta R \phi) = 0.6$~cm and $\sigma(\delta Z) = 1.0$~cm. If one constrains the two tracks to have a common vertex, one gets $\sigma(\delta R \phi) = 0.1$~cm. These resolutions allow one to make a distinction between two strips in a source foil in a given sector, which is crucial for sectors composed of different sources.
Finally, thanks to the very high statistics of collected events, the comparison of the experimental angular distribution between the two reconstructed conversion electrons and the expected one obtained with Monte Carlo, has shown a small discrepency. An effective correction obtained using these data was then applied to the measured $\beta \beta$ angular distribution, leading to a perfect agreement with the expected Monte Carlo $\beta \beta$ angular distribution, as shown in the next subsection.

\begin{figure}[!h]  
\centering
\includegraphics[scale=0.4]{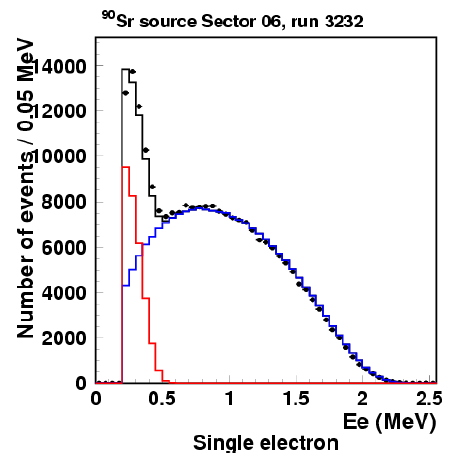}
\caption{Energy spectrum with a low active and calibrated $^{90}$Sr sources, deposited on small thin natural molybdenum foils: black dots corresponds to data, blue histogram to the Monte-Carlo $^{90}$Y $\beta$-decay (daughter of $^{90}$Sr), red histogram to the Monte-Carlo $^{90}$Sr $\beta$-decay, black histogram to the total Monte-Carlo spectrum (background is negligeable).}
\label{fig:nemo3-sr90}
\end{figure}

Finally data with two low active and calibrated $^{90}$Sr sources ($^{90}$Y, daughter of $^{90}$Sr, is a pure $\beta$ emitter), deposited on small thin natural molybdenum foils (60~mg/cm$^2$) and placed in the calibration tubes, have been used to test the energy losses of electrons and the bremsstralung $\gamma$ production in Molybdenum foils. The reconstructed activities of the two sources with NEMO-3 using single electron channel ($\beta$ event) are in perfect agreement with the activity previously measured in IRES (Strasbourg), but limited by the precision of the IRES measurements of only $10\%$. The reconstructed energy spectrum is shown in Figure~\ref{fig:nemo3-sr90} with a good agreement between data and Monte-Carlo simulations. A small systematic distorsion is observed but it is at the level of the Monte-Carlo uncertainty for the theoretical $^{90}$Sr $\beta$ spectrum.

\subsection{The background components}

The combined tracking-calorimetry technique used in NEMO-3 is a very powerfull technique to identify the origin of the different components of background and to measure the level. 
An article presenting the methods and results has been published in~\cite{nemo3-bkg}. I give here a summary.

\begin{figure}[!h]  
\centering
\includegraphics[scale=0.4]{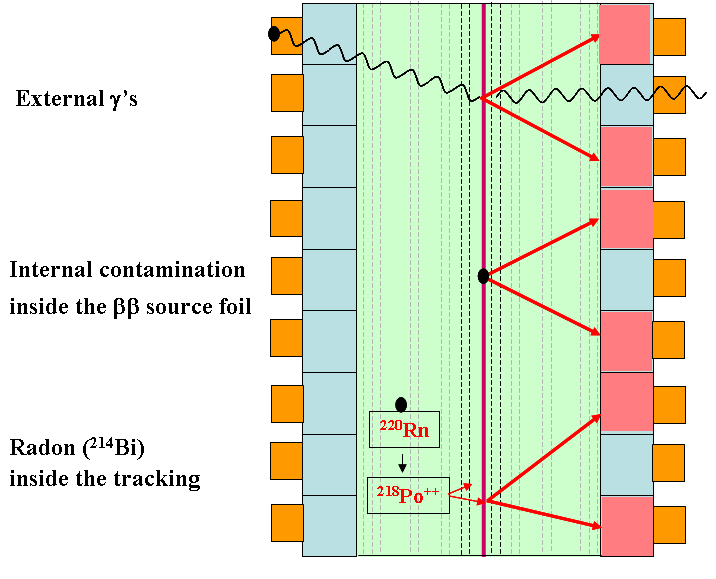}
\caption{Schematic view of the three types of NEMO-3 background, which mimic $\beta \beta$ events: external $\gamma$'s (up), radioactive contamination inside the $\beta \beta$ source foils (middle) and Radon contamination inside the wire chamber (down). See text for explanations.}
\label{fig:nemo3-bkg-process-1}
\end{figure}

The background present in NEMO-3 can be distinguished in four different components (as illustrated in Figure~\ref{fig:nemo3-bkg-process-1})

\begin{figure}[!h]  
\centering
\includegraphics[scale=0.4]{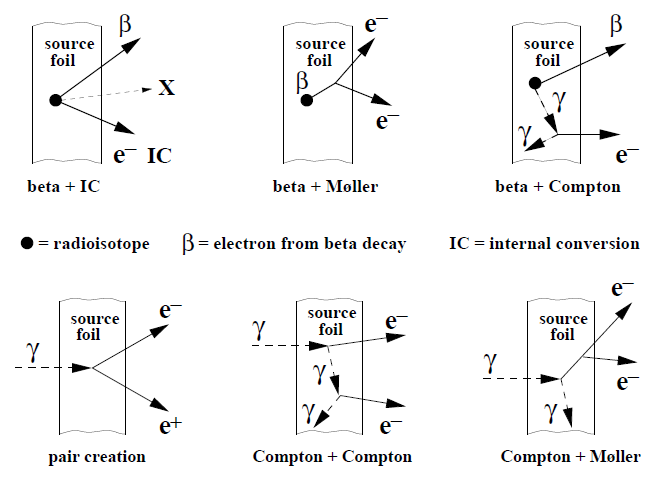}
\caption{Mechanisms to produce a $\beta \beta$-like event produced by $\beta$ decays of internal radioactive impurities inside the source foils (up) or by external $\gamma$'s (down).}
\label{fig:nemo3-bkg-process-2}
\end{figure}

\begin{figure}[!h]  
\centering
\includegraphics[scale=0.5]{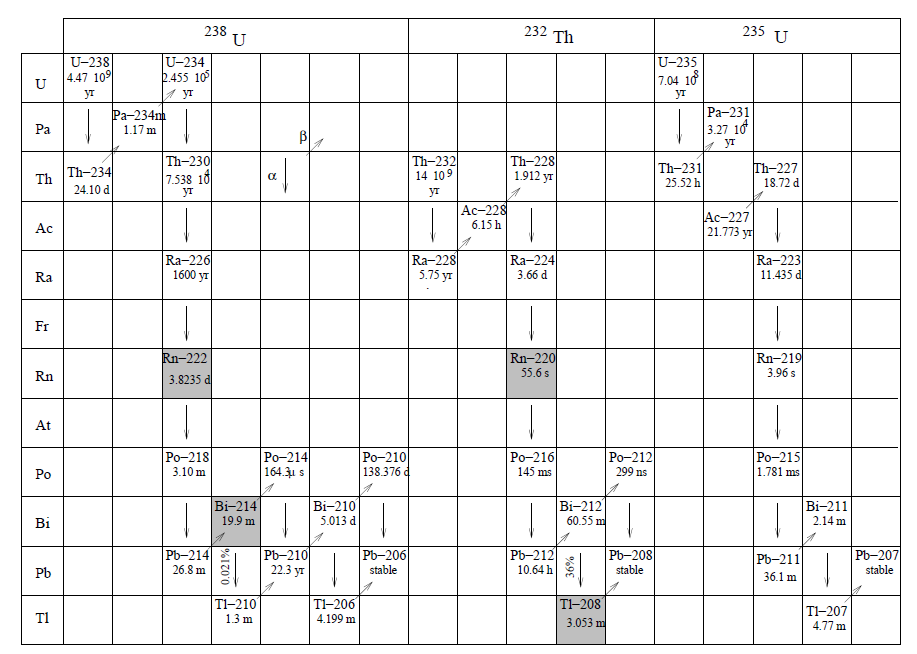}
\caption{Natural radioactivity chains.}
\label{fig:nat-rad-chains}
\end{figure}

\begin{figure}[!h]  
\centering
\includegraphics[scale=0.4]{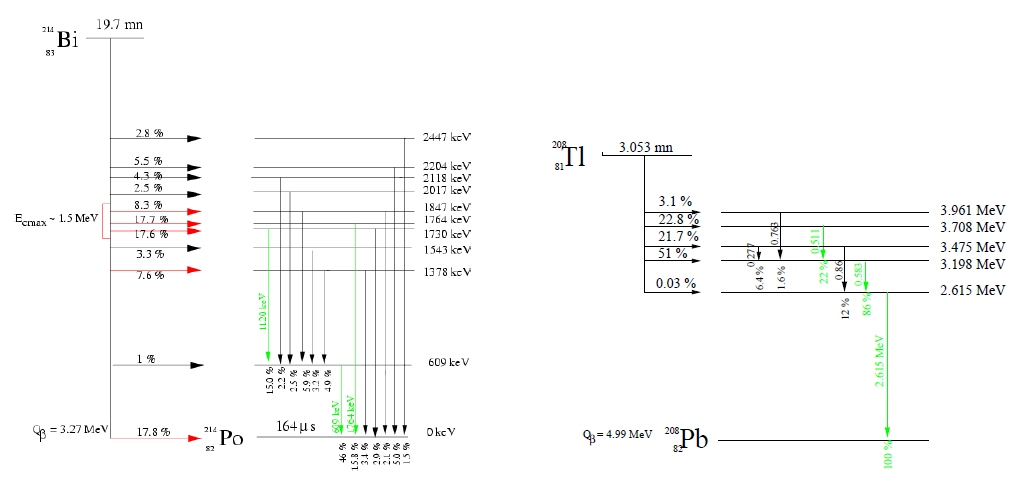}
\caption{$^{214}$Bi and $^{208}$Tl decay diagrams.}
\label{fig:tl208-bi214-diagrams}
\end{figure}

\begin{enumerate}

\item {\bf The external background} produced by the interaction of external $\gamma$'s originating from the natural radioactivity of the detector (outside of the source) or produced by external neutrons or cosmic rays from the LSM lab. If an external $\gamma$ is undetected by the crossed scintillator (50\% $\gamma$ tagging efficiency at 1 MeV, 33\% at 3 MeV), it can reach the $\beta \beta$ source foil without being tagged (gamma does not ionize the gas of the wire chamber). Thus the interaction of the  $\gamma$ can mimic $\beta \beta$  events by $e^+e^-$ pair creation, double Compton scattering or Compton followed by M\"{o}ller scattering (see Figure~\ref{fig:nemo3-bkg-process-2}). The external background is an important component in the $\beta \beta 2 \nu$ energy region but becomes negligible at $\approx$3~MeV in the $\beta \beta 0 \nu$ energy region for $^{100}$Mo and $^{82}$Se (with very rare $\gamma$-rays from $^{214}$Bi above 2.6~MeV).

\item {\bf The internal background} coming from radioactive contaminants inside the $\beta \beta$ source foils. The main internal background contribution due to natural radioactive impurities comes from the $\beta$-decay of $^{214}$Bi ($Q_{\beta}=3.27$~MeV) and $^{208}$Tl ($Q_{\beta}=4.99$~MeV) from $^{238}$U and $^{232}$Th decay chains respectively (see Figures~\ref{fig:nat-rad-chains} and \ref{fig:tl208-bi214-diagrams}). They can mimic $\beta \beta$ events by a $\beta$-decay accompanied by an electron conversion process, by a M\"{o}ller scattering of the $\beta$-decay electrons in the source foil, or by a $\beta$-decay emission to an excited state followed by a Compton scattered $\gamma$ (see Figure~\ref{fig:nemo3-bkg-process-2}). This process can be detected as a two electron events if the $\gamma$ is not detected. 

\item {\bf The Radon and Thoron contaminations inside the tracking detector}. Radon ($^{222}$Rn, $T_{1/2}=3.824$~days) and Thoron ($^{220}$Rn, $T_{1/2}=55.6$~s) are $\alpha$-decay isotopes, which have $^{214}$Bi and $^{208}$Tl as daughter isotopes respectively (see Figure~\ref{fig:nat-rad-chains}). Coming mainly from the rocks and present in the air, the $^{222}$Rn and $^{220}$Rn (rare gases) are very diffusion prone, and can enter the detector and contaminate the interior of the tracking chamber. Subsequent $\alpha$-decays of these gases give $^{218}$Po$^{++}$ and $^{216}$Po$^{++}$ ions respectively, which drifts mainly to the cathodic wires. If the deposition is close to the $\beta \beta$ source foils, this becomes like an internal background critical for the $\beta \beta 0 \nu$ search. The Thoron contamination is generaly much lower than Radon due to its short half-life which limits its diffusion capacity. 

\item {\bf The tail of the $\beta \beta 2 \nu$ energy spectrum} due to the limited energy resolution is the ultimate background. I would like to emphasize that the $\beta \beta 2 \nu$ background corresponds to a very steep energy spectrum, just below the $\beta \beta 0 \nu$ energy region. Thus any unstability of the energy response of a calorimeter block, which has not been identified correctly, may lead to an overestimated energy of the $\beta \beta 2 \nu$ events. Such event will ``fall like a rock'' from the $\beta \beta 2 \nu$ ``big wall'' into the $\beta \beta 0 \nu$ energy region. I like to call these $\beta \beta 2 \nu$ background events the {\it Yosemite} events. 

\end{enumerate}

\subsection{Measurement of the different background components using the NEMO-3 data}

Each component of background can be identified and measured separately using different dedicated topologies of events.

\subsubsection*{External background measurement}

\begin{figure}[!h]  
\centering
\includegraphics[scale=0.4]{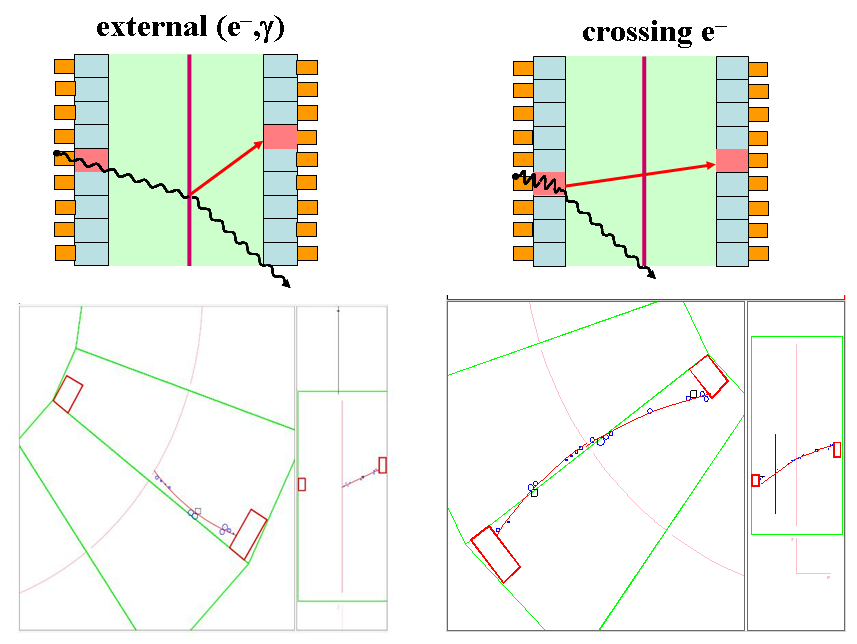}
\caption{The two event topologies used to measure the external background: (left) external $(e^-,\gamma)$ event; (right) crossing-$e^-$.}
\label{fig:nemo3-bkg-ext-event}
\end{figure}

\begin{figure}[!h]  
\centering
\includegraphics[scale=0.4]{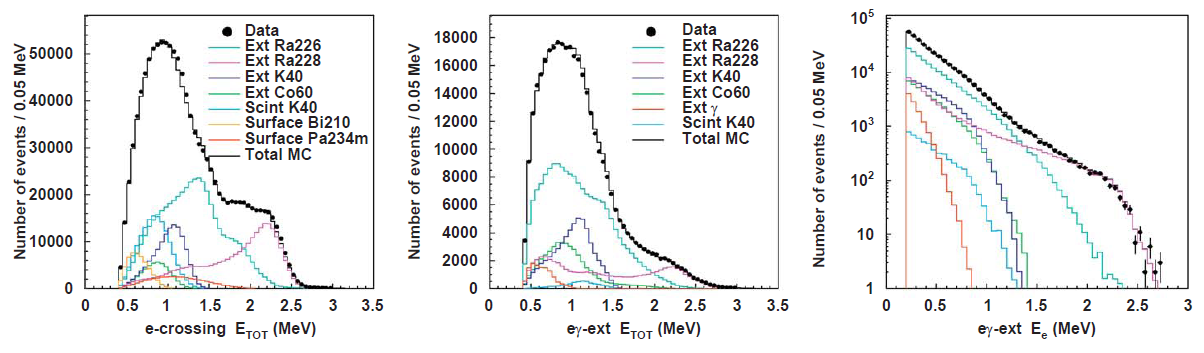}
\caption{Result of the fit of the external background model to the observed data (black dots) of the Phase~2 data, for the whole detector: (left) energy sum of crossing electrons, (center) energy sum of the electron and $\gamma$ for external $(e^-,\gamma)$ events and (right) detected energy of the electron for external $(e^-,\gamma)$ events.}
\label{fig:nemo3-bkg-ext-result}
\end{figure}

Two topologies of events are used to measure the external background (see Figure~\ref{fig:nemo3-bkg-ext-event}): 
\begin{enumerate}
\item External $(e^-,\gamma)$ events defined as one track coming from the source foil and associated with a scintillator hit (electron) and one isolated scintillator ($\gamma$) with a time of flight between the two fired scintillators in agreement with an external $\gamma$ hypothesis firing a first scintillator and producing a Compton electron in the source foil.
\item Crossing-$e^-$ defined as a track crossing the detector and associated on each ends to a fired scintillator with a time of flight in agreement with a crossing electron hypothesis.
\end{enumerate}
A very accurate external background model has been developped in order to fit the observed data in both channels of analysis (external $(e^-,\gamma)$ and crossing-$e^-$ events).
This model assumes contaminations in $^{238}$U, $^{232}$Ra and $^{40}$K  inside the PMT's, contamination in $^{60}$Co (cosmogenic) inside the mechanical structure, and external $\gamma$'s in LSM.
Activities of each components in the Monte-Carlo are fitted to the observed data using a global fit of several parameters (energy sum, energy of the $e^-$, energy of the $\gamma$, angle between $\gamma$ and $e^-$). 
Figure~\ref{fig:nemo3-bkg-ext-result} shows the excellent agreement of the energy sum distribution of the $(e^-,\gamma)$ events and the crossing-$e^-$ events obtained for the whole detector during Phase 2 and the result of the fitted background model in Monte Carlo. Morever, the activities of each components obtained from the global fit are in agreement with the previous HPGe radioactivity measurements of the detector materials.

One may expect that the background may vary from one sector to another due to possible inhomogeneities of the detector materials. Thus the background model is actually calculated separately for each isotopes (located in different sectors). 

Concerning the neutron background, it has been shown using dedicated runs with an AmBe neutron source outside of the shield that contribution of neutrons to the external background (via the neutron capture process resulting in emission of $\gamma$'s) is negligible even in the $\beta \beta 0 \nu$ energy region~\cite{nemo3-bkg} (the neutron background is measured using crossing electrons with a total energy above 4.5~MeV). 
This result has been verified by analysing the $e^+/e^-$ events emitted from the source foils with the  NEMO-3 physics data. Only 3 $e^+/e^-$ events with an energy above 3~MeV have been observed after $\approx 3$ years of data~\cite{arnold-note-2}. Their energy are contained between 3 and 3.8~MeV. Taking into account an unefficiency of $\approx 10\%$ to distinguish a positron to an electron with the track curvature, and assuming that these three events are produced by external neutrons, it would corresponds to an expected number of less than about 0.1~$(e^-,e^-)$ events per year (90\%~C.L.) in the $[2.8-3.2]$~MeV $\beta \beta 0 \nu$ energy window. 

\subsubsection*{Radon and Thoron measurements}

\begin{figure}[!h]  
\centering
\includegraphics[scale=0.4]{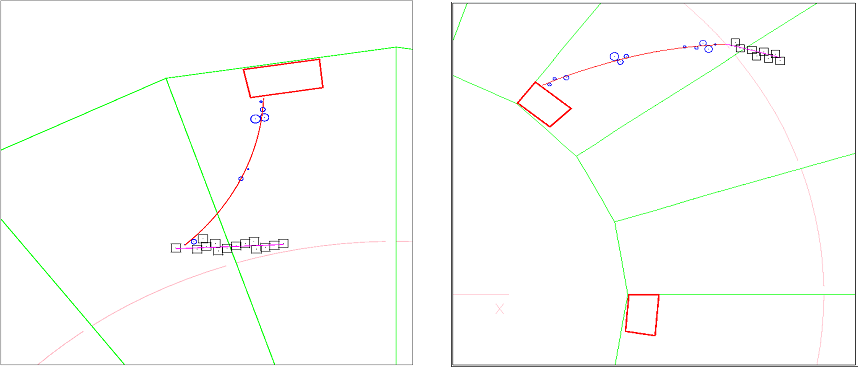}
\caption{Example of BiPo events observed in NEMO-3: (left) $^{214}$Bi decayed on a wire inside the tracking volume; (right) $^{214}$Bi decayed on the surface of the foil.}
\label{fig:nemo3-bkg-radon-event}
\end{figure}

The Radon contamination is measured by detecting inside its decay chain the electron from the $\beta$-decay of Bi followed by the delayed $\alpha$ from the $\alpha$-decay of Po with a short half-life of 164~$\mu$s, the so-called BiPo events. In order to detected the delayed $\alpha$, a dedicated electronic has been developped for the tracking detector which allows to readout any delayed Geiger hit inside the wire chamber with a delay up to 700~$\mu$s. 
A Bipo event is thus defined in NEMO-3 as an electron (a track inside the wire chamber associated with a fired scintillator) with at least one delay Geiger hit in the wire chamber close to the emission vertex of the electron. 
Displays of BiPo events detected in NEMO-3 are given in Figure~\ref{fig:nemo3-bkg-radon-event}. 
The delay is required to be greater than 90~$\mu$s and 30~$\mu$s for events with only one delayed Geiger hit and events with more than one delayed hit respectively, in order to suppress possible refiring of neighboring Geiger cells from the electron tracks. Applying these criteria, the mean efficiency to select a BiPo event produced on a wire surface has been estimated by Monte-Carlo simulation to be 16.5$\%$. 

\begin{figure}[!h]  
\centering
\includegraphics[scale=0.4]{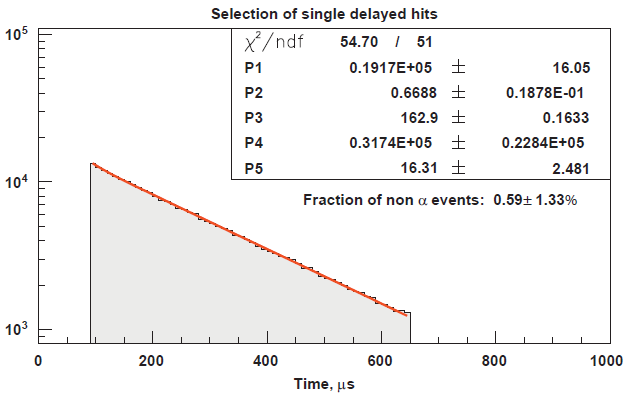}
\caption{The time distributions of events selected with single delayed signals. Each distribution was fit to the function: $f(t)=P1 e^{-ln2 t/P3} + P2 + P4 e^{-ln2/P5 }$ where $t$ is the event time, $P1$ and $P4$ are scaling constants, $P2$ is the amplitude of random coincidences, $P3$ the $^{214}$Po half-life in $\mu$s and $P5$ the time constant of the refirings.}
\label{fig:nemo3-deltat-radon}
\end{figure}

The time distribution of the delayed tracks (see Figure~\ref{fig:nemo3-deltat-radon} for single delayed hits) provides an efficient way to validate the quality of the event selection. The fitted half-life of the $^{214}$Po is T$_{1/2}=162.9 \pm 0.8 $(stat. only)~$\mu$s, and is in agreement with the table value of T$_{1/2}=164.3 \pm 2.0$~$\mu$s~\cite{po214}. 
The proportions of BiPo events due to refirings and random coincidences are found to be negligibly small, about $1\%$. 
This method allows to measure the Radon activity inside the wire chamber every day with a good accuracy. 
The average Radon and $^{214}$Bi activity measured in Phase~1 (high Radon activity) was $\approx 30$~mBq/m$^3$. 
The average Radon and $^{214}$Bi activity measured in Phase~2 (after installing the Radon-tight structure surrounding the detector and after flushing Radon-free air inside) was $\approx 5$~mBq/m$^3$ (a reduction factor $\approx 6$). 
The volume of the wire chamber is $\approx 30$~m$^3$, it corresponds to a total activity in Radon and $^{214}$Bi of $\approx 150$~mBq.

We also mention that the Radon contamination has been also measured separately using $(e^-,\gamma)$ events with both particles emitted from the same vertex inside the tracking chamber. Despite of a less sensitive analysis, this second measurement has given results in agreement with the analysis using $e^-$ delayed $\alpha$. 

The study of the spatial distribution of the $e^-/\alpha$ emission vertex in Phase~2 (low Radon activity) 
has shown that the  $^{214}$Bi activity is mainly on the wires which are on the edge of the large gas gaps. It confirms the drift of polonium ions to the wires. However higher activities have been observed either on the top or bottom of the wires for the internal or external part respectively of the tracking chamber. This asymetry has not yet been understood. 
The comparison of $^{214}$Bi distribution of Phase~2 relatively with Phase~1 has shown that the residual $^{214}$Bi activity is mainly near the scintillator walls and the end caps (top and bottom) of the wire chamber and, but to a lesser degree, near the $\beta \beta$ source foils. 

The study of possible origins of Radon contamination observed inside the tracking volume has been performed and is given in~\cite{nemo3-hdr-laurent}. It results in four possible origins of Radon contamination:

\begin{enumerate}
\item Emanation of Radon from both the aluminized mylar and the teflon used to wrap the scintillator blocks inside the tracking chamber. Only upper limits for the activity in $^{214}$Bi have been set with HPGe measurements: $<20$~mBq/kg (equivalent to $<$~414~mBq taking into account the total mass) for teflon and $<35$~mBq/kg (equivalent to $<126$~mBq taking into account the total mass) for aluminized mylar. 
\item The silicon seals (RTV~160) used to close the tracking chamber are porous to Radon. Any Radon contamination outside the tracking chamber might diffuse inside the detector. The HPGe measurements have shown that the RTV~160 itself is also contaminated in $^{214}$Bi with an activity of $35 \pm 10$~mBq/kg corresponding to a total activity of $95 \pm 27$~mBq. 
\item Outside of the tracking chamber, the dominant source of Radon is the PMT's glasses. The total activity in $^{214}$Bi of all the 5'' PMT's is about 240~Bq and $\approx 85\%$ of Radon is expected to emanate from the 1mm thick glass (the diffusion constant of Radon in the glass is $\approx 1.7 \ 10^{-11}$~m$^2$/s giving a diffusion length of 2.9~mm). A small fraction ($\approx \ 10^{-3}$) of the emanated Radon through the RTV seals can give the observed contamination of Radon inside the tracking chamber. 
\item The Radon measurement using the $(e^-,\gamma)$ events indicates a possible contamination from the calibration tubes with some hot spots. Possible external Radon leaks through the seals of the calibration tube or contamination of $^{238}$U inside the tubes could explain this Radon background. 
\end{enumerate}

Regarding Thoron background, the $^{208}$Tl activity inside the tracking chamber has been measured using $e^- \gamma \gamma$ events (see next section). The $^{208}$Tl activity measured in Phase~2 is $\approx 3$~mBq ($\approx 0.1$~mBq/m$^3$). Taking into account the 36$\%$ branching ratio of $^{208}$Tl in the $^{232}$Th chain, it corresponds to an activity in Thoron of $\approx 8$~mBq ($\approx 0.3$~mBq/m$^3$). Monte Carlo simulations show that this low level of Thoron is more than one order of magnitude lower than the background originating from Radon for the two-electrons events and is therefore negligeable in the $\beta \beta 0 \nu$ region.

Another source of background in the wire chamber is the contamination in $^{210}$Pb ($T_{1/2}=22.3$~y) on the surface of the wires. $^{210}$Pb belong to the $^{238}$U decay chain and its contamination comes from the deposition of Radon during the wiring of the wire chamber. $^{210}$Pb decays to $^{210}$Bi, which is a $\beta$ emitter with a $Q_{\beta}=1.16$~MeV. It is of no concern for $\beta \beta 0 \nu$ decay search, but it must be considered in the precise measurement of the $\beta \beta 2 \nu$ decay spectra, especialy for for $^{130}$Te.
One electron events with an energy greater than 600keV and their vertices associated with Geiger cells are selected to measure the $^{210}$Bi activity on the wire surfaces. 
A large variation of $^{210}$Bi activity from one sector to another is observed. The origin of the non-uniformity in $^{210}$Pb deposition on the wires is most probably due to the different histories of the wires and conditions during the wiring of the sectors.

\subsubsection*{Internal contaminations of the $\beta \beta$ source foils}

\begin{figure}[!h]  
\centering
\includegraphics[scale=0.4]{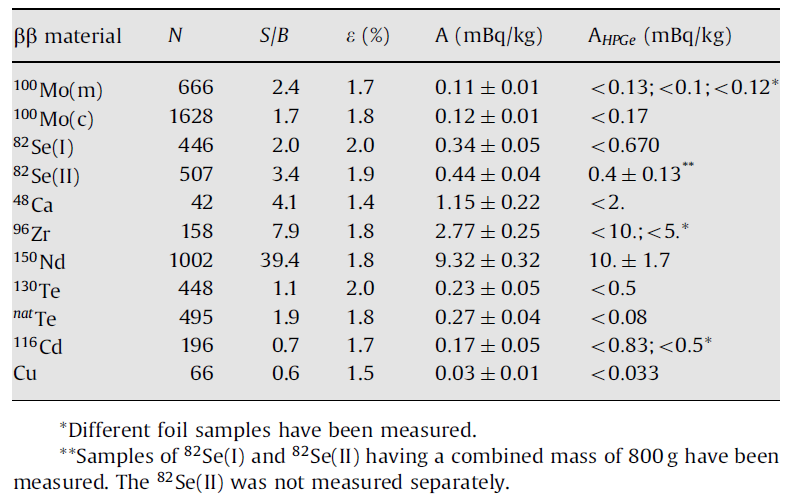}
\caption{Number $N$ of observed $(e^-,\gamma \gamma)$ and $(e^-,\gamma \gamma \gamma)$ events, signal-to-background ratio ($S/B$), signal efficiency ($\epsilon$) and results of the measurements of $^{208}$Tl activity of the source foils compared to the HPGe measurements.}
\label{fig:nemo3-bkg-tl208-result}
\end{figure}

The $\beta$-decay of $^{208}$Tl is mainly accompanied by two or three $\gamma$'s. Therefore its contamination  inside the sources foils is measured by using internal $(e^-,\gamma \gamma)$ and $(e^-,\gamma \gamma \gamma)$ events defined as one track coming from the source foil and associated with a scintillator hit (electron) and two or three isolated scintillators ($\gamma$'s) with a time of flight analysis in agreement with the hypothesis that all the particles involved have been emitted from the track vertex on the foil.
The most important background for this analysis is due to Thoron and Radon inside the tracking chamber. The two event topologies $(e^-,\gamma \gamma)$ and $(e^-,\gamma \gamma \gamma)$ give consistent results. The $^{208}$Tl activity in the $\beta \beta$ source foils and in the copper foils are presented in Figure~\ref{fig:nemo3-bkg-tl208-result}. 

The selenium foils had not been purified before their installation in the NEMO-3 detector, because of its relatively low quantity compared to molydbenum. It explains the relatively high activity of $\approx 0.4$~mBq/kg measured in NEMO-3. It is remarkable that this measurement is in agreement with the HPGe measurements done before their installation inside the NEMO-3 detector. Another positive HPGe measurement for $^{208}$Tl contamination  was obtained for the Nd foils and is also in agreement with the measured value in NEMO-3. 
It demonstrates the reliability of the $^{208}$Tl measurement. 

The $^{208}$Tl contaminations measured inside the Mo foils are the same for the two types of foils: $0.11 \pm 0.01$~mBq/kg for the composite foils (thin and chemically purified $^{100}$Mo powder  mixted with PVA glue and water and deposited between mylar foils) and $0.12 \pm 0.01$~mBq/kg for the metallic foils (pure $^{100}$Mo monocrystal heated and rolled in the form of foils). Both are in agreement with the upper limits obtained with HPGe (best limit is $<0.13$~mBq/kg). 

The copper foils appear to be highly radiopure although the $S/B$ signal-to-background ratio for the $^{208}$Tl measurement in NEMO-3 is quite low. The $^{208}$Tl activity is estimated to be $0.03 \pm 0.01$~mBq/kg, in agreement with HPGe upper limit of $<0.033$~mBq/kg.

The measurement of $^{214}$Bi contamination inside the source foils is limited by the Radon background in the tracking volume near the foil. However the distribution of the length of the delayed $\alpha$ tracks gives some indications of the possible fraction of Bi contamination inside tthe foil, relatively to the Radon background near, or on the surface of the foil. Only limit has been defined for Mo foils and in agreement with the upper limits obtained with HPGe measurements $<0.34$~mBq/kg for composite foils and $<0.39$~mBq/kg for metallic foils.

\vspace{1cm}

\begin{figure}[!h]  
\centering
\includegraphics[scale=0.4]{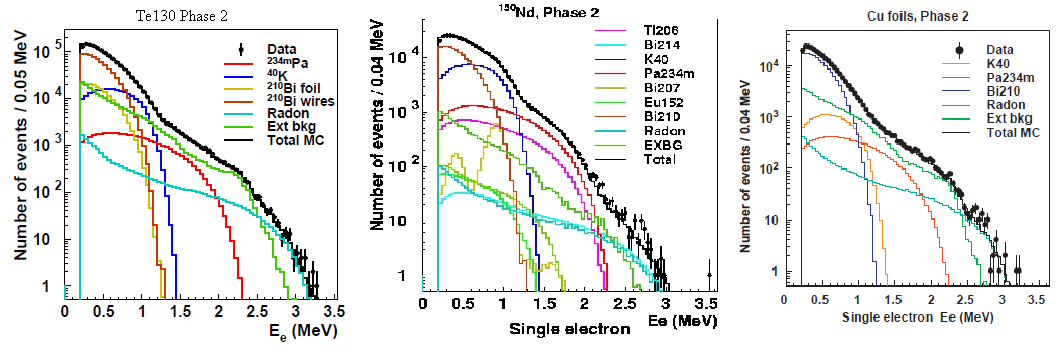}
\caption{Energy spectrum of single electrons emitted from the $^{150}$Nd source (left), the $^{130}$Te source (center) and the copper foil (right).}
\label{fig:nemo3-single-electrom-spectrum}
\end{figure}

We emphasize that NEMO-3 can also measure possible contamination of pure $\beta$ emitters inside the foil by analysing the single $e^-$ events coming from the foil. 
The energy spectra of single $e^-$ events emitted from the $^{150}$Nd source, the $^{130}$Te source and the copper foil are presented in Figure~\ref{fig:nemo3-single-electrom-spectrum}. 
The excellent agreement between the data and the fitted Monte Carlo background shows the capacity to distinguish and measure the $\beta$ emitters inside the source foils like $^{234m}$Pa from the $^{238}$U chain, $^{40}$K, $^{210}$Bi from the deposition of the Radon on the foil during the construction, or accidental contamination like $^{152}$Eu inside the $^{150}$Nd source foils.
However these $\beta$ contaminants represent a background only for the $\beta \beta 2 \nu$-decay measurement. 


\subsection{Test of the complete background model and background budget}

\begin{figure}[!h]  
\centering
\includegraphics[scale=0.4]{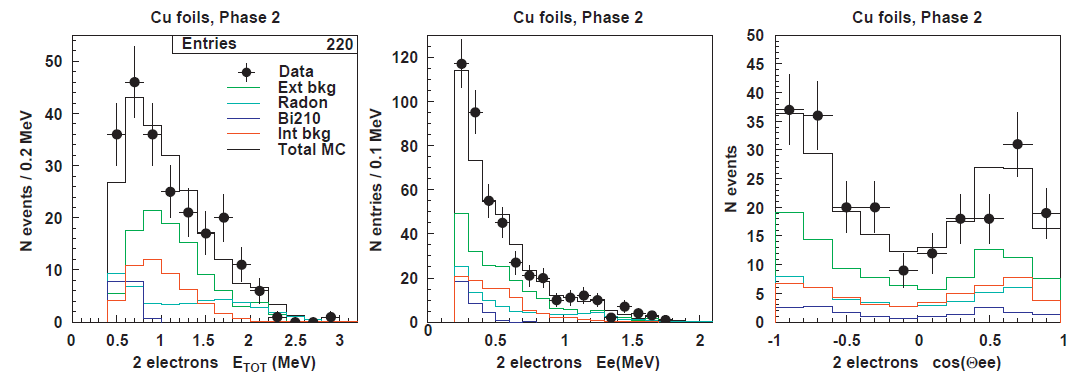}
\caption{Background model prediction compared to the data for two electrons events from the copper foils.}
\label{fig:nemo3-bkg-copper}
\end{figure}

The highly radiopure copper foils (1~sector) are used to compare the NEMO-3 data with the expected background calculated by Monte Carlo simulation using the complete background model described above. Two different event topologies have been tested: Single $e^-$ events emitted from the foil and internal ($e^-$,$\gamma$) events from the foil. Excellent agreements have been obtained in both channels. 
Finally, since copper is not a $\beta \beta$ emitter, it provides a test for the validity of the background model with the two electrons events where $\beta \beta$-decay is searched for. Figure~\ref{fig:nemo3-bkg-copper} shows the distributions of the energy sum of the two electrons, the single electron energies and angular correlation of two-electron events coming from the copper foils, observed in Phase~2. They are in excellent agreement with the expected background calculated by Monte Carlo simulation using the NEMO-3 background model. 

\begin{figure}[!h]  
\centering
\includegraphics[scale=0.4]{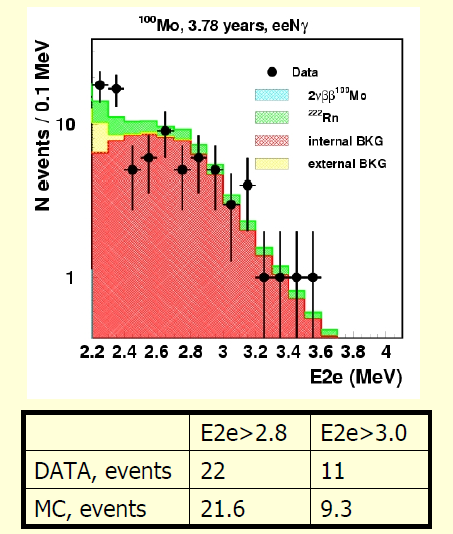}
\caption{Sum energy spectrum of the two electrons using $(e^- e^-, N \gamma)$ channel from $^{100}$Mo, with 3.78~years of collected data in Phase~2 and expected background from Monte-Carlo simulation.}
\label{fig:nemo3-eengamma}
\end{figure}

I also mention another test performed using the two electrons channel with at least one associated $\gamma$-ray emitted in time from the source foil $(e^- e^-, N \gamma)$. Above 2.5~MeV, this channel is dominated by $^{208}$Tl contamination inside the foil, and to a lesser extent, Radon. However $\beta\beta 2\nu$ is strongly supressed in this channel. Figure~\ref{fig:nemo3-eengamma} shows the sum energy spectrum of the two electrons in the $(e^- e^-, N \gamma)$ channel, from $^{100}$Mo source foils, with 3.78~years of collected data in Phase~2. Data are in good agreement with the expected background dominated by $^{208}$Tl contamination inside the foils. It validates the contribution of the $^{208}$Tl contamination.

The estimated background in the $\beta \beta 0 \nu$ energy region is summarized in Table~\ref{tab:nemo3-bkg-budget} for $^{100}$Mo. The relative contribution of each background components is given in terms of number of counts/(kg.y) in the $[2.8-3.2]$~MeV energy window where the $\beta \beta 0 \nu$ signal is searched for (it takes into account the NEMO-3 energy resolution). 
With 7~kg of $^{100}$Mo in NEMO-3, $\approx 3.5$~counts per year are expected in the $[2.8-3.2]$~MeV energy window. 

\begin{table}[!h]
\centering
\begin{tabular}{c|c|c|c}
 &  & cts/(kg.yr) &  Fraction\\
 &  & $[2.8-3.2]$~MeV  &  \\
\hline
\hline
Radon                  & $\approx$ 5 mBq/m$^3$      &  0.16           & 30$\%$\\
$^{208}$Tl in the foil & $\approx$ 100 $\mu$Bq/kg   &  0.11            & 20$\%$ \\
External bkg           &                            &  $\leq 10^{-3}$ & negligible \\
$\beta \beta 2 \nu$    & $T_{1/2} = 7 \ 10^{18}$~yr &  0.23           & 50$\%$\\
\hline
\hline
TOTAL                  &                            &  0.5            &  \\
\end{tabular}
\caption{Background budget in NEMO-3 calculated using the background simulations for $^{100}$Mo in terms of number of counts per kg and per year in the $[2.8-3.2]$~MeV energy window where the $\beta \beta 0 \nu$ signal is searched for. }
\label{tab:nemo3-bkg-budget}
\end{table}

\subsection{$\beta \beta$ results}

NEMO-3 has been running from February 2003 until January 2011.
During the Phase 1 (high Radon background until September 2004), about 1 year of data has been collected. 
During the Phase 2 (low Radon background from december 2004), about 4 years of data have been collected.
First results obtained only with Phase~1 and with Mo and Se have been published in~\cite{nemo3-prl}
 
\subsubsection*{$\beta \beta 2 \nu$ process}

\begin{figure}[!h]  
\centering
\includegraphics[scale=0.35]{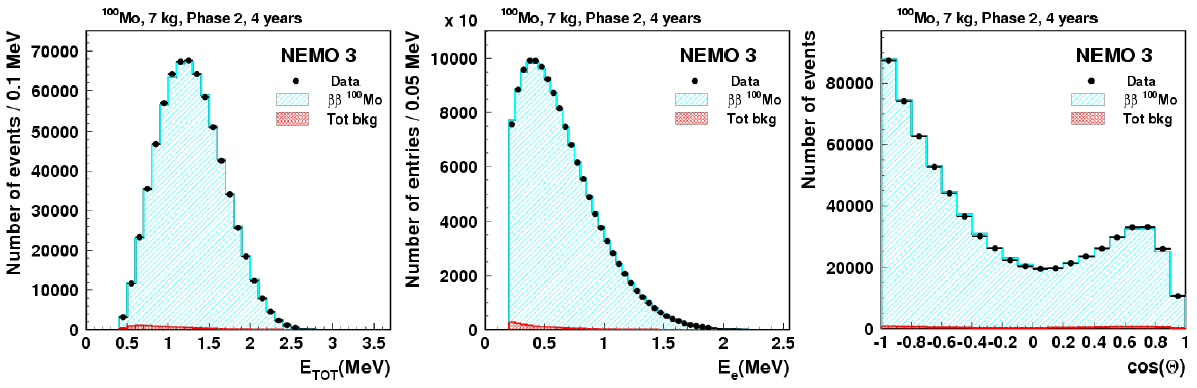}
\caption{Sum energy spectrum (left), single energy spectrum (center) and angular distribution (right) of the two electrons events, for $^{100}$Mo, obtained after 4~years of collected data in Phase~2.}
\label{fig:nemo3-bb2nu-mo}
\end{figure}

Figure~\ref{fig:nemo3-bb2nu-mo} shows the distribution of the energy sum of the two electrons emitted from the $^{100}$Mo foils, observed during Phase~2. The distributions of the minimum electron energy and the angular correlation between the two electrons are also presented. 
About 700000~$\beta \beta$~like events have been observed in 4 years with a very low background and a very high signal-to-background ratio of $S/B = 77$. 
The detection efficiency of the $\beta \beta 2 \nu$-decay for Phase~2 is $\epsilon(\beta \beta 2 \nu) = 4.2 \%$ for $^{100}$Mo (without applying laser survey rejection); and the systematic error on the efficiency is estimated to be $5\%$. 
It corresponds to the $\beta \beta 2 \nu$-decay half-life for $^{100}$Mo :

$$ \mathrm{T_{1/2}^{\beta\beta 2 \nu}(^{100}Mo) = \left( 7.16 \pm 0.01 (stat.) \pm 0.36 (syst.) \right) \ 10^{18} \ y }$$

The measured half-life and the energy spectrum are similar (within the accuracy) if the laser survey procedure is applied. 

The value of the $\beta \beta 2 \nu$ half-life $T_{1/2}^{\beta\beta 2 \nu}(^{100}Mo)$ has been used to study several possible extra systematics:
\begin{enumerate}
\item $T_{1/2}^{\beta\beta 2 \nu}(^{100}Mo)$ measured separately with composite or metallic Mo foils are in agreement within the $5\%$ systematic error (the half-life measured with composite foils appears to be slightly larger than with metalic foils)
\item $T_{1/2}^{\beta\beta 2 \nu}(^{100}Mo)$ has been measured year by year separately and appears to be stable within the statistic and systematic errors. 
\item $T_{1/2}^{\beta\beta 2 \nu}(^{100}Mo)$ has been measured using only the sector number 13 where the wire chamber was unstable (with some periodes of refiring geiger cells). The measured value is in agreement with the average value. It demonstrates the validity of the procedure to detect hot regions of the wire chamber and to supress them both in data and in simulations. 
\end{enumerate}

\begin{figure}[!h]  
\centering
\includegraphics[scale=0.35]{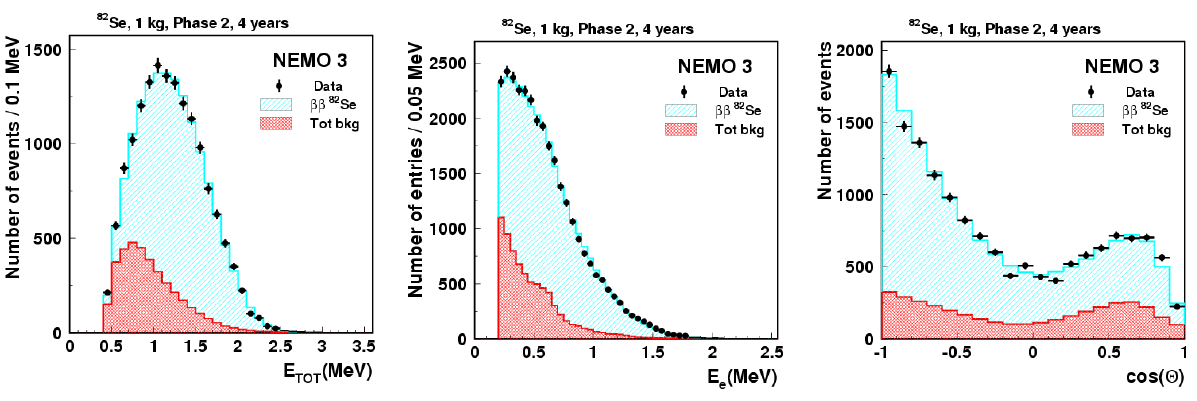}
\caption{Sum energy spectrum (left), single energy spectrum (center) and angular distribution (right) of the two electrons events, for $^{82}$Se, obtained after 4~years of data collection in Phase~2.}
\label{fig:nemo3-bb2nu-se}
\end{figure}

The distributions of the $\beta \beta 2 \nu$-decay measured with $^{82}$Se foils in Phase~2, are presented in Figure~\ref{fig:nemo3-bb2nu-se}.
15348~$\beta \beta$~like events have been observed in 4 years with a signal-to-background ratio of $S/B = 3.07$.
The measured efficiency of the $\beta \beta 2 \nu$-decay is $\epsilon(\beta \beta 2 \nu) = 5.8 \%$ for $^{82}$Se. 
It corresponds to the $\beta \beta 2 \nu$-decay half-life:

$$ \mathrm{T_{1/2}^{\beta\beta 2 \nu}(^{82}Se) = \left( 9.6 \pm 0.1 (stat.) \pm 0.5 (syst.) \right) \ 10^{19} \ y }$$

The $\beta \beta 2 \nu$-decay has been also measured in NEMO-3 with 5 other isotopes: $^{48}$Ca, $^{96}$Zr, $^{116}$Cd, $^{130}$Te and $^{150}$Nd. The measured half-lives are summarized in Table~\ref{tab:nemo3-bb-summary}. 

\begin{table}[!h]
\centering
\begin{tabular}{c|c}
Isotope & $ T_{1/2}(\beta\beta 2 \nu)$ \\
\hline
\hline
$^{100}$Mo &  $\left( 7.16 \pm 0.01 (stat.) \pm 0.36 (syst.) \right) \ 10^{18} \ y $~\cite{nemo3-prl} \\
$^{82}$Se  &  $\left( 9.6 \pm 0.1 (stat.) \pm 0.5 (syst.) \right) \ 10^{19} \ y $~\cite{nemo3-prl}    \\
$^{48}$Ca  &  $\left( 4.4 \pm 0.5 (stat.) \pm 0.4 (syst.) \right) \ 10^{19} \ y $    \\
$^{96}$Zr  &  $\left( 2.35 \pm 0.14 (stat.) \pm 0.16 (syst.) \right) \ 10^{19} \ y $~\cite{nemo3-zr}   \\
$^{116}$Cd &  $\left( 2.88 \pm 0.04 (stat.) \pm 0.16 (syst.) \right) \ 10^{19} \ y $  \\
$^{130}$Te &  $\left( 7.0 \pm  0.9 (stat.) \pm 0.9 (syst.) \right) \ 10^{20} \ y $~\cite{nemo3-te}    \\
$^{150}$Nd &  $\left( 9.11 \pm 0.25 (stat.) \pm 0.63 (syst.) \right) \ 10^{18} \ y $~\cite{nemo3-nd}    \\
\end{tabular}
\caption{Summary of $\beta\beta 2 \nu$-decay measured with  NEMO-3 for 7 isotopes, using Phase~1 and Phase~2 until end 2009 (4.5~years of data). Published results are indicated with references. Otherwise it corresponds to preliminary results }
\label{tab:nemo3-bb-summary}
\end{table}

\subsubsection*{$\beta \beta 0 \nu$ process}

The analysis for the search of the $\beta \beta 0 \nu$ process has been presented at the summer conferences in 2011, using the data of Phase~1 and Phase~2 until end of 2009 (4.51~years of data). Analysis of the data 2010 has not yet been presented. 

\begin{figure}[!h]  
\centering
\includegraphics[scale=0.4]{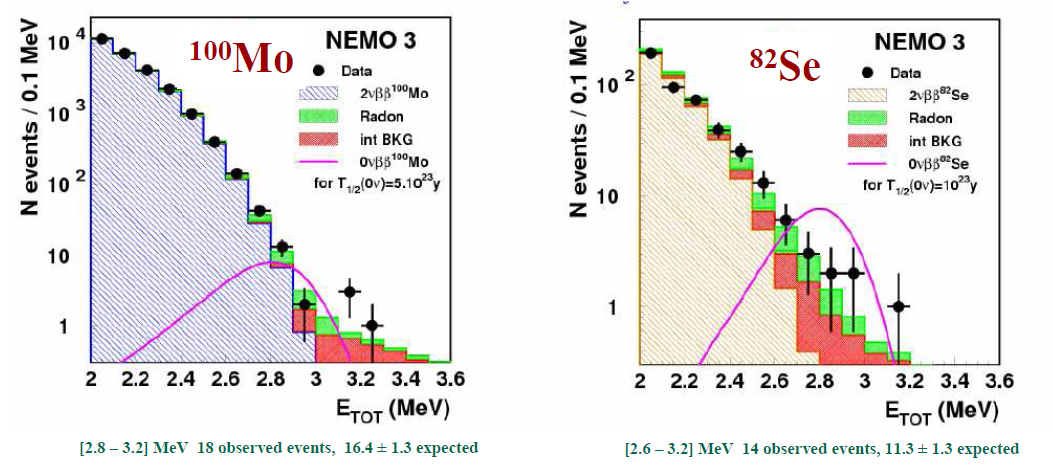}
\caption{Sum energy spectrum of the two electrons obtained after 4.5~y of collected data, (left) for $^{100}$Mo and (right) for $^{82}$Se.}
\label{fig:nemo3-bb0nu-mo}
\end{figure}

Figure~\ref{fig:nemo3-bb0nu-mo} present the energy sum spectra above 2~MeV for the two electrons emitted from the $^{100}$Mo foils or from the $^{82}$Se foils using data of Phase~1 and Phase~2 until end of 2009 (4.51~years of data) and using the procedure for laser survey corrections.

The limits set by Modified Frequentist Method (CLs) on the $\beta \beta 0 \nu$ process, using the full distribution and assuming that the  $\beta \beta 0 \nu$ process is governed by the exchange of a virtual Majorana neutrino, are for $^{100}$Mo ($\approx 31.2$~kg.yr) and for $^{82}$Se ($\approx 4.2$~kg.yr)~\cite{nemo3-taup11}:

$$ \mathrm{T_{1/2}^{\beta\beta 0 \nu}(^{100}Mo,~\langle m_\nu \rangle) > 1.0 \ 10^{24}~y~(90\%~C.L.)}$$
$$ \mathrm{T_{1/2}^{\beta\beta 0 \nu}(^{82}Se,~\langle m_\nu \rangle) > 3.2 \ 10^{23}~y~(90\%~C.L.)}$$ 

If one includes the systematic uncertainty of 5$\%$ on the $\beta \beta 0 \nu$ detection efficiency and a systematic background uncertainty of 10$\%$, the limits become

$$ \mathrm{T_{1/2}^{\beta\beta 0 \nu}(^{100}Mo,~\langle m_\nu \rangle) > 0.75 \ 10^{24}~y~(90\%~C.L.)}$$
$$ \mathrm{T_{1/2}^{\beta\beta 0 \nu}(^{82}Se,~\langle m_\nu \rangle) > 3.2 \ 10^{23}~y~(90\%~C.L.)}$$

However the background uncertainty  is still under study. Its reduction would improve the limit on $^{100}$Mo. 
The limit obtained with $^{82}$Se is unchanged if the systematic uncertainties are taken into account because of its much smaller background: its higher $\beta \beta 2 \nu$ half-life reduces the $\beta \beta 2 \nu$ background and its smaller available mass inside the detector and therefore its smaller surface of source foil reduces the Radon background.

The $\beta\beta 0 \nu$ detection efficiency in the full energy range is $\epsilon (\beta\beta 0 \nu) = 18 \%$.
After applying the laser correction (for PMT gain survey), the effective $\beta\beta 0 \nu$ efficiency becomes  $\epsilon (\beta\beta 0 \nu) = 12.5 \%$ because laser corrections can be applied to $\approx 70\%$ of the available data. 
We emphasize that the laser procedure is crucial. Indeed, without any PMT's gain stability survey, few counts are observed every year in the $\beta\beta 0 \nu$ energy region due to some unstable PMT gains. \\

Same limits on $T_{1/2}(\beta\beta 0 \nu)$ are obtained if one compares the number of events observed in the $\beta\beta 0\nu$ energy window with the number of expected background events. Numbers of observed and expected events are summarized in Table~\ref{tab:nemo3-counts-bb0nu}. The energy windows have been optimized with Monte Carlo simulation and are $[2.8-3.2]$~MeV for $^{100}$Mo and $[2.6-3.2]$~MeV for $^{82}$Se. The lower energy limit for $^{82}$Se is due to a lower $\beta\beta 2\nu$ background (higher half-life and lower available mass of source).

\begin{table}[!h]
\centering
\begin{tabular}{c|c|cc|cc}
        & & \multicolumn{2}{c}{Phase 1 (1 yr)}          & \multicolumn{2}{c}{Phase 2 (3.5 yrs)}      \\
Isotope & Energy & \# observed  & \# expected  & \# observed  & \# expected  \\
Isotope & window &  events &  background &  events &  background \\
\hline
\hline
$^{100}$Mo & [2.8-3.2] MeV & 6  & 5.3$\pm$0.6  & 12  & 11.1$\pm$0.9  \\
$^{82}$Se  & [2.6-3.2] MeV& 4  & 3.6          & 10  & 7.3           \\
\end{tabular}
\caption{Number of observed events and expected background.}
\label{tab:nemo3-counts-bb0nu}
\end{table}

\begin{itemize}
\item[\textbullet] For $^{100}$Mo: between 2.8 and 3.2~MeV, during Phase 1, 6 events have been observed for 5.3$\pm$0.6 expected background events, and during Phase~2 (until end of 2009), 12 events have been observed for 11.1$\pm$0.9 expected background events. 
\item[\textbullet] For $^{82}$Se: between 2.6 and 3.2~MeV, during Phase 1, 4 events have been observed for 3.6 expected background events, and during Phase~2 (until end of 2009), 10 events have been observed for 7.3 expected background events. 
\end{itemize}

\subsubsection*{$\beta \beta 0 \nu$ processes with V+A current or Majoron emission}

In the case of $\beta \beta 0 \nu$-decay with right-handed V+A weak current, the angular distribution between the two emitted electrons is expected to be different, compared to the ``standard'' $\beta \beta 0 \nu$-decay process.
Therefore NEMO-3 is the most sensitive $\beta\beta$ detector for the search of the V+A $\beta \beta 0 \nu$-process. 

A maximum likelihood fit using not only the energy sum of the two electrons but also the angular distribution and the single electron energy has been applied~\cite{nemo3-hdr-laurent}. No excess of events has been observed and limits on the half-life of the process have been derived for $^{100}$Mo and $^{82}$Se~\cite{nemo3-taup11}:
$$\mathrm{T_{1/2}^{\beta \beta 0 \nu,~V+A}(^{100}Mo) > 5.7~10^{23}~years~(90\%~C.L.) }$$
$$\mathrm{T_{1/2}^{\beta \beta 0 \nu,~V+A}(^{82}Se) > 2.4~10^{23}~years~(90\%~C.L.) }$$
It corresponds to a limit on the $\langle \lambda \rangle$ parameter of the right-handed V+A current 
$$\mathrm{ \langle \lambda \rangle < 1.4~10^{-6}~(90\%~C.L.)~~~~(^{100}Mo) }$$
$$\mathrm{ \langle \lambda \rangle < 2.0~10^{-6}~(90\%~C.L.)~~~~(^{82}Se) }$$

The $\beta\beta$ energy sum spectrum of the two electrons is expected to be distorded, in the case of the emission of a Majoron. 
Limits obtained only with Phase~1 data are sumarized in Table~\ref{tab:nemo3-majoron}~\cite{nemo3-majoron}.

\begin{table}[!h]
\centering
\begin{tabular}{ccccc}
\hline
\hline
Isotope    & $n=1$ & $n=2$ & $n=3$ & $n=7$  \\
\hline
$^{100}$Mo & $> 2.7 \ 10^{22}$ y & $> 1.7 \ 10^{22}$ y & $> 1.0 \ 10^{22}$ y & $> 7 \ 10^{20}$ y  \\
$^{82}$Se  & $> 1.5 \ 10^{22}$ y & $> 6.0 \ 10^{21}$ y & $> 3.1 \ 10^{21}$ y & $> 5 \ 10^{20}$ y  \\
\hline
\hline
\end{tabular}
\caption{Limits obtained with Phase 1 data on $T_{1/2}(\beta \beta 0\nu)$ in the case of $\beta\beta 0 \nu$-decay with a Majoron emission, as a function of the spectral index~\cite{nemo3-majoron}.}
\label{tab:nemo3-majoron}
\end{table}



\subsubsection*{$\beta \beta$-decay to excited states}

The tracko-calo technique allows to study the $\beta \beta$-decay for transitions involving excited states with a direct detection of the $\gamma$'s emitted by the excited state, associated to the two electrons. 

\begin{figure}[!h]  
\centering
\includegraphics[scale=0.3]{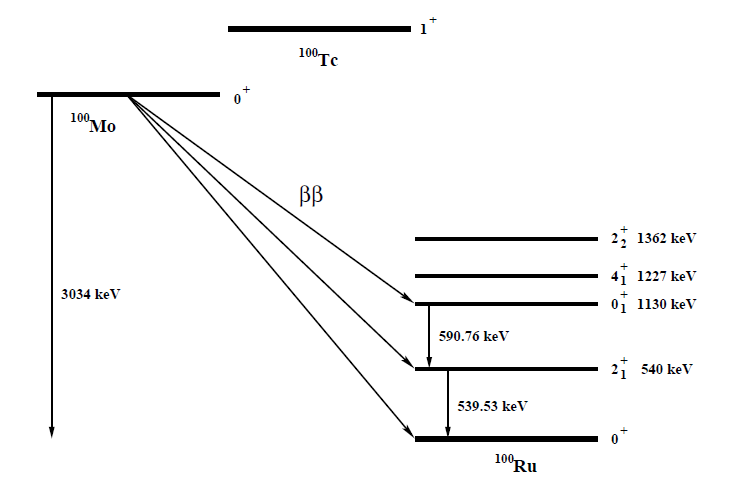}
\caption{A double beta decay scheme of $^{100}$Mo showing the decay to the ground and
excited states of $^{100}$Ru. The latter transitions are followed by de-excitation $\gamma$-rays.}
\label{fig:nemo3-excited-state}
\end{figure}

For the first time, the $\beta \beta 2 \nu$-decay of $^{100}$Mo to the excited state $0_1^+$ of $^{100}$Ru (see Figure~\ref{fig:nemo3-excited-state}) has been directly measured by selecting events containing two electrons and two $\gamma$'s (539.5 and 590.8~keV) with a time of flight analysis in agreement with the hypothesis that all the particles involved have been emitted from the same vertex on the foil and with a $\gamma$ energy distribution in agreement with expected Monte-Carlo simulations. The measurement of the half-life, using only Phase~1 data has been published in~\cite{nemo3-excited-state} and is:

$$ \mathrm{T_{1/2}^{\beta \beta 2\nu}(^{100}Mo \rightarrow ^{100}Ru,0_1^+) = 5.7^{+1.3}_{-0.9}(stat.) \pm 0.8(syst.)  \ 10^{20} \ y}$$

It corresponds to about 40 signal events observed with a signal-to-background ratio of about 4. Radon was the dominant background. 
Update analysis using Phase~2 data give a result in agreement with the published one but with a much larger signal-to-background ratio of about 30 (result not yet presented).

The limit on the $\beta \beta 0 \nu$-decay of $^{100}$Mo to the excited state $0_1^+$ of $^{100}$Ru has been also determined with Phase~1 data (the improved limit using Phase~2 is not yet public):

$$ \mathrm{T_{1/2}^{\beta \beta 0\nu}(^{100}Mo \rightarrow ^{100}Ru,0_1^+) > 8.9 \ 10^{22} \ y \ (90\%~C.L.)}$$

%
%

\section{SuperNEMO Experiment}

The SuperNEMO experiment is based on an extension and an improvment of the experimental technique
used in the NEMO-3 detector, by combining calorimetry and tracking. 
The goal is to accomodate about 100~kg of enriched $\beta\beta$ isotope in order to reach a sensitivity of 10$^{26}$~years on the $\beta\beta 0\nu$ half-life~\cite{supernemo-tdr}.

\begin{figure}[!h]  
\centering
\includegraphics[scale=0.4]{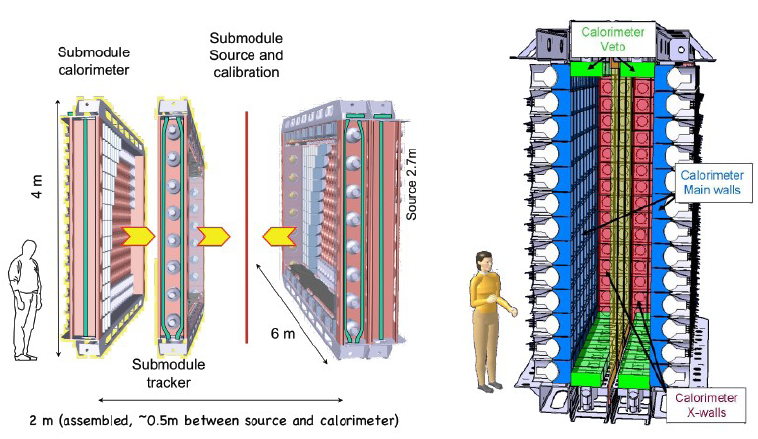}
\caption{Mechanical view of a SuperNEMO module.}
\label{fig:supernemo-layout}
\end{figure}

The SuperNEMO design is a planar geometry. The current design envisages about twenty identical modules, each housing 5~kg of enriched $\beta\beta$ isotope. 
The design of a SuperNEMO detector module is shown in Figure~\ref{fig:supernemo-layout}. 
The source is a thin (40 mg/cm$^2$) foil inside the detector. It is surrounded by a gas tracking chamber followed by calorimeter walls. 
The tracking volume contains 2000 wire drift cells operated in Geiger mode which are arranged in nine layers parallel to the foil. 
The calorimeter is divided into 500 to 700 plastic scintillator blocks (depending of the final size) which cover most of the detector outer area and are coupled to low radioactive PMT's.
A $\gamma$ veto is added on the top and bottom (calorimeter veto) and lateral sides (calorimeter X-walls, see Figure~\ref{fig:supernemo-layout}) of the tracking chamber, allowing a $4\pi$ $\gamma$ coverage.

The choice of isotope for SuperNEMO is aimed at maximising the $\beta\beta 0\nu$ signal over the background of $\beta\beta 2\nu$-decay and other nuclear decays mimicking the process.
Therefore the isotope must have a long two-neutrino half-life, a high endpoint energy and a large phase space factor. 
And of course, the possibility of isotopic enrichment on a large scale is also a factor for the selection of the isotope. 
Today, the isotope for SuperNEMO is $^{82}$Se. The $\beta\beta 2\nu$ half-life of Se is about 14 times higher than for Mo. Therefore, for a constant energy resolution, the contribution of the $\beta\beta 2\nu$ background in the $\beta\beta 0\nu$ energy region with 100~kg of $^{82}$Se (case of SuperNEMO) is identical to the $\beta\beta 2\nu$ background with 7~kg of $^{100}$Mo (NEMO-3 case)~! In parallel, SuperNEMO plans to improve by a factor of two the energy resolution in order to reduce by a factor two the $\beta\beta 2\nu$ background. 
The SuperNEMO collaboration is also investigating the possibility of enriching large amounts of $^{150}$Nd via the method of atomic vapour laser isotope separation (AVLIS). 
AVLIS is based on isotope-selective optical pumping by very narrow bandwidth laser followed by efficient infrared photo-ionization. 
A 50$\%$ rate of enrichment seems to be feasible in a reasonable time scale. 
One could argue that the $\beta\beta 2\nu$ half-life for Nd and Mo are almost equal. However, the large number of valence electrons in Nd produces a Coulomb screening which disfavours the emission of the two $\beta$'s with maximum available energy $Q_{\beta\beta}$, and consequently reduces the number of  $\beta\beta 2\nu$ background events around $Q_{\beta\beta}$ in the $\beta\beta 0\nu$ energy region. 
Thus $\beta\beta 2\nu$ background is expected to be roughly the same with $^{150}$Nd or $^{82}$Se. 
Finally, new collaborators from South Korea are in charge of providing significant amount of $^{48}$Ca (1~kg within 3 years). Enrichment is performed using the same AVLIS method as $^{150}$Nd~\cite{ca48-enrichment}.

The expected sensitivity of SuperNEMO on $T_{1/2}(\beta \beta 0\nu)$ at 90$\%$~C.L. is given in Table~\ref{tab:supernemo-sensitivity} for the three possible isotopes. 
With 100~kg of Nd (and 50$\%$ enrichment in $^{150}$Nd), the SuperNEMO sensitivity becomes $3 \ 10^{25}$~years (90$\%$~C.L.). It is still very competitive to other isotopes because of the large phase factor for $^{150}$Nd. Morever, the $\beta\beta$ transition energy for Nd is $Q_{\beta\beta}=3.367$~MeV, above the transition energy of $^{214}$Bi $Q_{\beta}=3.27$~MeV. Thus $^{214}$Bi and Radon cannot contribute anymore to the  background in the $\beta \beta 0\nu$ energy region. It makes the life much easier !
Life is even much easier if one uses $^{48}$Ca. Its $\beta\beta$ transition energy is $Q_{\beta\beta}=4.274$~MeV, near the transition energy of $^{208}$Tl $Q_{\beta}=4.99$~MeV. The requirement for the $^{208}$Tl radiopurity of the $\beta\beta$ foils is thus much less stringent ($< 100 \mu$Bq/kg) than for $^{82}$Se and $^{150}$Nd ( $< 2 \mu$Bq/kg, see below).

\begin{table}[!h]
\centering
\begin{tabular}{c|c}
Isotope & Limit on $T_{1/2}(\beta \beta 0\nu)$ (90$\%$~C.L.)\\
\hline
\hline
  $^{82}$Se                         &   $> 10^{26}$ y       \\
  $^{150}$Nd (50$\%$ enrichement)   &   $> 0.3 \ 10^{26}$ y       \\
  $^{48}$Ca  (50$\%$ enrichement)   &   $> 0.75 \ 10^{26}$ y       \\
\end{tabular}
\caption{Sensitivity of SuperNEMO on $T_{1/2}(\beta \beta 0\nu)$ at 90$\%$~C.L. for various isotopes. See text for background discussion.}
\label{tab:supernemo-sensitivity}
\end{table}

To reach the sensitivity of $10^{26}$~years with 100~kg of $^{82}$Se and 5~years of data, the required level of background in the $[2.8-3.2]$~MeV energy window\footnote{It corresponds to the energy window which optimizes the sensitivity.} must be around 1~count per year with the following background budget: a  $\beta \beta 2 \nu$ background less than $\approx$~0.25~count/year, a background due to Radon contamination inside the tracking chamber less than $\approx$~0.25~count and a background due to $^{208}$Tl and $^{214}$Bi contamination inside the $\beta \beta$ source foils less than $\approx$~0.5~count. 

\begin{figure}[!h]  
\centering
\includegraphics[scale=0.35]{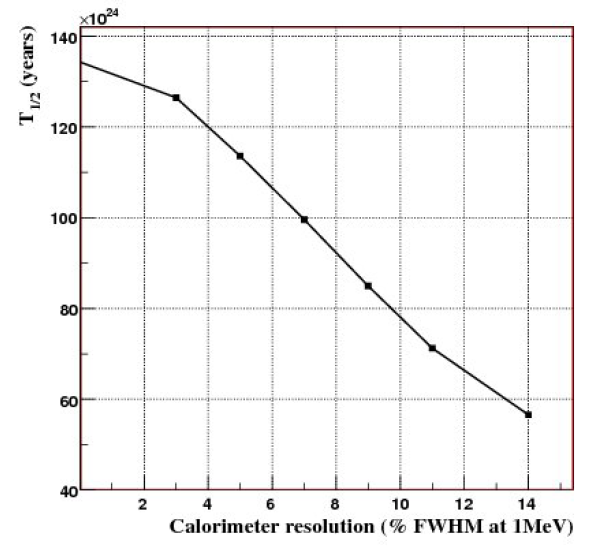}
\includegraphics[scale=0.45]{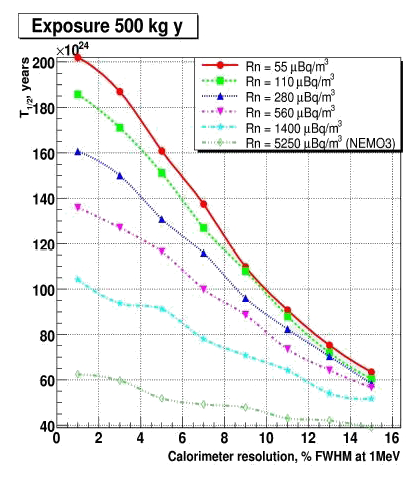}
\includegraphics[scale=0.35]{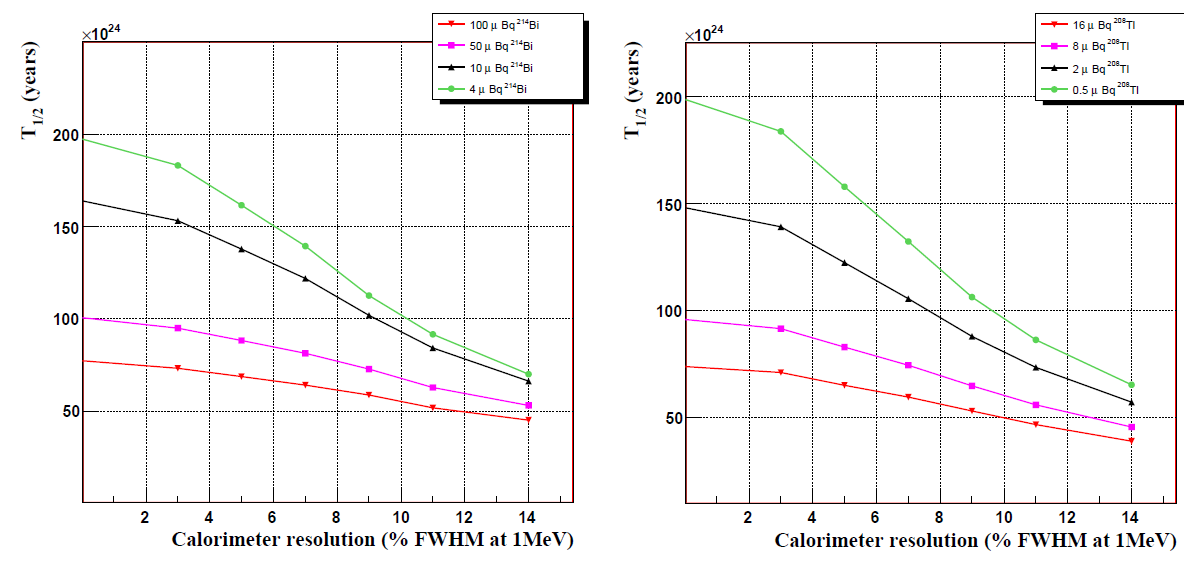}
\caption{Sensitivity of SuperNEMO to $T_{1/2}(\beta\beta 0\nu)$ at (90$\%$~C.L.) as a function of the energy resolution (FWHM at 1 MeV) assuming $^{82}$Se foils, for an exposure of 500~kg.year. Upper left: with 10~$\mu$Bq/kg of $^{214}$Bi and 2~$\mu$Bq/kg of $^{208}$Tl and no Radon; Upper right: assuming several Radon activity (but without any $^{214}$Bi nor $^{208}$Tl internal contamination); Lower left: for 4~$\mu$Bq/kg (green), 10~$\mu$Bq/kg (black), 50~$\mu$Bq/kg (cyan) and 100~$\mu$Bq/kg (red) of $^{214}$Bi (assuming no $^{208}$Tl contamination and no Radon); Lower right: for 0.5~$\mu$Bq/kg (green), 2~$\mu$Bq/kg (black), 8~$\mu$Bq/kg (cyan) and 16~$\mu$Bq/kg (red) of $^{208}$Tl (assuming no $^{214}$Bi contamination and no Radon).}
\label{fig:supernemo-bkg-simu}
\end{figure}

Therefore, the three main improvements required for SuperNEMO (compared to NEMO-3) are
\begin{enumerate}
\item the energy resolution and the energy calibration and survey (in order to reduce the $\beta \beta 2 \nu$ background)
\item the radiopurity of the $\beta \beta$ source foils (in order to reduce the $^{208}$Tl and $^{214}$Bi background)
\item the Radon purity inside the tracking chamber.
\end{enumerate}
These parameters have been studied extensively during the current design study phase with Monte-Carlo simulations. The SuperNEMO sensitivity calculated by Monte Carlo simulations  as a function of these parameters, is summarized in Figure~\ref{fig:supernemo-bkg-simu}. Their requirements (to reach a sensitivity of 10$^{26}$~years) are summarized in Table~\ref{tab:nemo3-supernemo-compare} and compared to NEMO-3 characteristics. 
The energy resolution must be improved by a factor 2. 
We will see that it has already been obtained with prototypes. 
The Radon and $^{208}$Tl contaminations measured in NEMO-3 must be reduced by a factor $\approx$~50 in SuperNEMO.
I will present the achievements and current status of this three issues. All other details are given in~\cite{supernemo-tdr}.

\begin{table}[!h]
\centering
\begin{tabular}{c|c|c}
 & NEMO-3 & SuperNEMO\\
\hline
\hline
 Isotope &  $^{100}$Mo & $^{82}$Se ($^{150}$Nd, $^{48}$Ca)\\
 Mass    &  7 kg       & 100 kg   \\
\hline
 Energy resolution   & 15$\%$  & 7$\%$\\
 (FWHM @ 1 MeV)      &   & \\
\hline
 Source radiopurity  &   & \\
 $^{208}$Tl          & $\approx 100 \mu$Bq/kg  & $<2 \mu$Bq/kg \\
 $^{214}$Bi          & $<300 \mu$Bq/kg         & $<10 \mu$Bq/kg\\
\hline
 Radon               & $\approx 5$ mBq/m$^3$                & $<0.1$ mBq/m$^3$\\
\end{tabular}
\caption{Comparison of the NEMO-3 characteristics with the SuperNEMO requirements.}
\label{tab:nemo3-supernemo-compare}
\end{table}

\subsection{Energy resolution}

An energy resolution of FWHM$=7\%$ at 1 MeV is required to reach a sensitivity of $T_{1/2}(\beta \beta 0\nu) > 10^{26}$~years for an exposure of 500~kg.year with $^{82}$Se, and assuming a radiopurity of the foils of 2~$\mu$Bq/kg in $^{208}$Tl and 10~$\mu$Bq/kg in $^{214}$Bi (see Figure~\ref{fig:supernemo-bkg-simu}). 

An extensive R$\&$D program has been performed in order to study how to reach an energy resolution of FWHM$=7\%$ at 1 MeV. 
Only organic scintillators have been studied because it limits the backscattering of electrons and guarantee a very high radiopurity.
The energy resolution has been improved thanks to three modifications:
\begin{enumerate}
\item The surface of the PMT photocathode has been increase (in comparison to the entrance surface of the scintillator)  by  using larger 8'' PMT's
\item The PMT has been coupled directly to the scintillator in order to avoid any extra light loss through the optical guide
\item The characteristics of the scintillator blocks have been improved by selecting the best available organic scintillating material and by testing the best geometry and wrapping of the scintillators to optimize the collection of the scintillating photons to the photocathode.
\end{enumerate}

Other issues have been also considered.
The granularity of the calorimeter has been optimized in order to minimize the $\gamma$ pile-up in the scintillator blocks with respect to cost considerations. 
The final choice of the geometry layout is the result of a compromise between the geometry efficiency and the number of channels. 
The thickness of the scintillator block is constrained by the $\gamma$ tagging efficiency, which must be greater than 50\% at 1 MeV

\begin{figure}[!h]  
\centering
\includegraphics[scale=0.4]{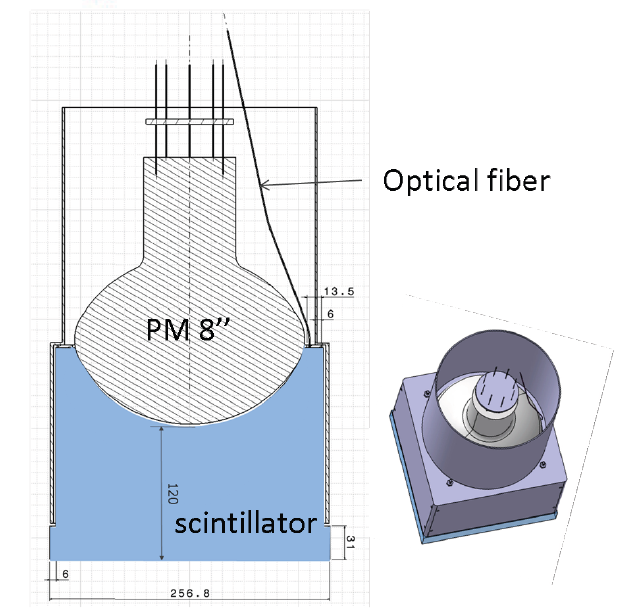}
\caption{Mechanical view of an optical module of the calorimeter: a scintillator block coupled to a 8'' PMT and surrounded by a magnetic shield.}
\label{fig:supernemo-calo-block}
\end{figure}

A mechanical view of an optical module of the calorimeter is presented in Figure~\ref{fig:supernemo-calo-block}. 

\subsubsection*{Photomultipliers}

Today the 8'' photomultipliers developed for the SuperNEMO project have quantum effciency (QE)
around 35\% at 420 nm. Such performances have been achieved after developments on smaller size of tubes.
Tens of 3'' and 5'' Photonis PMT's have been tested at CENBG before extrapolating the high QE
process for 8'' tubes. First measurements on the energy resolution of such large tubes highlighted
defects on the vacuum quality of the tubes and on the collection
of photoelectrons. Such defects have been solved by Photonis and Hamamatsu on their new
tubes registered as XP1886 for Photonis and R5912MOD for Hamamatsu.
Studies and developments have been necessary to guaranty the linearity of the 8'' PMts. The linearity
of these tubes has been tested with calibrated electrons up to 2 MeV and satisfies the requirements. 
Tests with LED light are under progress to confirm these performances up to 8 MeV.
The performances on the energy resolution depend also on the homogeneity of the PMt's cathode
as scintillating photons illuminate the entire photocathode. Important improvements have been also realized by
Hamamatsu to increase the quantum efficiency on the sides of the phototocathode. 
Finally, Hamamatsu PMT's have been selected, since Photonis decided to stop producing PMT's (a strange and unintelligible capitalism decision...).

\subsubsection*{Scintillator blocks}

The design study of the scintillators is based on organic materials which limit the backscattering of electrons and guarantee a very high radiopurity. The goal of the study was to define the best material, the geometry and the wrapping of the scintillators to optimize the collection of the scintillating photons to the photocathode.
Several geometries have been designed and studied by GEANT4 simulations and measurements with the electron spectrometers and $^{207}$Bi sources. These studies and the mechanical constraints led to the choice of a cubic block with two possible entrance surfaces of 257~mm$^2$ or 308~mm$^2$ and a minimal thickness of 120~mm to detect the $\gamma$-rays . 
The light collection is improved by a wrapping of Teflon on the sides of the blocks and 6~$\mu$m of aluminized Mylar on the entrance face.
The main walls of the calorimeter consist in 374 or 520 optical modules depending on the scintillator size. 

Two materials have been selected and tested: polystyrene-base scintillator produced in JINR Dubna and Poly-Vinyl-Toluene (PVT). 
The two different geometries of blocs have been characterized with the electron spectrometers for PS and PVT scintillators coupled to the Hamamatsu R5912MOD PMT (baseline PMT for SuperNEMO).
Table~\ref{tab:supernemo-calo-resol} presents the energy resolution obtained at 1~MeV. 
PVT has a larger light yield and provides a higher energy resolution. However its production is much more expansive than standard polystyren scintillator.  Larger scintillator block allow to reduce the number of channels. However, the energy resolution is also reduced and there is a larger risk of $\gamma/\beta$ pile-up, which reduces the capability to reject $\gamma$-rays from background.

\begin{table}[!h]
\centering
\begin{tabular}{c|c|c}
 & Cubic block & Cubic block \\
 & $257 \times 257$~mm$^2 $ & $308 \times 308$~mm$^2$  \\
\hline
\hline
 Polystyrene &  $8.7 \pm 0.1 \%$ & $9.8 \pm 0.1 \%$ \\
\hline
 Poly-Vinyl-Toluene &  $7.3 \pm 0.1 \%$ & $8.3 \pm 0.1 \%$ \\
\end{tabular}
\caption{Energy resolution (FWHM) for 1~MeV electrons with scintillators of different sizes and
materials.}
\label{tab:supernemo-calo-resol}
\end{table}

\begin{figure}[!h]  
\centering
\includegraphics[scale=0.6]{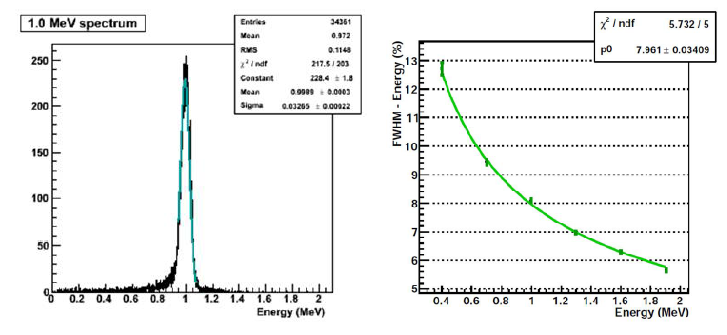}
\caption{(Left) Charge of 1~MeV electron measured with a PVT cubic block (256mm$^2$ entrance surface)
coupled to a R5912MOD PMT. The energy resolution has to be corrected by -0.5$\%$ due to the energy
losses of the electrons in the trigger system. (Right) Energy resolution as a function of the energy of the electron and fit of the square root function.}
\label{fig:supernemo-calo-result}
\end{figure}

These results have been confirmed with several blocks and phototubes, and with different test benches.
The energy spectrum measured with the 257~mm$^2$ PVT bloc is given in Figure~\ref{fig:supernemo-calo-result}.
It has been verified that the energy resolution of the optical module is inversely proportional to the square root of the energy of the electron, as shown in Figure~\ref{fig:supernemo-calo-result}.

\subsection{Radiopurity of the source foils}

Figure~\ref{fig:supernemo-bkg-simu} shows the SuperNEMO sentitivity for various levels of internal contaminations in inside the source foils. Selenium foil radiopurities of 2~$\mu$Bq/kg in $^{208}$Tl and 10~$\mu$Bq/kg in $^{214}$Bi are required to reach a sensitivity of $10^{26}$~years after 500~kg.yr with $^{82}$Se.
Thus enriched selenium must be first purified, then foils must be produced and finally its radiopurity must be controlled. 
Nevertheless, the best detection limit that can be reached by $\gamma$ spectrometry with ultra low background HPGe detectors is around 50~$\mu$Bq/kg for $^{208}$Tl, which is one order of magnitude less sensitive than the required value. 
In order to achieve the required  sensitivity for SuperNEMO, the collaboration has decided to develop a BiPo detector (under the responsability of LAL Orsay), dedicated to the measurement of ultra-low levels of contamination in $^{208}$Tl and $^{214}$Bi in the foils of SuperNEMO. The development of the BiPo detectors is presented separately in the second part of this document. I present here the status of the selenium purification.

\subsubsection*{Purification}

The selenium can be purified by two different processes: the chemical process and the distillation process.

The principle of the chemical purification is to remove long lived radioactive isotopes of the $^{238}$U and $^{232}$Th decay chains while filling Radium sites with Barium by spiking samples during the processing.
The process takes advantage of an equilibrium break in the $^{238}$U and $^{232}$Th decay chains, which can selectively transform these chains to non-equilibrium states in which only short lifetime daughters exist. So, the objective is to remove $^{226}$Ra in the $^{238}$U chain, and $^{228}$Ra and $^{228}$Th in the $^{232}$Th chain.

This technique had been used by INL (Idaho National Laboratory, USA) to purify enriched $^{100}$Mo powder for NEMO-3~\cite{nemo3-purif} and an extraction factor of 100 for $^{226}$Ra was measured. 
The selenium purification process is similar to the one used for molybdenum and very high extraction factor can be also achieved.
The critical point of this method is to use very pure chemical products (water and acid have to be distillate), especially barium chloride. NEMO-3 purification succeeded by using an old and ultra radiopure barium chloride reagent. 
About 0.5~kg of enriched $^{82}$Se has been purified by INEL with this technique. Table~\ref{tab:supernemo-purif-chemical} shows  the chemical extraction factors deduced from the activities measured by $\gamma$-ray spectroscopy before and after purification. Only lower limits have been obtained due to the limited sensitivity of the HPGe measurement. However a factor at least 50 has been obtained for $^{226}$Ra. 
We also mention that the sample was slightly contaminated in Ruthenium with a measured extraction ratio of 162.

\begin{table}[!h]
\centering
\begin{tabular}{c|c|c|c|c|c|c|c|c}
Sample   & Mass & Time     & $^{40}$K & $^{60}$Co & $^{226}$Ra & $^{228}$Ra & $^{228}$Th & Ru \\
(mBq/kg) & (g)  & (h)      &  &  &  &  &  &  \\
\hline
\hline
 $^{82}$Se & 196  &  447  & $668 \pm 31$ & $<1$ & $46 \pm 2$ & $13 \pm 2$ & $11 \pm 2$ & 485\\
 Purified $^{82}$Se& 563  &  436  & $<20$ & $<0.7$ & $<0.9$ & $<2.4$ & $<1.6$ & 3\\
\hline
 Reduction &   &   & $>33$ &   & $>51$ & $>5.4$ & $>6.9$ & 162 \\
 Factor    &   &   &   &   &   &   &   & \\
\end{tabular}
\caption{Table of the chemical extraction factors obtained for an enriched $^{82}$Se sample purified in INEL. The purification factors are deduced from the activities (in mBq/kg) measured by $\gamma$-ray spectroscopy before and
after purification. Only lower limits are obtained for K, U and Th chains.}
\label{tab:supernemo-purif-chemical}
\end{table}

Another radiochemical process is currently under R$\&$D in LSM in collaboration with JINR Dubna. 
It proposes to use resin to retain specific radio impurities.
This technique is very efficient because in aqueous media, selenium is dissolved as
anions and impurities such as Thorium, Radium and Uranium are cations. The use of cationic resin (for instance
Dowex1X50) would retain only impurities and would keep selenium free. 
The advantage of this method is that chemical products are purified as well as selenium. 
The addition of a reverse cleaning of the resin allows to use continuously the same resin with a constant purification efficiency. 
There is a possibility to automatize the method in order to purify large amounts of enriched isotopes and this is under investigation. 
Such a method could be used also for neodynium purification. 

Selenium can also be purified by distillation. Melting and boiling points depend on the structures of the raw selenium compounds. For instance, the RaSe structure has to be taken into account in the case of radium contamination. Distillation of 2~kg of natural selenium has been performed by the Institute of Chemistry of High Purity Substances (IChHPS) in Russia. However, the final products are under the form of metallic selenium ingots and must be first transformed in the form of powder before measuring it. New distillation process leading to selenium powder with small grain has to be developped.

\subsubsection*{Foil production}

The SuperNEMO source foils will be produced using the method developped for NEMO-3. 
First selenium powder with diameters smaller than $\approx 50 \mu$m is required. 
Then the powder is mixed with an organic binder (PolyVinyl Alcohol, named PVA) and ultra pure water. 
The binder contribute to $\approx 10\%$ in weight to the final product.
This paste is uniformly ($\pm 10\%$) deposited between two holey Mylar strips, called baking film. The thickness of the Mylar is 10~$\mu$m (corresponding to about 1.4~mg/cm$^2$). 
For the SuperNEMO demonstrator, the thickness of the source foils is 55~mg/cm$^2$ (comparable to NEMO-3) in order to accomodate 7 kg of source foil. For the final SuperNEMO detector, the thickness will be 40~mg/cm$^2$.
In this process the losses of enriched product are less than 1$\%$ and it is relatively easy to recover the enriched material. 
Another advantage of this method is the possibility to use any metallic powder (Nd, Ca,...).

One of the issue is to obtain very radiopure holey Mylar foils. 
Mylar must undergo special treatment in which a large number of microscopic holes are produced in order to ensure a good bond with the PVA glue and also to allow the water evaporation during the drying of the glue.
For NEMO-3, the holes have been produced by irradiating the Mylar at JINR with an ion beam and then by etching the Mylar in a Sodium hydroxide solution. 
Although all the chemical products used in this process have been selected for their radiopurity they
are probably a source of residual contamination of the NEMO-3 source foils.

For SuperNEMO, a new approach based on laser drilling is currently developped by CPPM Marseille in collaboration with LP3 (Lasers, Plasmas et Proc\'ed\'es Photoniques, Universit\'e de la M\'editerran\'e), in order to reduce the risk of radio contamination. They have demonstrated that a picosecond UV laser can produce holes of 100$\mu$m diameters in a mylar foil of 20~$\mu$m thick. 
First samples of holey Mylar (few hundreds cm$^2$) have been produced using this method. The production of a natural selenium foil with this Mylar sample is under test. The main advantage of this laser technique is that there is no risk of radio contamination.

The use of polyesther mesh is also studied by the collaboration.

The other requirements to produce an ultra radiopure selenium foils are:
\begin{itemize}
\item Select a radiopure mylar at the level of $\approx 15 \mu$Bq/kg\footnote{We assume here that both the Mylar and PVA contributions to the source foil contamination are each at the level of 1~$\mu$Bq/kg in $^{208}$Tl and 10~$\mu$Bq/kg in $^{214}$Bi.} in $^{208}$Tl and $\approx 75 \mu$Bq/kg in $^{214}$Bi, since Mylar is $\approx 7\%$ in weight of the complete source foil.
\item Select PVA at the level of $\approx 10 \mu$Bq/kg in $^{208}$Tl and $\approx 50 \mu$Bq/kg in $^{214}$Bi,  since the PVA concentration is $\approx 10\%$ in weight of the complete source foil;
\item If the grain sizes of the purified selenium (at the end of the purification process) are too large (diameter be less than $\approx 50 \mu$m), selenium must be grained using a radiopure technique. 
\end{itemize}

\subsubsection*{Radiopurity measurement}

A first radiopurity measurement of purified selenium samples can be performed by $\gamma$ spectrometry with ultra low background HPGe (High Purity Germanium) detectors. Nevertheless, the best detection limit that can be reached with this technique for $^{208}$Tl is around 50~$\mu$Bq/kg, which is one order of magnitude less sensitive that the required value. 
In order to achieve the required  sensitivity for SuperNEMO, we are developping the BiPo-3 detector dedicated to the measurement of ultra-low levels of contamination in $^{208}$Tl and $^{214}$Bi in the foils of SuperNEMO.  
This detector will be able to qualify the radiopurity of the foils in their final form,  before installation into the SuperNEMO detector. It can also validate the radiopurity of the selected Mylar and PVA. 
The BiPo-3 detector will be installed in Canfranc Underground Laboratory (Spain) in summer 2012. 
A complete description of the BiPo detector is presented in the second part of this document.

\subsection{Strategies against Radon contamination inside the gas of the tracking chamber}

Figure~\ref{fig:supernemo-bkg-simu} presents the sensitivity of the full SuperNEMO detector with $^{82}$Se sources  as a function of the Radon activity, assuming a null activity of $^{214}$Bi of the source foils. 
Typically, a Radon contamination of 280~$\mu$Bq/m$^3$ is equivalent to a contamination of 10~$\mu$Bq/kg in $^{214}$Bi inside the source foils.
For the full SuperNEMO detector, the sensitivity gain on $T_{1/2}(\beta\beta 0\nu)$ obtained by reducing
the Radon activity below 0.1 mBq/m$^3$ is not significant (from $1.1 \ 10^{26}$~years to $1.4 \ 10^{26}$ years for a calorimeter resolution of 7$\%$ FWHM at 1 MeV).

We remind that the three most probable origins of Radon inside the NEMO-3 tracking chamber were the  emanation from materials used to wrap the scintillator blocks inside the tracking chamber, the poor tightness against Radon of the seals closing the chamber, and the diffusion of Radon from PMT's emanation or from external Radon in the lab. Taking into account this knowledge, the strategies developped against Radon for SuperNEMO are the following

\begin{enumerate}

\item The tracker will be isolated from the rest of the detector using a thin foil, very tight to Radon
and not degrading significantly the energy resolution performance of the calorimeter . 
With a 30$\mu$m thickness of EVOH foil, the Radon reduction rate measured by the UTEF-CTU laboratory is 31000. However pure EVOH foils are sensitive to humidity, which produces mechanical cracks.
Selection of appropriate films is under tests. The use of composite materials including EVOH or nylon is under study.

\item All materials inside the tracker have to be very radiopure. 
The Radon emanation of these materials has also to be measured. 
Radon emanation measurements will also be performed after the assembly of the tracker, to be sure that no mechanical process during assemblying induces a significative Radon contamination. 
A Radon concentration line, similar to the one developped in Heidelberg and used for BOREXINO, has been constructed by the UCL group to measure this emanation with a sensitivity of 0.2 mBq/m$^3$. 
The idea is to concentrate Radon emanation into a Radon trap like active charcoal or organic zeolith and then to measure Radon from the trap. 

\item Radon-free air will be flushed around the PMT glass, to avoid a too high Radon concentration
close to the PMT. The emanation of the PMT glass has been measured by the UNIBA Bratislava
laboratory at the level of $(5.7 \pm 2.1) 10^{-7}$~Bq/s. With the flushing of Radon-free air close to
the PMT, the contribution of the PMT emanation to the Radon level in the tracking chamber
of the demonstrator should be well below the required activity limit (contrary to the NEMO-3
detector, the calorimeter of the SuperNEMO demonstrator is installed outside the tracking
chamber). Morever, like in NEMO-3, the detector (tracker and calorimeter) will be surrounded by a Radon-tight structure. Radon-free air will be flushed inside this tent in order to prevent external Radon to diffuse inside the tracking chamber. 

\item The incoming gas used inside the tracking chamber (mixture of 95$\%$ of Helium, 4$\%$ of alcohol, 1$\%$ of Argon) needs also to be as low radiopure as possible in Radon. It should be purified and measured. The  usual technique (Radon trap by active charcoal at $-$50 degrees) can be applied to purify Helium gas. However, alcohol and Argon can be trapped and can saturate the Radon trap, which will become inefficient. The CPPM group in cooperation with chemistry laboratories, is studying the best adapted charcoal or organic zeolith able to trap Radon and not Argon nor alcohol.

\item The possibility to increase the gas flux should be kept under study. A typical flux of 3~m$^3$/h for the demonstrator volume of 20~m$^3$ would provide a Radon reduction factor greater than 7, if Radon contamination inside the incoming gas stays negligible. But recycling of the gas, gas purification, and helium-alcohol separation could be necessary for the whole SuperNEMO detector and must be studied seriously.

\end{enumerate}

\subsection{Demonstrator module}

A first SuperNEMO module, named SuperNEMO demonstrator, will be installed in Modane Underground Laboratory. 
The main goals of the demonstrator module are:
\begin{itemize}
\item[\textbullet] Demonstration of the feasibility of a large scale detector production with the requested
performances (calorimeter energy and time resolution, tracker efficiency and radiopurity).
\item[\textbullet] Measurement and validation of the Radon background contribution especially from internal materials outgasing.
\item[\textbullet] Measurement and validation of the background contribution from the detector components
\item[\textbullet] Finalization of the technical choices.
\end{itemize}

The size and the design of the demonstrator correspond to the final design of the SuperNEMO modules. The only difference is the thickness of the source foils which is 55~mg/cm$^2$ for the demonstrator (comparable to NEMO-3), instead of 40~mg/cm$^2$ for the baseline design, in order to accomodate 7~kg of source foil (instead of ~5~kg for standard SuperNEMO modules) and to optimize the physical performances of the demonstrator. 
With 7~kg of $^{82}$Se and 2.5 years of data taking, the sensitivity of the demonstrator will be 
$$\mathrm{ T_{1/2}^{0 \nu}(^{82}Se) > 6.6 \ 10^{24} \ years (90\% C.L.)}$$
It is equivalent to the expected sensitivity of GERDA Phase~1 of $3 \ 10^{25}$~years obtained with $^{76}$Ge, assuming equal nuclear matrix elements and using the phase space ratio for these two isotopes. 
Thus the sensitivity of the SuperNEMO demonstrator to the effective Majorana neutrino mass will be similar to GERDA Phase 1.

The construction of the tracking detector and the $\gamma$ veto (``calorimeter X-walls'' and ``calorimeter veto'') has been funded by UK and is under progress (UCL, MSSL, Manchester Univ.), for a delivery in LSM Modane in 2013 and a first underground commissioning. 
The calorimeter is under the French responsability and must be funded in the next year in order to be installed in 2014 in Modane. 
5~kg of enriched $^{82}$Se are available. Radiopurity measurement of the source foils is foreseen in 2013. 
First data of the SuperNEMO demonstrator could be taken in 2015.

%
%

\chapter{Germanium detectors}
\label{chap:germanium}

Ultra radiopure Germanium semiconductor diodes have been used historicaly as one of the first detectors for the direct search of $\beta\beta$-decay using source as detector~\cite{fiorini-1967}.
The $^{76}$Ge $\beta\beta$ emitter has unfortunately a relatively low transition energy $Q_{\beta\beta}=2039$~keV~\cite{douysset-2001}. Therefore germanium experiments are very sensitive to 2.6~Mev $\gamma$-rays from $^{208}$Tl, to standard cosmogenics~\footnote{Cosmogenics are long-lived cosmic ray induced isotopes produced during the germanium enrichment process or during the detector fabrication above ground} like $^{60}$Co ($T_{1/2}=5.272$~years) and $^{68}$Ge ($T_{1/2}=270$~days), and other possible contaminations like $^{42}$K recently observed in GERDA.
However Germanium detectors are today well established detectors, manufactured in large amount, and excellent energy resolution of few keV at 2~MeV is obtained when diodes are used at Liquid Nitrogen temperature. Also recent development of Broad-Energy Germanium detectors provides superior pulse shape discrimination performances for the reduction of multi-Compton background.


I will first briefly present the results obtained 10 years ago by Heidelberg-Moscou and IGEX experiments. I will then present the new experiment GERDA recently installed in LNGS and I will summarize its preliminary results and possible improvements.

\section{Heidelberg-Moscou and IGEX experiments}

The Heidelberg-Moscou experiment has been running during 13 years from 1990 until 2003 in Gran Sasso Underground Laboratory (LNGS, Italy). The experiment was carried out with 5 high purity germanium detectors with Ge enriched to 86$\%$ in $^{76}$Ge corresponding to a total of $\approx 10$~kg of $^{76}$Ge and a total exposure of 71.7~kg~year. The energy resolution integrated over 8 years and over the 5 detectors was 3.3~keV at $Q_{\beta\beta}=2039$~keV. 


Pulse shape analysis (PSA) has been developped to suppress the $\gamma$'s background events which correspond mainly to multi-sites events (MSE) due to multiple Compton interactions. In contrast, $\beta \beta$ events are confined to a few mm region in the detector (corresponding to the track length of the emitted electrons), and consequently appear as single-site events (SSE). The PSA has been tested with an external $^{228}$Th source. Standard $\gamma$ lines are reduced by a factor $\approx 5$ (20\%) while the known SSE line at 1593~keV, corresponding to double escape peak of the 2615~keV\footnote{The double escape peak correspond to the peak produced when incident gamma-rays of a certain energy interact with the detector by pair production and deposit all of their energy in the detector except 1022 keV, the difference in energy being the result of the escape of both of the 511 keV photons that are produced when the positron from the electron-positron pair is annihilated.}, is only reduced by a factor $\approx 2$ (62$\%$). The same order of reduction is found for the  $\gamma$ lines observed in a calibration measurement with a Ra source and in the full energy of the background spectrum.  

\begin{figure}[!h]  
\centering
\includegraphics[scale=0.3]{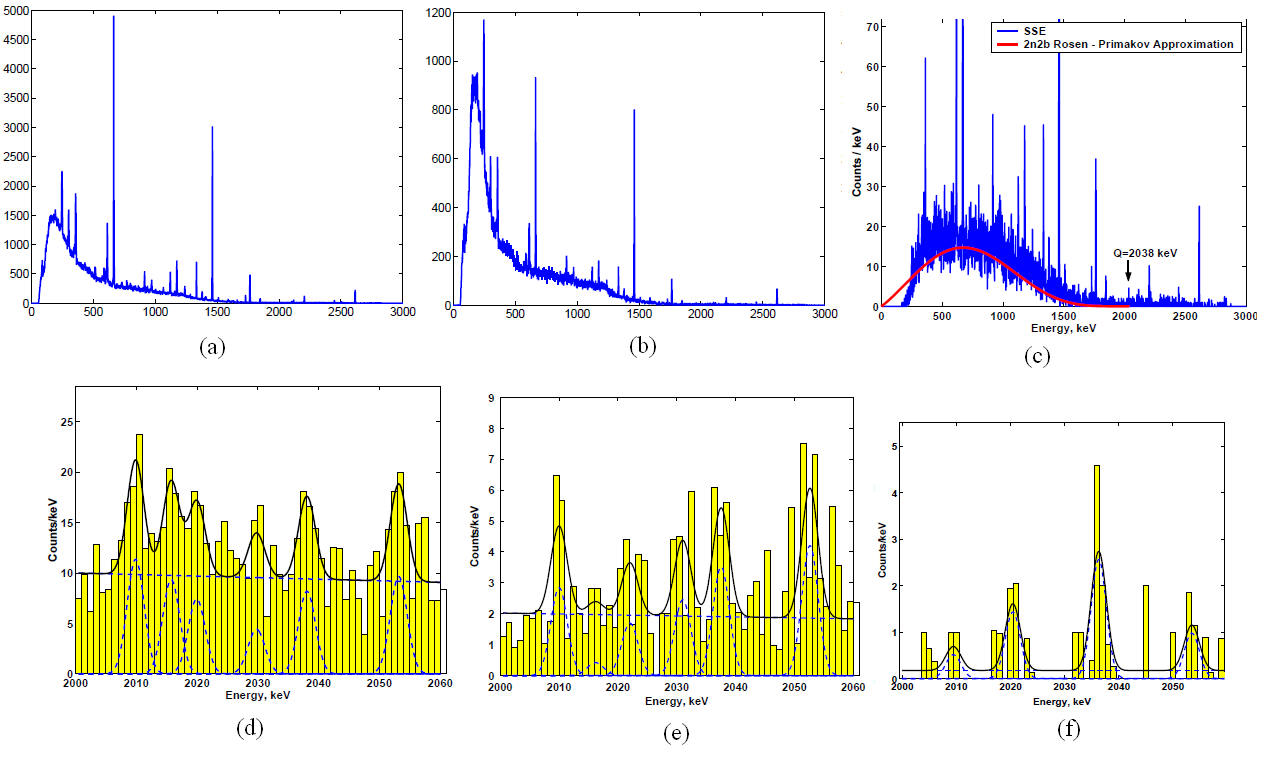}
\caption{Energy spectrum measured with 4 detectors from 1995 until 2003: (a) and (d) without pulse shape analysis (PSA); (b) and (e) with PSA Single Site events selection; (c) and (f) after specific $\beta \beta$ PSA~\cite{klapdor-2004}.}
\label{fig:heidelberg-moscow}
\end{figure}

Part of the collaboration has claimed a $3.3 \sigma$ indication of $\beta\beta 0\nu$ signal after analysing the full statistic and after applying the pusle shape analysis (often called {\it Klapdor's claim})~\cite{klapdor-2004}. When applying a specific pulse shape analysis using reference signal shapes created by double beta decay, the confidence level becomes $4.2 \sigma$. Figure~\ref{fig:heidelberg-moscow} shows the published energy spectrum measured with 4 detectors from 1995 until 2003, without PSA, after PSA SSE selection and after specific $\beta \beta$ PSA. 
We mention also that  the four lines 2010, 2016, 2022 and 2053 observed near the $Q_{\beta\beta}$ region, are consistent in intensity with normal known $\gamma$-lines. Possible origin of the unknown line at 2030~keV is discussed in ~\cite{klapdor-2004} and might be understood as originating from electron conversion of the $\gamma$-line at 2118~keV originating from $^{214}$Bi. 

This positive result has not been admitted by the rest of the collaboration.
The result indeed depends on the background description and on the reliability of the pulse shape analysis. 

One member of the collaboration has recently performed a complete analysis of the origins and descriptions of the background using measurements with sources and MC simulations. The model accounts for the $^{226}$Ra, $^{232}$Th and $^{60}$Co contaminations, the muon induced neutrons and the $\beta\beta 2\nu$ decay. The obtained value of the background using this model is ($11.8 \pm 0.5$) counts/keV, corresponding to 0.17~counts/(keV.kg.y), which is higher than the background used in the publication ($10.0 \pm 0.3$~counts/keV). 
Using this background and without PSA analysis, the peak significant at $Q_{\beta\beta}$ is only 1.3$\sigma$~\cite{chkvorets-2008} and the 68$\%$ C.L. interval of $T_{1/2}^{\beta\beta 0\nu}(^{76}Ge)$ is $[0.4 - 1.4] \ 10^{25}$~y with a central value of 2.2~10$^{25}$~y.

\begin{figure}[!h]  
\centering
\includegraphics[scale=0.4]{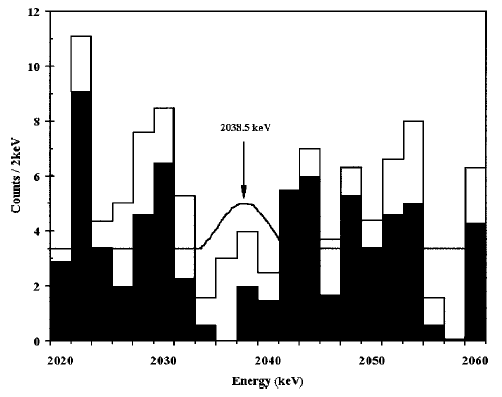}
\caption{Energy spectrum measured by the IGEX experiment with 8.87~kg.y of $^{76}$Ge exposure. The darkened spectrum results from the application of PSD to $\approx 45\%$ of the total data set. The Gaussian curve represents the 90$\%$~C.L. excluded $\beta \beta 0 \nu$ signal~\cite{igex-2}.}
\label{fig:igex}
\end{figure}

At the same time, another germanium experiment, named IGEX~\cite{igex-1}, has been carried out with three enriched $^{76}$Ge detectors of 2 kg each operating in the Canfranc Underground Laboratory (LSC, Spain) with pulse-shape analysis electronics, and three other smaller detectors in Baksan Low-Background Laboratory. A total exposure of 8.87~kg.y has been collected. The level of background was $\approx 0.2$~cts/(keV.kg.y). No excess of signal has been observed (see Figure~\ref{fig:igex} correponding to a limit of $T_{1/2}^{\beta\beta 0\nu}(^{76}Ge) > 1.57 \ 10^{25}$~years (90$\%$ C.L.)~\cite{igex-2}.

\section{The GERDA experiment}

The GERDA experiment consists of using bare Germanium detectors directly immersed inside Liquid Argon which acts both as cryogenic liquid and shield against external $\gamma$'s. 
Morever, the detection of the scintillation light emitted in Argon can be used as an active shield against external $\gamma$'s.

\begin{figure}[!h]  
\centering
\includegraphics{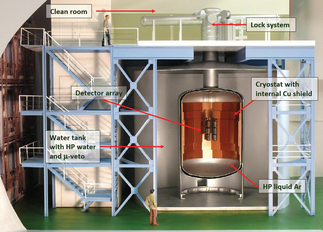}
\caption{Schematic view of the GERDA experiment.}
\label{fig:gerda-setup}
\end{figure}

\begin{figure}[!h]  
\centering
\includegraphics[scale=0.37]{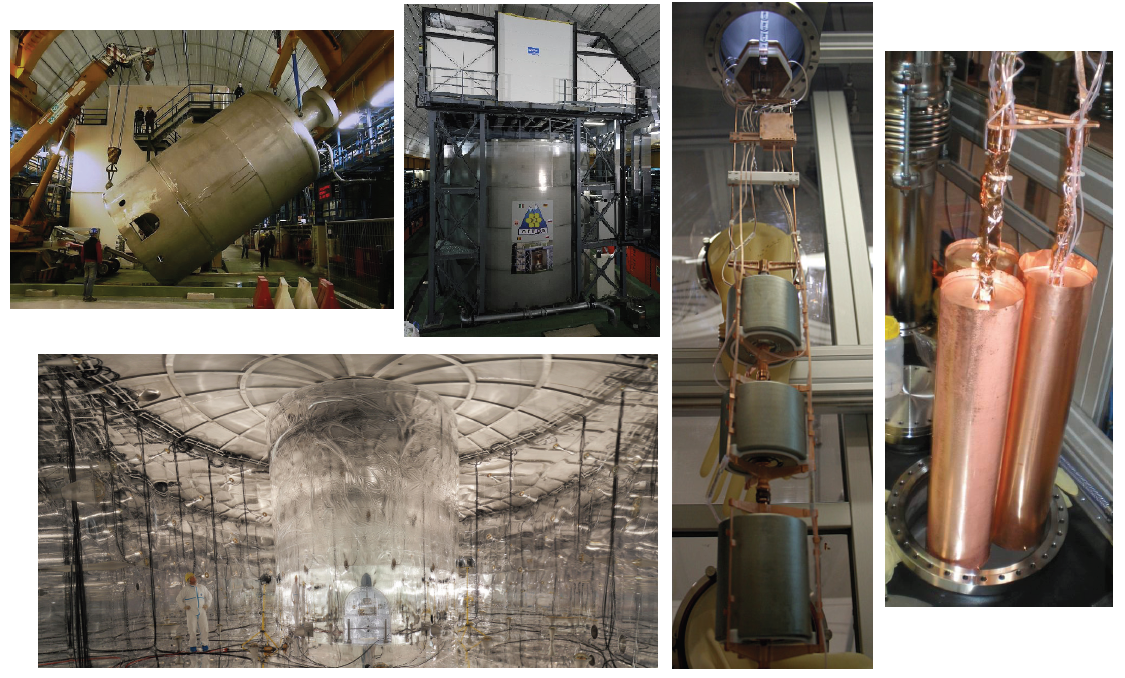}
\caption{Some pictures of the GERDA assembly: the stainless steal cryostat; the water tank and the clean room above; inner view of the water tank; deployment of the first array with three natural germanium detector; deployment in November 2011 of the 3 arrays with 8 enriched germanium detectors (17.7~kg of $^{76}$Ge, GERDA Phase 1).}
\label{fig:gerda-photos}
\end{figure}

The cryostat is a stainless steel tank ($\approx 25$~tons) with internal copper shield ($\approx 20$~tons) on the inner surface of the cylindrical part of the cryostat to suppress background contribution of the stainless steel. Its dimensions are an inner diameter of 4~m, and a total height of $\approx 4$~m. 
It contains about 100~tons of liquid Argon.
The cryostat is installed inside a larger water tank for extra $\gamma$ and neutrons shield (diam=10m, height=8.5m) (the attenuation length of a 2.615~MeV $\gamma$ in water is $\approx 0.04$~cm$^{-1}$.
The Ge detector array is made up of individual detector strings and is situated in the central part of the cryostat. This strings design allows to deploy crystals progressively inside the experiment (see Figures~\ref{fig:gerda-setup} and \ref{fig:gerda-photos}). Huge efforts and work have been done to ensure a ultra radiopure liquid Argon and selection of radiopure materials. 

The  background due to external $\gamma$'s (from the lab or from the cryostat) at $Q_{\beta\beta}=2039$~keV has been calculated by a Monte-Carlo simulation and is expected to be at the level of $10^{-4}$~cts/(keV.kg.y)~\cite{barabanov-2009}.

The first phase of data taking (Phase 1) has started in November 2011. The 8 enriched germanium detectors used previously in Heidelberg-Moscow (5) and IGEX (3) experiments and one extra  natural germanium crystal have been installed inside the cryostat. It corresponds to about 18~kg of enriched $^{76}$Ge. The goal is to reach a level of background of 0.01~cts/(keV.kg.y). It would correspond to a sensitivity of $\approx 3. \ 10^{25}$~y in 1 year of measurement. 

In the second phase (Phase 2), new broad-energy germanium (BEGe) detectors will be deployed. These detectors are good candidates to supress multi-site background events via a pulse shape analysis.  
The detection of the scintillation light from argon as external background veto will be also installed. 
The goal is to reach a level of background of 0.001~cts/(keV.kg.y). In total 35.5~kg of germanium enriched in the isotope $^{76}$Ge at 88$\%$ will be used to produce BEGe detectors. Adding the 18~kg of Phase~1, GERDA could measure up to $\approx 50$~kg of enriched $^{76}$Ge. A level of background of 0.001~cts/(keV.kg.y) in Phase 2 will lead to a sensitivity of $\approx 2. \ 10^{26}$y in 2 years of measurement. 

Depending of the achieved background, extra enriched germanium detectors could be produced and installed inside GERDA for a third phase. The goal is to reach a sensitivity of $\approx 10^{27}$y with 100~kg of enriched Ge and 10~years of measurement, with a background lower than 0.001~cts/(keV.kg.y).

\subsection{Phase 1}
\label{sec:gerda-phase1}

The liquid Argon cryostat is running since December 2009 and the operation of the cryostat is very stable from the begining. 

A single string supporting three natural germanium detectors has been first installed inside the liquid Argon cryostat for background measurement. And in June 2011, 3 enriched germanium detectors have been deployed. 

The energy resolutions measured with calibration runs are good. Preliminary values are FWHM~$\approx 4$~keV at 2615~keV for natural germanium detectors and 4 to 5~keV for enriched germanium detectors.

\begin{figure}[!h]  
\centering
\includegraphics[scale=0.35]{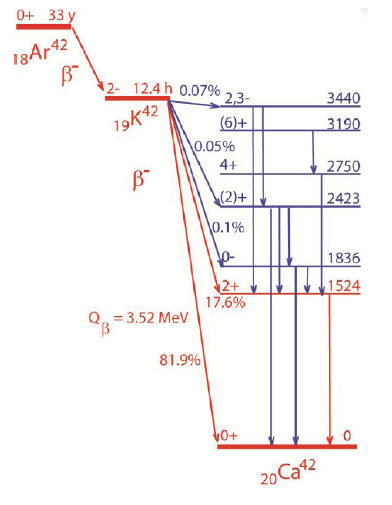}
\caption{Deacy diagram of $^{42}$Kr.}
\label{fig:gerda-kr42-scheme}
\end{figure}

However one prominent and unexpected background has been observed: the contribution of $^{42}$K, the progeny of $^{42}$Ar, about 20 times higher than expected. 
$^{42}$K is mostly a pure $\beta$ emitter (81.9$\%$ with $Q_{\beta}=3525$~keV and a half-life $T_{1/2}=12.36$~h, with a most intense $\gamma$-ray at 1524.7~keV (18.1$\%$) (see Figure~\ref{fig:gerda-kr42-scheme}). 
If $\beta$ and $\beta / \gamma$  from the $^{42}$K beta decay are emitted close to, or on the surface of the germanium crystals, they can produce events with energies above 1525~keV and ranging into the energy region of interest of $Q_{\beta\beta}$. 

The measured counting rate in the line 1525~keV ($\approx 1$~cts/(kg.day)) was more than one order of magnitude higher than when adopting the 90$\%$ upper limit of $^{42}$Ar/$^{40}$Ar~$< 4.3 \ 10^{-21}$ given by the DBA experiment assuming a homogeneous distribution in the liquid argon.
The reason is that $^{42}$K is positively charged after its production. It is then drifted by the moderate electrical fields produced by the germanium diodes bias towards the germanium detectors, which leads to a background enhancement. It has been verified with the LArGe facility that the counting rate of the $^{42}$Ar ($^{42}$K) signal depends on the electric fields in the liquid argon in the vicinity of the germanium detectors. 
In order to avoid $^{42}$K ions to drift towards the germanium crystals, two copper electrostatic shields have been installed inside the cryostat: a small closed cylinder surrounding the detectors array and a larger one, outside it, to protect against the convective liquid argon flow. The reduction of the 1525~keV $\gamma$ line has been successfully observed with the shrouds. 

In November 2011 (1.11.11 !), the 8 available enriched germanium detectors and 1 natural germanium have been deployed inside GERDA cryostat in the form of 3 arrays, each surrounded by a mini-shroud. It corresponds to the GERDA Phase 1 data taking. In the $\beta\beta 0\nu$ energy region, the preliminary observed level of background is $\approx 1.7 \ 10^{-2}$~counts/(keV.kg.y). It corresponds almost to the required level of background of the Phase 1.
A single dominant contribution to the background at $Q_{\beta\beta}$ cannot be identified and its understanding is statisticaly limited and will require longer measurements. Possible origins of background are the 2615~keV $\gamma$ line from $^{208}$Tl, the $^{214}$Bi $\gamma$ lines, some residual contribution from $^{42}$K, and degraded $\alpha$'s and long-lived cosmogenic isotopes.

\subsection{Phase 2: BEGe detectors and LAr scintillating veto}

\subsubsection{Broad Energy Germanium detectors}

\begin{figure}[!h]  
\centering
\includegraphics[scale=0.35]{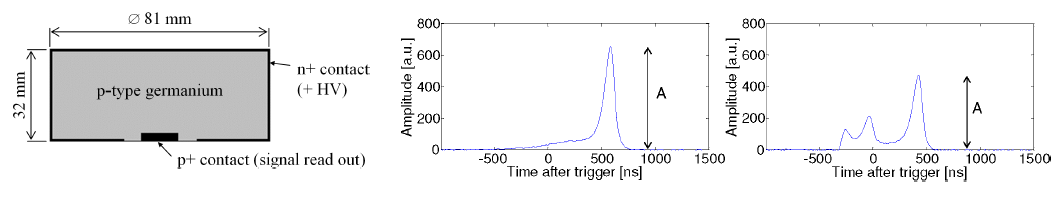}
\caption{(Left) Schematic view of a point-contact Broad Energy Germanium (BEGe); (Center) Single Site Event ($\beta\beta 0\nu$ topology); (Right) Multi Site Event (Multi Compton background topology)}
\label{fig:gerda-bege-principle}
\end{figure}

When the background events are produced by multiple Compton interaction of $\gamma$'s,  their topology is characterized by multiple interaction sites inside the crystal (MSE). 
Such events are produced either by external $\gamma$'s emitted by $^{238}$U and $^{232}$Th external contamination or by internal $\gamma$'s emitted by $^{60}$Co and $^{68}$Ge cosmogenics. 
Such events are different from $\beta \beta$ events for which the topology is a single site event (SSE). Pulse shape analysis allows to distinguish partially SSE and MSE events. 
Recently it was realized that commercially available Broad Energy Germanium detectors (BEGe) produced by Canberra Company exhibits superior pulse shape discrimination performances. 
The crystal is made of p-type HPGe (with point-contact) with the Li-drifted n$+$ contact (0.5~mm thickness) covering the whole outer surface, including most of the bottom part (see Figure~\ref{fig:gerda-bege-principle}). 

\begin{figure}[!h]  
\centering
\includegraphics[scale=0.4]{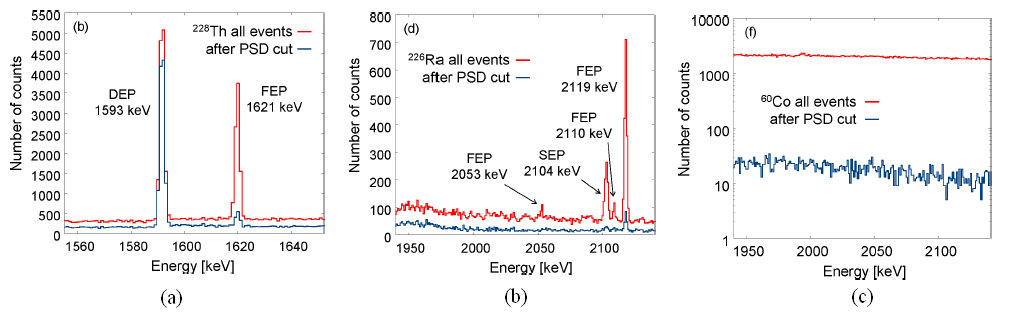}
\caption{Result of the pulse shape discrimination with BEGe detectors, using $^{228}$Th, $^{226}$Ra and $^{60}$Co sources. The double escape peak (DEP) events at 1592~keV are accepted in 89$\%$ of the case, while the Compton continuum events in the $Q_{\beta\beta}$ energy region are supressed by a factor 2.5 ($^{228}$Th) up to $\approx 100$ ($^{60}$Co).}
\label{fig:gerda-bege-spectre}
\end{figure}

The GERDA collaboration has carried out several tests and measurements of large BEGe detectors in vacuum and then in liquid argon cryostats~\cite{bege-2009}~\cite{bege-2010}~\cite{bege-2011}. 
Pulse shape analysis has been tested by using $^{228}$Th, $^{226}$Ra and $^{60}$Co source measurements (Figure~\ref{fig:gerda-bege-spectre}). Two types of samples have been used to define the PSA parameters: (i) double escape peak (DEP) at 1519~keV from the 2615~keV $^{208}$Tl $\gamma$-ray has been used as a substitute for $\beta\beta$ events, since it corresponds to pure $e^+ e^-$ pairs produced in a single site (SSE); (ii) single escape peak (SEP, where only one of the two 511~keV $\gamma$ annihilation escapes, the second deposits its full energy) and full energy absorption (FEA, where the initial $\gamma$ deposits all its energy) have been used as representative samples with dominant MSE events. 
After applying the PSA to the energy spectrum obtained with the source measurements, it has been demonstrated that the DEP events at 1592~keV are accepted in 89$\%$ of the case (Monte Carlo simulation confirmed the $90\%$ efficiency to keep $\beta\beta 0\nu$ events) while the Compton continuum events in the $^{76}$Ge $Q_{\beta\beta}$ energy region are supressed by factors $\approx 100$ for $^{60}$Co, $\approx 5$ for $^{226}$Ra ($^{214}$Bi) and $\approx 2.5$ for $^{228}$Th ($^{208}$Tl). 

Morever, it has been also observed that BEGe detectors are very stable, with no charge collection losses, and an excellent energy resolution of $\approx 2$~keV, achieved with two different large detectors (630 and 900~g). It is almost a factor of two better than germanium detectors used in previous experiments or in GERDA Phase 1.

BEGe detectors are thus excellent candidates for the second phase of GERDA. The collaboration has decided to use the 36.5~kg available enriched germanium (in form of ingots) to produce $\approx 20$ BEGe detectors in Canberra (in Oak Ridge, TN, US) for a total mass of enriched Ge $\approx 20$~kg. A second production campaign is planned with the 13~kg rest of germanium (to be reprocessed chemicaly)

\subsubsection{Liquid argon scintillating veto}

A liquid argon scintillating detector, called LArGe, has been developped by the GERDA collaboration and installed in LNGS, in order to study the efficiency of the liquid argon scintillation to tag an external $\gamma$'s (active LAr shield). 
The scintillation light yield $\approx 40$~photons/keV for ultra pure Argon.
In LArGe, the number of collected photoelectrons is only 0.05~$\gamma e$/keV.
BEGe detector has been installed inside the LArGe cryostat and background suppression measurements have been carried out using external and internal  $^{226}$Ra ($^{214}$Bi) and $^{228}$Th ($^{208}$Tl) sources.


Table~\ref{tab:gerda-large} summarizes the  supression factors of the continuum Compton background, observed in the $\beta\beta 0\nu$ energy region (2004-2074~keV) using the LAr scintillating veto and the pulse shape discrimination. 
The LAr scintillation veto is very efficient for $^{208}$Tl ($^{228}$Th) but relatively less for $^{214}$Bi ($^{226}$Ra). This is because, the most intense $\gamma$-ray emitted by $^{214}$Bi, above $Q_{\beta\beta}$, is the 2204~keV (4.86$\%$ branching ratio), corresponding to a maximum deposited energy in argon of 130 to 200~keV in order to be able to deposit 2004 to 2074~keV in the BEGe detector. With 0.05~$\gamma e$/keV, it corresponds to an average of 6 to 10 photoelectrons only. 
However, in the case of $^{208}$Tl, the excess energy deposited in liquid argon by the 2615~keV $\gamma$-ray is 610-680~keV, high enough to produced a significantly large number of photoelectrons well above the detection threshold.

Based on the successful background suppression achieved with LArGe and the current background counting rate in GERDA, the collaboration decided to prepare a design to instrument the LAr in Gerda. 
Different options are currently under development: scintillation readout using photomultiplier tubes (PMT's), scintillating and wave length shifting (WLS), fibers coupled to SiPMs, and large-area SiPMs or APDs.

\subsection{Comments}

I would like to emphasize that the pulse shape discrimination and the LAr scintillation veto are only efficient to reduce external $\gamma$ background. However, background produced by $\beta$-decay or degraded $\alpha$ from potential surface contaminations of the detectors (like $\beta$ from $^{42}$K), or in the vicinity of the germanium crystals, cannot be tagged by the LAr veto and are not rejected by the pulse shape discrimination since they are single site events. The thick dead layers of Li (0.5~mm thickness) on the surface of the BEGe detectors contribute to protect from $\alpha$'s and $\beta$'s provided that lithium layers are radiopure.

\begin{table}[!h]
\centering
\begin{tabular}{c|c|c|c|c}
Source & Position & \multicolumn{3}{c}{Rejection factor}  \\
       &          & LAr veto &   PSD  &  Total \\
\hline
\hline
$^{226}$Ra &  Internal  &   4.6   &    4.1   &  45 \\
           &  External  &   3.2   &    4.4   &  18 \\
\hline
$^{228}$Th &  Internal  &   $1175 \pm 200$  &    2.4   &  $5000 \pm 1500$ \\
           &  External  &    25  &    2.8   &   $130 \pm 15$ \\
\end{tabular}
\caption{Background supression factors observed in the $\beta\beta 0\nu$ energy region (2004-2074~keV) obtained using, either the LAr scintillating veto, the pulse shape discrimination and by combining both.}
\label{tab:gerda-large}
\end{table}

\section{The MAJORANA experiment}

\begin{figure}[!h]  
\centering
\includegraphics[scale=0.4]{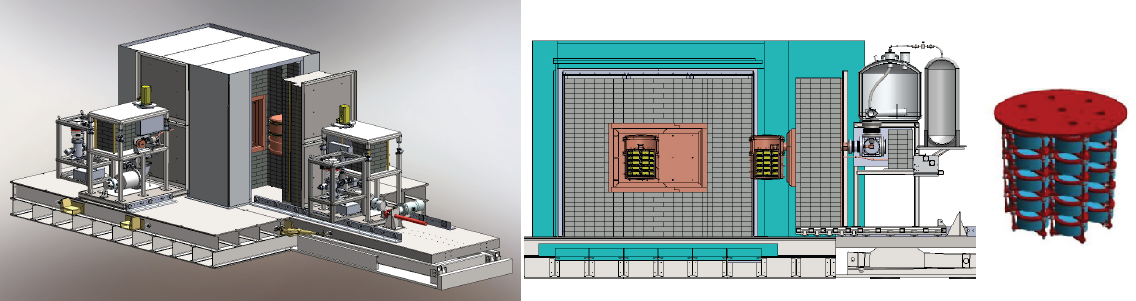}
\caption{Schematic view of the LArGe Facility and some pictures.}
\label{fig:gerda-majorana}
\end{figure}

Another germanium experiment, named MAJORANA experiment, is under development in the USA. 
The construction of the MAJORANA demonstrator has been funded by DOE and NSF. In contrast with GERDA, the germanium detectors are runned in vacuum Nitrogen cryostat with  standard low-background passive lead and copper shield and an active muon veto (see Figure~\ref{fig:gerda-majorana}).
40~kg of germanium detectors will be measured with the demonstrator: 20~kg of 86$\%$ enriched Ge and 20~kg of natural germanium crystals. 
Two independent cryostats will be constructed using ultra-clean, electroformed copper (the electroforming will be performed underground). 
As GERDA, p-type and point-contact detectors (BEGe detectors) will be used in order to obtain a good pulse shape discrimination against multi-site events produced by external $\gamma$'s. The detectors will be also deployed in strings. Each cryostat contains 35 detectors assembled in 7 strings (detectors are relatively smaller than GERDA).

The experiment will be located in the so-called Davis Campus in the Sanford Underground Laboratory (SUL, 4850~feet deep) in the Homestake mine (South Dakota, USA). The outfitting is underway and the lab will be ready in spring 2012. The 20~kg enriched Ge have been delivered and first batch of BEGe detectors with natural germanium are available. 

The plan is first to test 2 or 3 strings (10-15 detectors) in a prototype cryostat (with some non-electroformed components) in summer 2012. Then it is planed to run the first electroformed cryostat (with 3 strings natural Ge and 4 strings of enriched Ge) in spring 2013 in SUL and the second cryostat (up to 7 strings $^{enr}$Ge) in fall 2014. 

Since there is no possible active shield (no LAr veto as GERDA), the main issue is to demonstrate the capability to develop an ultra radiopure cryostat and an ultra radiopure inner shield in order to reach the target background of $10^{-2} - 10^{-3}$~counts/(keV.kg.y).

\section{Towards the ton scale: which limitations ?}

B. Majorovits has done a very interesting study~\cite{majorovits-dbd2011} of the possible limitations of the next generation of experiments measuring, not anymore $\approx 100~kg$ but a ton of germanium, with a required background  of $\approx 10^{-5}$~counts/(kg.y.keV) in the $\beta\beta 0\nu$ energy region. Three limiting origins of background have been listed: the crystal metalization (``the good''), the Pb on surfaces (``the bad'') and the Ge cosmogenic isotope (``the ugly''). 

\begin{enumerate}

\item The metalization of the surface of germanium detectors is in general done with aluminum. The contaminations introduced by this aluminium in cosmogenics $^{26}$Al ($\beta^+$-decay, $Q_{\beta}=4$~MeV, $T_{1/2}=7.4 \ 10^{5}$~years) and $^{22}$Na ($Q_{\beta}=2.84$~MeV, $T_{1/2}=2.6$~years) or in natural radioactivity $^{226}$Ra and $^{228}$Th, are a possible source of background. 
Maximal radio-impurities of less than 3.3~mBq/kg, 0.6~mBq/kg, 0.2~mBq/kg and 0.2~mBq/kg are required in order not to exceed individual background contributions of larger than 10$^{-6}$ counts/(kg.y.keV) from $^{26}$Al, $^{22}$Na, $^{226}$Ra and $^{228}$Th, respectively. Several sample of aluminium have been measured with ultra low background HPGe detectors with this required sensitivity and it has been verified that one sample was pure enough ($<0.15$, $<0.26$, $<0.28$ and $<0.58$~mBq/kg in $^{26}$Al, $^{22}$Na, $^{226}$Ra and $^{228}$Th, respectively)~\cite{majorovits-alu}. 

\item The usual contamination in $^{210}$Pb, observed on the crystal surface produces a limiting background from degraded $\alpha$'s. To illustrate, 100 of $^{210}$Pb nuclei per m$^2$ would produce $\approx 10^{-5}$~counts/(kg.y.keV) in the $^{76}$Ge $\beta\beta 0\nu$ energy region. Experimental tests of etching of germanium crystals have been recently performed with Canberra in order to measure the removal and deposition efficiencies of $^{210}$Pb. Reduction factors of 1000 or higher have been obtained. But deposition by non pure solvent has been also observed. It means that $^{210}$Pb screening methods and clean solvents ($<10 \mu$Bq/m$^3$) are needed. 

\item The most dangereous background is the bulk contamination of the germanium crystals in $^{68}$Ge cosmogenic isotope ($T_{1/2}=3.84$~years)~\cite{elliot}. About 0.5 $^{68}$Ge nucleus per kg would produce $10^{-5}$~counts/(kg.y.keV) in the $^{76}$Ge $\beta\beta 0\nu$ energy region. 
The depletion of $^{68}$Ge in enriched $^{76}$Ge is not known and measurements for $^{70}$Ge depletion are inconsistents (from $2 \ 10^{-2}$ to $7 \ 10^{-5}$). It means that next gereneration experiments require R$\&$D on deenrichment of $^{68}$Ge with a deenrichment factor of at least $2 \ 10^{-5}$ to get less than 0.5 $^{68}$Ge nucleus per kg. 

\end{enumerate}

To conclude, I would like to add another possible limitation, which illustrates the difficulty and the danger to scale a double beta experiment to the ton scale. The Pb(n,$\gamma$'s) reactions\footnote{neutron capture on Pb leading to the emission of $\gamma$-rays} can produce the 2041~keV $\gamma$-rays (only 2~keV above $Q_{\beta\beta}=2039$~keV) and the 3062~keV $\gamma$-rays~\cite{mei}. The former is also a dangerous background for $^{76}$Ge experiments, because the double escape peak energy ($3062 - 2 \times 511 = 2040$~keV) coincides with the  $Q_{\beta\beta}$  value. The measured rate of this transition is too small to explain the Klapdor's claim. This background is dangerous for the MAJORANA experiment, which uses traditional lead shield, but it might be significantly reduced using depth and/or an inner layer of Cu within the shield.

With this example, we see that very rare or even unknown $\gamma$-rays may appear at the $Q_{\beta\beta}$  value and dedicated extra neutrons capture measurements are probably needed.

%
%

\chapter{Bolometer detectors}
\label{sec:bolometer}

The use of bolometers for $\beta \beta 0 \nu$-decay searches was suggested by Fiorini and Niinikoski~\cite{fiorini-1984} and applied first by the Milano group in the MIBETA experiment~\cite{arnaboldi-2003} using natural Te0$_2$ crystals as bolometers. 

The main advantage to use Te0$_2$ crystal is that the natural abundance of $^{130}$Te (the $\beta \beta$ emitter) in tellerium is high, about 34~$\%$. However its $Q_{\beta\beta}=2530.3\pm2.0$~keV~\cite{wapstra-2003} is lower than the 2615~keV $\gamma$ ray from $^{208}$Tl. Experiments using TeO$_2$ crystals are therefore sensitive to Compton electron background due to $^{232}$Th contamination.

In order to avoid to enrich $\beta \beta$ isotopes, the Milano group decided in the 90's to continue using natural Te0$_2$ crystals and to developp a larger detector, CUORICINO, which has been running from 2003 until 2008 in Gran Sasso, and measured about 10~kg of $^{130}$Te. It is a pilot detector for a larger experiment, CUORE, currently in construction, which will measure up to 200~kg of $^{130}$Te.

I will first detail the CUORICINO experiment and I will present its main results and limitations. Then I will present the current development of the large CUORE experiment. Finally, I will present the current developments and projects of scintillating bolometers. I will detail it because I think that it is one of the most promising technique for the $\beta \beta 0 \nu$ search.

\section{The CUORICINO experiment}


\subsection{Description of the detector}

\begin{figure}[!h]  
\centering
\includegraphics{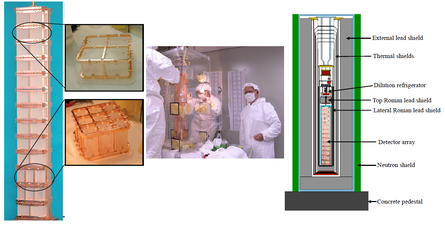}
\caption{Pictures of the CUORICINO tower, and layout of its assembly inside the cryostat and shields.}
\label{fig:cuoricino-setup}
\end{figure}

CUORICINO is an array of 62 TeO$_2$ crystals, a total mass of crystals of 40.7~kg, corresponding to a total mass of 11.64~kg for enriched $^{130}$Te isotope. As shown in Figure~\ref{fig:cuoricino-setup}, the CUORICINO detector is a tower composed of 11~layers of 4~crystals, each one with a dimension $5\times5\times5$~cm$^3$ and a mass of 790~g, and two layers of 9 smaller crystals, $3\times3\times6$~cm$^3$ and 330~g. Two small crystals were enriched in $^{130}$Te at 75~$\%$ and two other ones in $^{128}$Te at 82.3~$\%$. 
All crystals were grown with pre-tested low radioactivity material by the Shanghai Institute
of Ceramics. They were lapped with specially selected low contamination polishing compound.
The mechanical structure was made of oxygen-free high-conductivity copper and Teflon.
A sketch of the tower assembly inside the low radioactive shield is shown in Figure~\ref{fig:cuoricino-setup}.
The CUORICINO cryostat operated at a temperature of about 8 mK.
Thermal pulses were measured with NTD Ge thermistors. 
The average FWHM energy resolutions at $Q_{\beta\beta}=2530$~keV are 7 and 9~keV, for the $5\times5\times5$~cm$^3$ and $3\times3\times6$~cm$^3$ crystals, respectively. The spread in the FWHM is about 2~keV in both case.



The excellent energy resolution has a sense only if the energy calibration of the bolometers is stable and if any drift is well controlled. 
The gain of each bolometer is stabilized by means of a Si resistor, attached to each bolometer that acts as a heater. 
Heat pulses are periodically (every 300~s) supplied by a calibrated ultrastable pulser~\cite{arnaboldi-2003-ieee}. 
Any variation in the voltage amplitude recorded from the heater pulses indicates that the gain of that bolometer has changed.
The heater pulses are used to measure (and later corrected offline) for the gain drifts. 


\subsection{$\beta \beta 0 \nu$ result and background study}

\begin{figure}[!h]  
\centering
\includegraphics[scale=0.35]{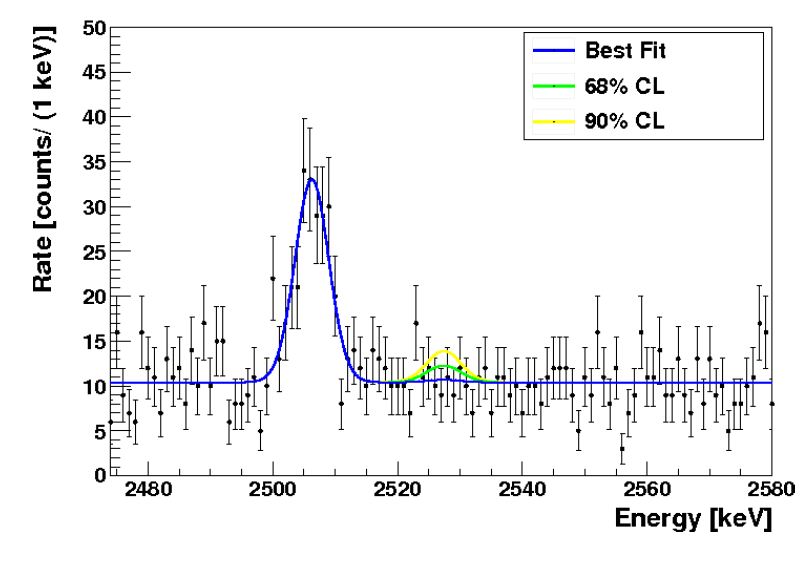}
\caption{Energy spectrum in the $\beta\beta 0\nu$ region~\cite{cuoricino-final-result}}
\label{fig:cuoricino-result}
\end{figure}

The total exposure obtained with CUORICINO was 19.75 kg.y of $^{130}$Te.
The $\beta \beta 0 \nu$ detection efficiency is 0.83$\%$. 
Figure~\ref{fig:cuoricino-result} shows the energy spectrum around the $Q_{\beta\beta}$ value. No excess of events has been observed at $Q_{\beta\beta}$, resultingg to a lower limit on the $\beta \beta 0 \nu$-decay half-life of~\cite{cuoricino-final-result}:

$$   \mathrm{T_{1/2}(\beta \beta 0 \nu) > 2.8 \ 10^{24} \ y~(90\%~C.L.)}$$

The observed peak at 2506~keV corresponds to the pile-up of the two external $\gamma$-rays (1173.2 and 1332.5~keV) emitted by external $^{60}$Co. It gives a posteriori a good cross-check of the energy calibration and the energy resolution. 

We also mention that the searches for both 0$\nu$ and 2$\nu$ double-beta decay to the first excited 0$^+$ state in $^{130}$Xe were performed recently by studying different coincidence scenarios~\cite{andreotti-2011}. No evidence for a signal was found. The resulting lower limits on the half-lives are $\mathrm{T_{1/2}^{2\nu}(^{130}Te \rightarrow ^{130}Xe^*) > 1.3 \ 10^{23} \ y~(90\%~C.L.)}$ and $\mathrm{T_{1/2}^{0\nu}(^{130}Te \rightarrow ^{130}Xe^*) > 9.4 \ 10^{23} \ y~(90\%~C.L.)}$

The level of background in the $Q_{\beta\beta}$ region is 0.16~counts/(keV.kg.yr). Taking into account the energy FWHM resolution of 7~keV at $Q_{\beta\beta}$, it corresponds to a level of background in the $\beta \beta 0 \nu$ region of interest of 1.12 counts/(FWHM.kg.yr). 

\begin{figure}[!h]  
\centering
\includegraphics[scale=0.4]{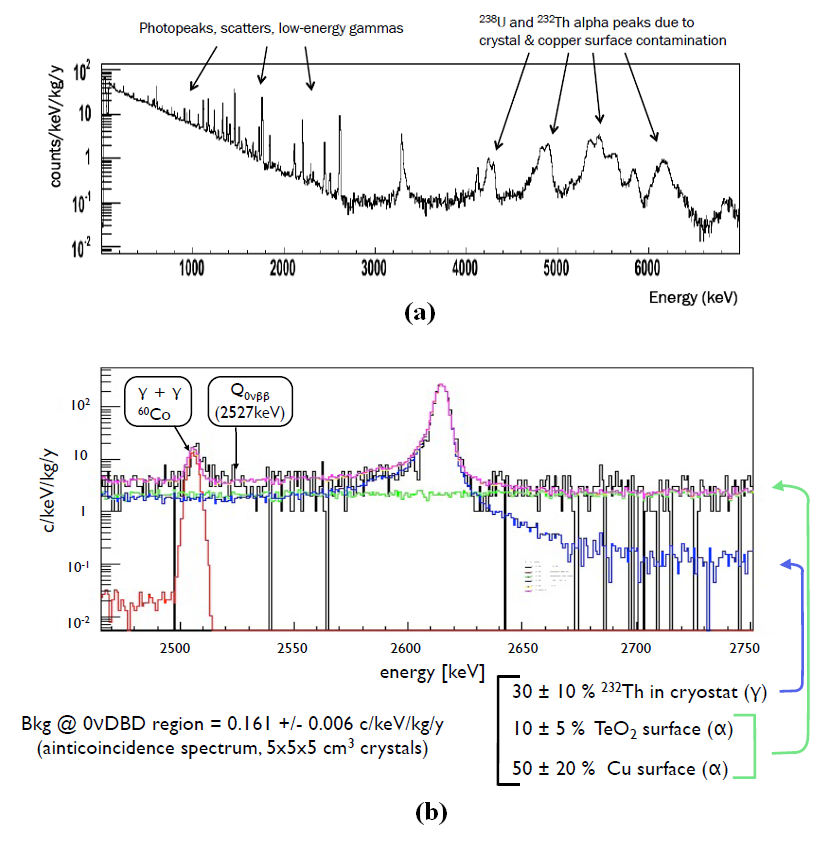}
\caption{(a) Energy spectrum in the full energy range; (b) Energy spectrum in the $\beta \beta 0 \nu$ region and fitted Monte-Carlo background model.}
\label{fig:cuoricino-bkg}
\end{figure}

The origin of the observed background has been studied carefully by the collaboration in order to define a strategy to reduce it for a larger detector like CUORE.
Figure~\ref{fig:cuoricino-bkg} shows the energy spectrum in the full energy range. The background is dominated by $\gamma$'s below the 2615~keV $\gamma$ ray line from $^{208}Tl$, while $\alpha$'s are the main origin of background above 2615~keV. 
In the $\beta \beta 0 \nu$ energy region, both $\gamma$'s and degraded $\alpha$'s contribute to the observed background. The $\alpha$-particles are produced from surface contaminations in $^{238}$U, $^{232}$Th and $^{210}Pb$ (from Radon deposition during the construction) on the surfaces of the crystals or the surface of the copper holders facing the crystals. The $\alpha$-particles can mimic a $\beta \beta 0 \nu$ event if they deposit a part of their energy elsewhere and about 2530~keV ($Q_{\beta \beta}$-value) in the crystal.

The measurement of the different origins of the background was done by analysing both the coincidence and anti-coincidence spectra. 
Final spectrum and fitted Monte-Carlo background spectrum in the $\beta \beta 0 \nu$ region is shown in Figure~\ref{fig:cuoricino-bkg}. 
The conclusion of this analysis was the identification of three main contributions of background in the $\beta \beta 0 \nu$ energy region: 
\begin{itemize}
\item[\textbullet] $30\pm10\%$ of the background is due to multi-Compton events due to the 2615~keV gamma ray from the decay chain of $^{232}$Th from the contamination of the cryostat shields; 
\item[\textbullet] $10\pm5\%$ is due to surface contamination of the TeO$_2$ crystals with $^{238}$U and $^{232}$Th (primarily degraded alphas from these chains);
\item[\textbullet] $50\pm20\%$ is ascribed to similar surface contamination of inert materials surrounding the crystals, most likely copper. 
\end{itemize}
The level of backrgound due to degraded $\alpha$'s particles from the crystal and copper surfaces is measured in the 2.7 to 3.2~MeV energy window and is $0.104 \pm 0.002$~counts/(keV.kg.yr)~\cite{alessandria-2011}. 
The $\gamma$ of 2615~keV from the $^{232}$Th contamination of the cryostat increases the backrgound level in the $\beta \beta 0 \nu$ region to 0.16~counts/(keV.kg.yr).

The contribution of cosmic ray muons to the CUORICINO background has been recently measured~\cite{andreotti-2010}. From these measurements, an upper limit of 0.0021 counts/(keV.kg.yr) (95$\%$~C.L.) was obtained on the cosmic ray-induced background in the $\beta \beta 0 \nu$ energy region.

We mention that the response of a TeO$_2$ bolometer to $\alpha$ particles has been investigated using a TeO$_2$ crystal containing $^{147}$Sm ($\alpha$ emitter)~\cite{bellini-2010-1}. 
Signal pulse shape shows no difference between $\alpha$ and $\beta$/$\gamma$ particles. Therefore no pulse shape analysis can be applied to recognize an $alpha$ interaction.

%
%

\section{The CUORE experiment}


\subsection{Description of the detector}

\begin{figure}[!h]  
\centering
\includegraphics[scale=0.4]{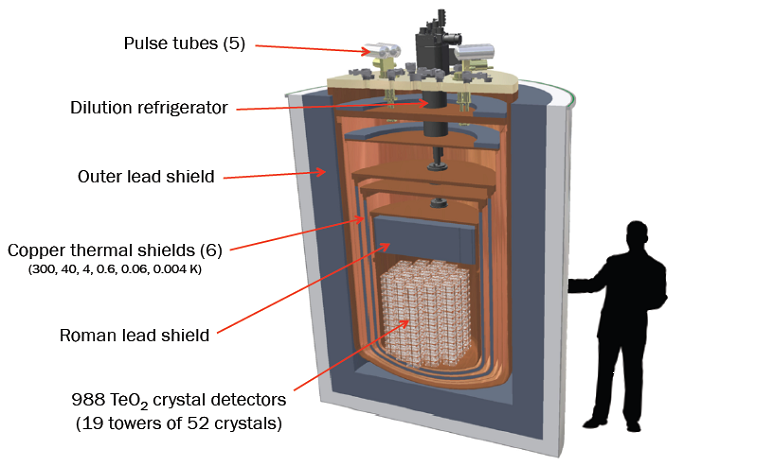}
\caption{Schematic view of the CUORE experiment.}
\label{fig:cuore-setup}
\end{figure}

CUORE will consist of an array of 988 TeO$_2$ cubic detectors, similar to the $5\times5\times5$~cm$^3$ Cuoricino crystals described above. 
The total mass of the crystals will be 741~kg, corresponding to 206~kg of $^{130}$Te. 
The detectors will be arranged in 19 individual towers and operated at 10~mK in the Gran Sasso underground laboratory in a new large cryostat with improved external shield (see Figure~\ref{fig:cuore-setup})
The expected energy resolution FWHM of the CUORE detectors is 5~keV at the $Q_{\beta\beta}$ transition energy. 
This resolution represents an improvement over that seen in Cuoricino and has already been achieved in tests performed in the CUORE R$\&$D facility at LNGS with first delivered crystals~\cite{arnaboldi-2010}. 
CUORE is expected to accumulate data for about 5~years of total live time. 
The experiment is currently being constructed and first data-taking is scheduled for 2014.
The construction of the new large low radioactive cryostat is a delicate and long task and inner vessels should be produced in 2012.

\begin{figure}[!h]  
\centering
\includegraphics{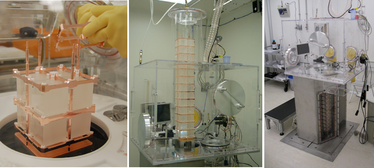}
\caption{Pictures of the CUORE-0 tower during its assembly in LNGS.}
\label{fig:cuore-photo-1}
\end{figure}

The CUORE collaboration will first operate a single CUORE-like tower in the former Cuoricino cryostat. 
This configuration, named CUORE-0, will validate the assembly procedure and the readiness of the background reduction measures.
The assembly of CUORE-0 is now completed (see Figure~\ref{fig:cuore-photo-1}) and data collection will start in 2012 for two years of running. 

\subsection{Background reduction}

On the basis of the CUORICINO background analysis, the R$\&$D for CUORE has pursued two major complementary issues: the reduction of surface contamination of crystals and copper in order to reduce the contribution of degraded $\alpha$'s particles, and the creation of an experimental setup in which potential background contributions are minimized by the selection of extremely radio-pure construction materials and the use of highly efficient shields, in order to reduce the contribution of external 2615~keV $\gamma$'s. 


CUORE crystals are produced following a controlled protocol~\cite{arnaboldi-2010}.  
Several surface-treatment techniques were developed to clean the surface of crystals and copper~\cite{pavan-2008}. 
Since the required surface contamination levels are extremely low, of the order of 10 to 100 $\mu$Bq/m$^2$, undetectable with any standard technique used in surface analysis, the surface radiopurity of the crystals have been measured in the CUORICINO cryostat by assembling them as an array of 3$\times$4 crystals (Three Tower Test, TTT). Results show that the best results were obtained either with passive chemical cleaning and polyethylen barrier, or with plasma cleaning. 
The level of background measured in the anticoincidence spectrum in the 2.7 to 3.9~MeV energy window (contribution of degraded $\alpha$'s) is statisticaly limited and is about 0.05~counts/(keV.kg.yr) corresponding to a reduction factor of about 2 compared to CUORICINO. If we extrapolate CUORICINO background analysis result, half of this background is dominated by copper surface contamination.


The bulk contamination was measured (with Cryogenic validation runs CCVR) to be lower than 5.3~10$^{-14}$~g/g for $^{238}$U (0.67~$\mu$Bq/kg) and 2.1~10$^{-13}$~g/g for $^{232}$Th (0.84~$\mu$Bq/kg) (90$\%$~C.L.)~\cite{alessandria-2011}. These values are well below the concentration limits of 3~10$^{-13}$~g/g in $^{238}$U and $^{232}$Th, requested for TeO$_2$ crystals to be used in CUORE experiment. 


The final background budget expected in the $\beta \beta 0 \nu$ region is sumarized in Table~\ref{tab:cuore-bkg-budget}, both for CUORE-0 and CUORE~\cite{gorla-taup11}. The background is dominated by degraded $\alpha$'s from surface contaminations of the copper facing the crystals, or by external $\gamma$'s from cryostat or shield in the case of CUORE-0 (which will be installed inside the CUORICINO cryostat). A higher radiopurity cryostat facility and a higher shield efficiency is expected for CUORE. The CUORE collaboration expects to reach a level of background of 0.01~cts/(keV.kg.yr). Background suppression could be improved by installing suface sensitive bolometers (SSB, see next section). But only first data collection in 2014 will tell us the real level of background !

\begin{table}[!h]
\centering
\begin{tabular}{c|c|c|c}
Bkg source & CUORE-0 & CUORE & Source of data \\
\hline
\hline
External ($\mu$+n+$\gamma$)     & $<$ 1.0~10$^{-2}$ & $<$ 2.0~10$^{-3}$  & measured flux + MC \\
$\gamma$ from cryostat + shield & 0.05            & $<$ 1.0~10$^{-3}$  & material selection + cuoricino \\
Cu det. holder (bulk)       & $<$ 2.0~10$^{-3}$ & $<$ 2.0~10$^{-3}$  & HPGe + NAA + MC \\
Cu det. holder (surface)    & $\leq$ 4.0~10$^{-2}$ & $\leq$ 2.5~10$^{-2}$  & TTT + MC \\
TeO$_2$ (bulk)                  & $<$ 4.0~10$^{-4}$ & $<$ 1.1~10$^{-4}$  & CCVR \\
TeO$_2$ (surface)               & $<$ 7.3~10$^{-3}$ & $<$ 5.5~10$^{-3}$  & CCVR + MC \\
\end{tabular}
\caption{Background budget in counts/(keV kg yr) expected in the $\beta \beta 0 \nu$ region, both for CUORE-0 and CUORE.}
\label{tab:cuore-bkg-budget}
\end{table}

\subsection{Expected sensitivity for CUORE-0 and CUORE}

An optimistic scenario is to consider that the irreducible background for CUORE-0 comes from the 2615~keV $^{208}$Tl line due to $^{232}$Th contaminations in the cryostat, in the case that all other background sources (i.e., surface contaminations) have been rendered negligible. 
This would imply a lower limit of 0.05~cts/(keV.kg.yr) on the expected background in CUORE-0. 
The expected sensitivity is $T_{1/2}(\beta\beta 0 \nu) > 6 \ 10^{24}$~yr~(90$\%$~C.L.) after 2 years of measurement.

Similarly, an upper limit of 0.11~cts/(keV.kg.yr) follows from scaling the Cuoricino background in the conservative case, described above, of a factor of 2 improvement in crystal and copper contaminations.
The expected sensitivity becomes $T_{1/2}(\beta\beta 0 \nu) > 4 \ 10^{24}$~yr~(90$\%$~C.L.) after 2 years of measurement.

\begin{figure}[!h]  
\centering
\includegraphics[scale=0.5]{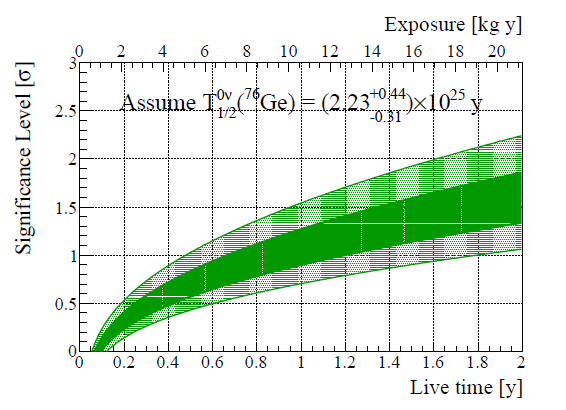}
\caption{Significance level at which CUORE-0 can observe a signal corresponding to the $^{76}$Ge claim, assuming the best expected background of 0.05 cts/(kev.kg.y). The inner band assumes the best-fit value of the $^{76}$Ge claim, and its width arises from the 1$\sigma$ range of QRPA-F NMEs calculated in~\cite{faessler-2009}. The outer band accounts for the 1$\sigma$ uncertainty on the $^{76}$Ge claim in addition to the range of NMEs.}
\label{fig:cuore-sensitivity-klapdor}
\end{figure}

The significance level at which CUORE-0 can observe a signal corresponding to the $^{76}$Ge claim, assuming the best expected background of 0.05~cts/(kev.kg.y), is shown in Figure~\ref{fig:cuore-sensitivity-klapdor}.

For CUORE, with 19 complete towers, an energy resolution of 5~keV, and a level of backrgound of 0.01~cts/(keV.kg.yr), the expected sensitivity is $T_{1/2}(\beta\beta 0 \nu) > 10^{26}$~yr~(90$\%$~C.L.) after 5 years of measurement.


\subsection{Surface-sensitive bolometers}

In order to reduce the background due to degraded $\alpha$'s from the surface of the copper facing the crystals, it was proposed to develop a surface sensitive bolometer (SSB)~\cite{foggetta-2011}. The basic idea of the SSB is to surround the TeO$_2$ crystal (main absorber) with thin auxiliary absorbers which function as active shields and are operated as bolometers. The main absorber and the shields are thermally connected, and a near 4$\pi$ coverage from external charged particles can be achieved. 
Charged particles from materials outside the SSB, such as $\alpha$-particles, will be stopped and
tagged by one of the active shields (surface events). They release part of their energy in
the shield, but raise the temperature of all the detector elements as they are, in fact, all thermally
connected. Because of the small heat capacity of the shield due to its small mass, the signal
read by its thermistor will have a higher amplitude and faster rise time than the signal read by
the thermistor attached to the main absorber. If, on the other hand, an energy deposition occurs
inside the main absorber (bulk event), all of the thermistors will read pulses with comparable
amplitudes and rise times.
It is therefore possible to separate bulk and surface events by comparing
amplitude and shape of pulses among the diferent thermistors. No degradation in energy
resolution of the main absorber is expected since the device is still operating in the mode where
phonon thermalization is nearly complete. 
Thus, the whole system does not act as an usual veto but acts as a pulse shaper.  

Recently three prototype detectors have been tested~\cite{foggetta-2011}. The surface event rejection capability has been clearly demonstrated and preliminary results show that it is possible to detect
energy depositions that occurred on the shields without separate readout channels for them. 
However, a real reduction of background without adding new contaminations from the SSB has still to be demonstrated with long background runs.

%
%

\section{Scintillating bolometers}

The two dominant sources of background in CUORE (using TeO$_2$ crystals) are external 2615~keV $\gamma$'s from possible $^{232}$Th ($^{208}$Tl) contamination inside the cryostat or the shield, and degraded $\alpha$'s from the surface of the copper facing the crystals:
\begin{itemize}
\item[\textbullet] If the crystal is based on a double beta isotope with a $Q_{\beta \beta}$ transition value above 2615~keV, the external $\gamma$ background is strongly suppressed. 
\item[\textbullet] In the case of a scintillating bolometers, based on scintillating crystals, the simultaneous detection of heat and scintillation light allows the supresssion of the background due to degraded $\alpha$ particles, thanks to the different scintillation quenching factor between $\alpha$ and $\beta$/$\gamma$, 
\end{itemize}

\begin{figure}[!h]  
\centering
\includegraphics[scale=0.4]{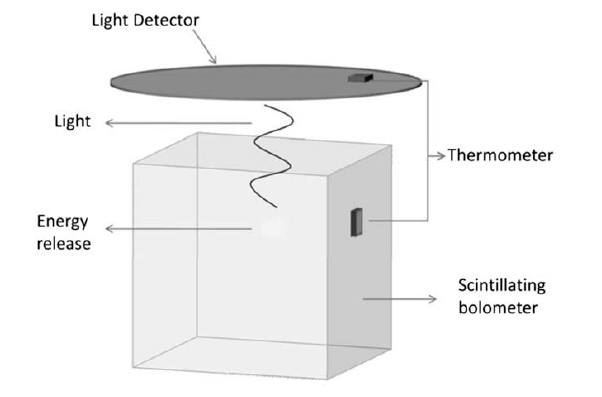}
\caption{Operating principle of scintillating bolometers. The release of energy inside a scintillating crystal follows two channels: light production and thermal excitation. The heat is read out by a temperature sensor (NTD) glued on the primary crystal while the light is read by a second thin crystal (light detector) where it is completely converted into heat.}
\label{fig:boloscint-schema}
\end{figure}

Thus scintillating bolometers appear to be a very promising technique to suppress the background. 
Its operating principle is illustrated in Figure~\ref{fig:boloscint-schema}. 

The first light/heat measurement with a background discrimination for $\beta \beta 0 \nu$ DBD searches was performed by the Milano group in 1992~\cite{alessandrello-1992} but was not pursued due to the difficulties of running such a light detector at ultra low temperatures (10~mK). Scintillating bolometers were then developed~\cite{bobin-1994} and optimized for Dark Matter searches~\cite{angloher-2005}\cite{cebrian-2003}. 
Starting from that work, a long R$\&$D program, coordinated by S. Pirro and mainly funded by ILIAS-IDEA (European FP7 program), investigated several scintillating bolometer candidates for a double beta experiment. 
The crystals are CdWO$_4$ ($^{116}$Cd, $Q_{\beta \beta}=2809$~keV), ZnSe ($^{82}$Se, $Q_{\beta \beta}=2995$~keV), ZnMoO$_4$, PbMoO$_4$ or MgMoO$_4$ ($^{100}$Mo, $Q_{\beta \beta}=3034$~keV) and CaMoO$_4$ ($^{100}$Mo and $^{48}$Ca, $Q_{\beta \beta}=4271$~keV). 

Let's mention that a TeO$_2$ crystal cannot be operated as scintillating bolometer due to its extremely low light yield measured with pure and also doped samples~\cite{coron-2004}.

The different experimental studies have demonstrated the feasibility of scintillating bolometers to discriminate $\alpha$ to $\beta$/$\gamma$ events. Morever these studies have shown that a discrimination based only on the pulse shape analysis is achievable for some of the tested crystals. This feature would open the possibility of realizing a bolometric experiment that can discriminate among different particles, without the need of a light detector coupled to each bolometer. 

I will summarize here the main results obtained with the different crystals and I will discuss the advantages and limitation of each one.

\subsection{$^{116}$Cd: CdWO$_4$ crystal}

CdWO$_4$ is a well established scintillator crystal with a large light yield. It was the first crystal studied for $\beta \beta 0 \nu$ search. 
We mention that four enriched $^{116}$CdWO$_4$ crystals (total mass of 330~g of $^{116}$Cd) were already used in the 90's, as standard scintillator at room temperature for the search of $\beta \beta 0 \nu$-decay in the Solotvina Underground Laboratory~\cite{danevich-2003}. 

\begin{figure}[!h]  
\centering
\includegraphics[scale=0.4]{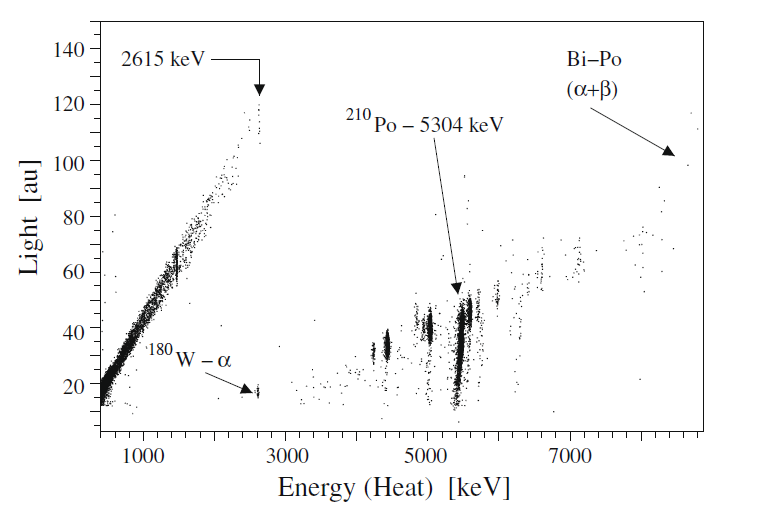}
\caption{Result of a background measurement performed with a 3$\times$3$\times$6~cm$^3$ CdWO$_4$ crystal (426~g) during 1066~hours. No $\beta$/$\gamma$ events have been observed above the 2615~keV $^{208}$Tl line.}
\label{fig:boloscint-cdwo4-bkg}
\end{figure}

Several tests with various CdWO$_4$ crystals as scintillating bolometers were carried out in the last few years~\cite{pirro-2006}~\cite{arnaboldi-2010-2}. We show in Figure~\ref{fig:boloscint-cdwo4-bkg} the result of the longest background measurement (1066~hours) performed with a large 3$\times$3$\times$6~cm$^3$ 426~g crystal. It was operated in a cryostat in Gran Sasso with a dedicated Roman lead and neutron shield in order to reduce as far as possible the environmental background.
This plot shows how it is possible to separate very well the background due to $\alpha$ particles from the $\beta/\gamma$ region. In particular it should be noticed that $\alpha$ continuum is completely removed thanks to the combined measurement of heat and scintillation. In the region above the 2615~keV line, $\beta / \gamma$ events have not been observed, demonstrating the performance of this technique.


Since degraded $\alpha$'s are totaly suppressed, the residual possible background is either a $^{238}$U and $^{232}$Th bulk contamination inside the crystal or external $\gamma$'s:

\begin{itemize}

\item[\textbullet] $^{238}$U and $^{232}$Th bulk contamination: The only high energy $\beta+\gamma$ decays above 2.6~MeV present in the $^{238}$U and $^{232}$Th decay chain are shown in Figure~\ref{fig:nat-rad-chains}. The $\beta + \gamma$ decays are always followed (or preceded) by an $\alpha$ emission. 
In the case of $^{238}$U decay chain, the half life of the delay $\alpha$-decay of $^{214}$Po (164~$\mu$s) is shorter than the time response of the bolometer (of the order of the ms). Thus the $\beta + \gamma$ and $\alpha$ signals are piled-up and can be easily rejected (see Figure~\ref{fig:boloscint-cdwo4-bkg}). In the case of $^{232}$Th, rejection is more delicate. Indeed, the $^{208}$Tl $\beta+\gamma$ decay is preceded by the $\alpha$ emission of $^{212}$Bi with a half life of 3.05~mn. This pre-$\alpha$ can be used to reject the background event in the condition that internal contamination is not too large in order to avoid dead time.

\item[\textbullet] External $\gamma$ background: The contribution of extremely rare high-energy $\gamma$-rays from $^{214}$Bi in the $\beta \beta 0 \nu$ energy region for $^{116}$Cd is negligible. The main source of  $\gamma$ background due to external contamination is induced by the $^{208}$Tl decay. In fact, if contaminations are sufficiently close to the detectors, the probability of spurious counts between 2615~keV and 3198~keV, due to coincidences betwen 2615~keV and 583~keV gamma emitted in the $^{208}$Tl decay, may not be negligible. Assuming the limit of contamination level measured in the copper structure surrounding the detector in CUORICINO, a GEANT4 simulation gives a limit of 10$^{-4}$~counts/(keV.kg.yr) in the  $\beta \beta 0 \nu$ energy region due to this main contribution.

\end{itemize}


\begin{figure}[!h]  
\centering
\includegraphics[scale=0.35]{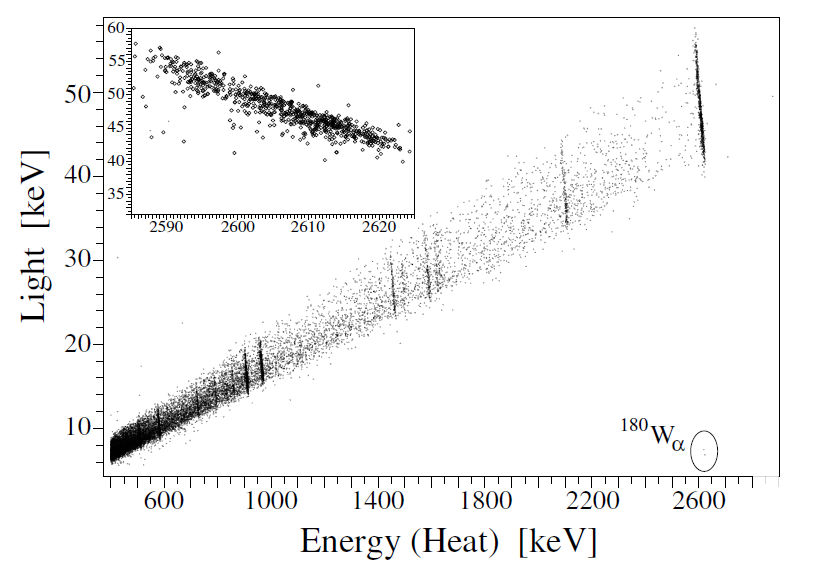}
\caption{Scatter plot Light versus Heat obtained with a large CdWO$_4$ crystal (4~cm diameter and 5~cm height, 0.51~kg) in a 96 h calibration using an external $^{232}$Th source. In the inset the highlight of the $^{208}$Tl line..}
\label{fig:boloscint-cdwo4-calib}
\end{figure}

It has been also demonstrated~\cite{arnaboldi-2010-2} that  the energy resolution is similar to standard TeO$_2$ bolomters, even with large CdWO$_4$ crystals.  Figure~\ref{fig:boloscint-cdwo4-calib} shows an example of Lights versus Heat scatter plot of $\beta/\gamma$ events collected with a calibration run using an external $^{232}$Th source. 
The observed $\gamma$ lines appear as negative slope lines, due to energy anticorrelation between the heat and scintillation signals. This anticorrelation is a well known phenomena and already observed for other scintillating bolometers. The important feature here is that the slopes of each $\gamma$ line are the same. Therefore an energy anticorrelation correction can be applied offline in order to improve the energy resolution. 
Without correction, the energy resolution would be 16~keV at 2615~keV. After energy anticorrelation correction, the FWHM energy resolution becomes 6~keV at 2615~keV. 

\subsubsection*{Possible limitations}

A possible limitation to use CdWO$_4$ crystals as scintillating bolometers is the presence of $^{113}$Cd (12.2$\%$ in natural cadmium) which can produce high energy $\gamma$'s up to 10~MeV in the case of neutron background, due to its large $(n,\gamma)$ cross section (about 20.000~b). 
A future experiment using CdWO$_4$ would require an efficient neutron shield.

Production of scintillating crystals with enriched isotope might be a critical issue because it is necessary to avoid contaminations in transition metals, which would deteriorate optical and scintillation quality. 
Recently large enriched $^{116}$CdWO$_4$ crystals (82$\%$ of $^{116}$Cd enrichment) have been produced~\cite{barabash-2011}. A raw crystal with mass of 1868~g was grown from 2139~g of the initial $^{116}$CdWO$_4$ compounds. Then 3 near cylindrical shape $^{116}$CdWO$_4$ crystals (586~g, 589~g and 325~g) have been cut from the raw crystal. It has been also demonstrated that enriched $^{116}$Cd can be recovered from $^{116}$CdWO$_4$ crystalline residue. Thanks to the careful purification of the initial components, the crystal has a high transmittance and a high light yield at room temprature. A background measurement over 1727~h in Gran Sasso, using sincillation light detection at room temperature allowed to measure radioactive contaminations: the activities of $^{226}$Ra ($^{214}$Bi) and $^{228}$Th ($^{208}$Tl) are $<5 \mu$Bq/kg and $\approx 60 \mu$Bq/kg, respectively which is low enough for a cryogenic experiment. Isotopic composition of cadmium in the $^{116}$CdWO$_4$ crystals has been also measured. The isotope abundance for $^{113}$Cd is 2.1$\%$, about 6 times smaller than in natural cadmium (12.2$\%$). 

Finally, a limiting feature is the relative high cost for  $^{116}$Cd enrichement compared to $^{82}$Se or $^{100}$Mo. 
I note that the 600~g of enriched $^{116}$Cd available from NEMO-3 could be used to developp extra $^{116}$CdWO$_4$ crystals.

\subsection{$^{82}$Se: ZnSe crystal}

ZnSe scintillating bolometer is another good candidate for $\beta\beta 0 \nu$ search and $^{82}$Se can be easily enriched by centrifugation. 
Experimental studies of several ZnSe crystals as scintillating bolometers have been performed in the last few years~\cite{arnaboldi-znse}. 
A relatively large light yield and good energy resolution have been obtained. 

\begin{figure}[!h]  
\centering
\includegraphics[scale=0.4]{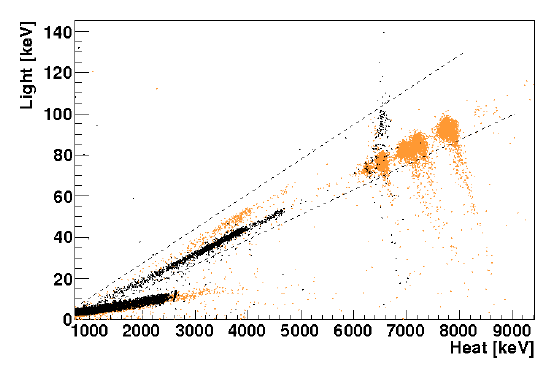}
\caption{Scatter plot of Light vs. Heat recorded with a large ZnSe crystal in two different runs with alpha sources at different positions on the crystal. The dotted lines identify the opening of the alpha band on a wide angle, due to the variation of the Quenching Factor upon source position~\cite{arnaboldi-znse}.}
\label{fig:znse-calib}
\end{figure}

However an unexpected characteristic has been observed with ZnSe: the sintillation light yield for $\alpha$ is larger than for $\beta/\gamma$ of the same energy, equivalent to a quenching factor larger than one (see Figure~\ref{fig:znse-calib}). No other scintillator with a quenching factor higher than 1 for $\alpha$ particles have ever been reported before in literature. 
Possible effects of scintillation light self-absorption or transparency of the light detector to the scintillation photons have been investigated but they cannot explain the anormal quenching factor.

The problem for $\beta \beta 0 \nu$ search is that the $\alpha$ band is spread to lower light signal and higher heat signal. Some $\alpha$ particles can then be misidentified as $\beta/\gamma$ events and thus produce a background in the $\beta \beta 0 \nu$ region of interest. 
This problem appears to be connected to the quality of crystal surface or to the presence on the surfaces of residuals of the abrasive, non-scintillating, powder used for polishing. 

\begin{figure}[!h]  
\centering
\includegraphics[scale=0.4]{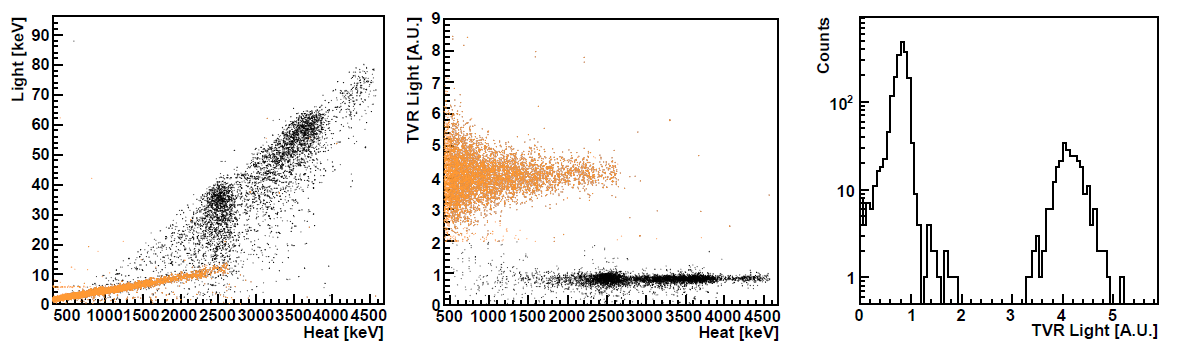}
\caption{Large ZnSe crystal (41~mm diameter and 17~mm height) exposed to $\alpha$ and $\beta/\gamma$ source: (Center) Distribution of the TVR pulse shape parameter on light pulses versus the amplitude of the heat pulses. Black points are $\alpha$ particles and colored points are $\beta/\gamma$'s; (Right) TVR distribution for light pulses between 2.4 and 2.6~MeV, the nearest energy window to the $Q_{\beta\beta}$ value (3~MeV); (Left) Scatter plot of Light versus Heat.}
\label{fig:znse-psa}
\end{figure}

This effect might be a limitation for $\alpha$ background rejection and makes ZnSe a not as good candidate.
However recently a high $\alpha$ rejection efficiency has been obtained by using a pulse shape analysis of the light signal instead of the standard amplitude. A dedicated run with a large ZnSe crystal was performed where the Light Detector was operated at a slightly higher temperature than usual, in order to have faster response and the signal sampling was increased to improve the resolution on the signal shape.
A light pulse shape parameter (here named Test Value Right, TVR) has been developped. 
Figure~\ref{fig:znse-psa} shows the distribution of TVR versus the heat energy obtained with this dedicated run with sources producing $\beta/\gamma$'s up to 2.6~MeV and $\alpha$'s up to 5~MeV. A high $\alpha$ rejection efficiency has been obtained near the $Q_{\beta\beta}$ energy region. 

The most critical point for ZnSe crystal is the risk of a unreproductibility of its characteristics with different samples of crystal growing, due to contaminations in transition metals. 
An important work has been performed in order to define a protocole for ZnSe growing, avoiding any metals contamination. Recent preliminary cryogenic tests of new crystals grown with this protocole, are very promizing. 

Two years ago, the LUCIFER project has been funded by the european grant ERC. The goal is to build a demonstrator with few tens of kg of enriched Zn$^{82}$Se and to reach a level of background of $10^{-3}$~cts/(keV.kg.yr).  
The design of the LUCIFER experiment is similar to CUORICINO. It consists of a tower of 12~modules, each elementary module corresponds to an array of 4 ZnSe crystals (each crystal 5$\times$5$\times$5~cm$^3$, 660~g) read out by a single light detector.
The setup could house up to $\approx$~20~kg of $^{82}$Se isotope corresponding to an expected sensitivity of  $\approx 10^{26}$~yr~(90$\%$~C.L.) after 5 years of running and a level of background of $10^{-3}$~cts/(keV.kg.yr).
Today the R\&D efforts in LUCIFER are the production of pure natural ZnSe crystals without metalic contaminations in order to obtain good heat and light performances and good reproductibility of the crystals growing. 
10~kg of enriched Se has been already purchased (URENCO). Studies are in progress to purify Se, to define protocole to obtain good beads for crystal production (powder is not good for crystals production) and to obtain a good efficiency for the process of crystallization. The optical quality and radiopurity of the obtained enriched Zn$^{82}$Se crystal will be also studied carefully.

\subsection{$^{100}$Mo: ZnMoO$_4$ and CaMoO$_4$ crystals}

ZnMoO$_4$ and CaMoO$_4$ crystals are two promising candidates, which have been recently tested as scintillating bolometers~\cite{pirro-2006}\cite{gironi-znmoo4}\cite{arnaboldi-novel-tech}\cite{pirro-znmoo4-2012}. 

The main advantage of these two crystals is that the pulse shape analysis of the heat signal alone, without any need of light readout, is sufficient to discriminate $\beta/\gamma$ to $\alpha$ particles. 
It has been initialy observed~\cite{arnaboldi-novel-tech} that the thermal signal induced by $\beta/\gamma$ and $\alpha$ particles shows a slightly different time development. This feature seems to be explained by the relatively long scintillation decay time (hundreds of $\mu$s) observed in these crystals. This long decay, combined with a high percentage of non-radiative de-excitation of the scintillation channel, will transfer phonons (heat signal) to the crystal. This extra phonon signal is extremely small but is different in amplitude for $\beta/\gamma$ and $\alpha$ particles, because of a different scintillation light yield.

Recently two different ZnMoO4 crystals, grown by the Nikoleav Institute of Inorganic Chemistry (NIIC, Novosibirsk, Russia), $\approx 30$~g each, have been tested in LNGS~\cite{pirro-znmoo4-2012} to validate this feature. A $\beta/\gamma$ to $\alpha$ discrimination of 20~$\sigma$ has been obtained at 2.6~MeV, using only a pulse shape analysis of the heat signal. The crystals are also very radiopure: the limits obtained after a long background measurement (407~hours) are $< 32 \mu$Bq/kg for $^{228}$Th and $^{226}$Ra. 

However if we take into account the slow response of the detector, the pile-up of two successive $\beta \beta 2 \nu$ decays becomes the limiting background ! 
With the $\beta \beta 2 \nu$ half-life of $T_{1/2}^{2\nu}=7 \ 10^{18}$~years for $^{100}$Mo, and assuming a time window of 5~ms in which one cannot distinguish two different pulses, this background has been estimated by Monte-Carlo simulations at the level of $2 \ 10^{-3}$~cts/(keV.kg.yr).

This background is still negligeable for an experiment with few tens of kilograms of crystal. For instance, assuming the CUORICINO cryostat equipped with ZnMoO4 crystals, equivalent to $\approx 30$~kg of crystal and 13~kg of Mo, and a typical energy resolution of FWHM$=5$~keV, the expected background is 0.3~counts/year in the FWHM energy window at $Q_{\beta\beta}$. It would correspond to a sensitivity of $T_{1/2}^{0\nu} > 0.7 \ 10^{26}$~yrs in 5 years of running.
However, for larger isoptope mass, developments of new NTD readout, delivering faster phonon signal, are required.

Recently a project of a $\beta \beta$ pilot experiment~\cite{znmoo4-giuliani-2012}, has been proposed by french, ukrainian and russian collaboration. It is a step-by-step program and the pilot experiment consists of first assembling 4 ZnMoO$_4$ crystals with enriched $^{100}$Mo. The mass of each crystal would be around 400~g, corresponding (with the four crystals) to a total mass of 675~g of $^{100}$Mo and an expected sensitivity on the $\beta\beta 0\nu$ half-life of $10^{24}$~y after 1 year of data collection and an expected zero background. A cryogenic test of a large ZnMoO$_4$ crystal of $\simeq 300$~g is foreseen in the next months in CSNSM Orsay.

I note that the 7~kg of enriched $^{100}$Mo available from NEMO-3 could be used to developp  Zn$^{100}$MoO$_4$ crystals.

CaMoO4 is also a good candidate. Unfortunately, the $\beta\beta 2\nu$-decay of $^{48}$Ca, with $Q_{\beta\beta} = 4.27$~MeV, although only 0.18\% of natural isotopic abundance, will result in an irreductible background
in the $\beta\beta 0\nu$ region of $^{100}$Mo. This background can be easily evaluated as $\approx 0.01$~counts/(keV.kg.yr). 
A new collaboration, named AMORE, propose to use $^{40}$Ca$^{100}$MoO$_4$ with depleted $^{40}$Ca~\cite{amore-taup11}. Indeed, about 30~kg of $^{40}$Ca ($^{48}$Ca~$< 0.001\%$) is available today (from the enrichment in the 80's in Russia of the $\approx 30$~g of $^{48}$Ca used in NEMO-3 and TGV experiments). Recently three large crystals of $^{40}$Ca$^{100}$MoO$_4$ with enriched $^{100}$Mo and depleted $^{40}$Ca have been produced and should be tested soon in the Yang Yang Underground Laboratory in South Corea (700~m depth).

\chapter{Other experiments}

%
%

\section{Liquid scintillators experiments}

Recently two large existing liquid scintillator detectors, Kamland and SNO, initialy developped for neutrino oscillation measurements, have been reused as $\beta\beta$ detectors by adding $\beta\beta$ isotope inside the liquide scintillator. It allows to reach relatively quickly a large amount of isotope $\approx 100$~kg.
I will review these two projects: Kamland-Zen and SNO$+$.

\subsection{KAMLAND-Zen experiment}

KAMLAND-Zen proposed to measure $^{136}$Xe isotope. 
This is because $^{136}$Xe is the simplest and least costly $\beta\beta$ isotope to enrich, its high $\beta\beta 2\nu$ half-life ($\simeq 2 \ 10^{21}$~y) reduces naturally the $\beta\beta 2\nu$ background, and it is relatively easy to dissolve Xenon gas in liquid scintillator with a mass fraction of a few $\%$. 
The original idea to dissolve Xenon gas in liquid scintillator installed in a very low background detector has been initially proposed by R.S. Raghavan in 1994~\cite{raghavan}. 
And KAMLAND-Zen collaboration did it !

KAMLAND~\cite{kamland} is an ultra low background detector, installed 1000~m deep in the Kamioka mine (Japan) and built in 1998-2001. It was initialy designed to confirm the solar neutrino oscillation by detecting the reactor anti-neutrino deficit with an oscillation length $\approx 175$~km. 
KAMLAND consists of a sphere (13~m diameter) filled with ultra radiopure liquid scintillator (1200~m$^3$, $\approx 1$~kton) contained in a second sphere filled of inert buffer oil (shield), surrounded by a water Cerenkov outer detector. 
The energy resolution of the liquid scintillator detector is FWHM~$\approx 6.5\%/\sqrt{E(MeV)}$ with 34$\%$ of photo-coverage (1325 17'' PMT's + 554 20'' PMT's). 
A long R$\&$D program has provided a purification of the liquid scintillator at a level of $0.2 - 2 \ 10^{-18}$~g/g in $^{238}$U and $1.9 - 4.8 \ 10^{-17}$~g/g in $^{232}$Th. 

\begin{figure}[!h]
\centering
\includegraphics[scale=0.5]{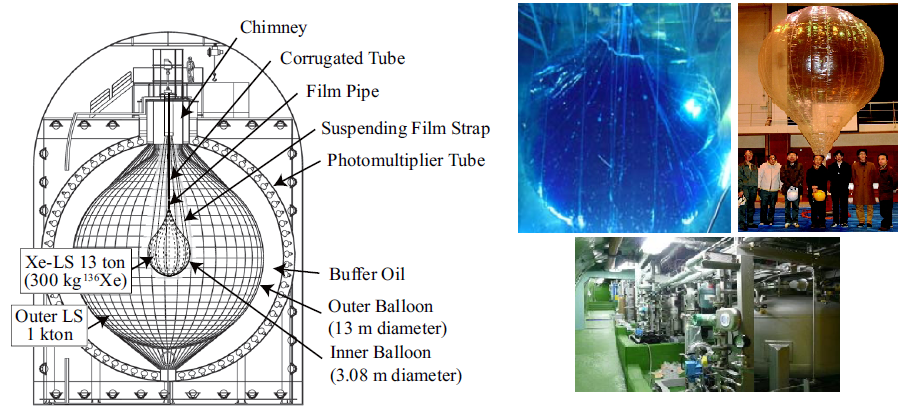}
\caption{(Left) Schematic view of the Kamland-Zen detector; (Right) Pictures of the mini ballon during tests in air and in water; View of the Xenon distribution.}
\label{fig:kamland-setup}
\end{figure}

The major modification of KAMLAND-Zen~\cite{kamland-zen} to the KamLAND experiment was the installation in summer 2011 of an inner, very radiopure ($2 \ 10^{-12}$~g/g in $^{238}$U and $3 \ 10^{-12}$~g/g in $^{232}$Th), very thin (25$\mu$m) and very transparent balloon ($\approx 3.1$~m diameter), suspended at the center of the KamLAND detector (see Figure~\ref{fig:kamland-setup}). This ballon contains 13 tons of Xe-loaded liquid scintillator (Xe-LS). The external liquid scintillator acts as an active shield against external $\gamma$'s. 

In August and September 2011, 330~kg of xenon gas, 91$\%$ enriched in $^{136}$Xe, has been dissolved in liquid scintillator and filled inside the balloon in KAMLAND. Data taking started in September 2011. 
The $2.5\%$ (mass ratio) of Xenon dilution in the liquid scintillator allows to maintain the KAMLAND energy resolution of FWHM~$\approx 10\%$ at $Q_{\beta\beta}(^{136}Xe)$.
It is also important to emphasize that the systematic variation of the energy reconstruction over the Xe-LS volume is less than 1$\%$ and the detector energy response is stable to within 1$\%$ also.

In order to reduce the background from the ballon material, the reconstructed vertex of events must be within 1.2~m of the ballon center, defining the fiducial volume with an effective mass of $^{136}$Xe of 129~kg. 
The residual $^{238}$U and $^{232}$Th concentration internal to the Xe-LS are measured using sequential decays of $^{214}$Bi$-^{214}$Po and $^{212}$Bi$-^{212}$Po and estimated to be $\approx 3 \ 10^{-16}$~g/g and $\approx 2.2 \ 10^{-15}$~g/g, respectively. It corresponds to a negligeable background in the $\beta\beta 0 \nu$ energy region.
One possible important background is the production of $^{10}$C($\beta^+$, $T_{1/2}=29.4$~mn, $Q_{\beta}$=3.65~MeV) by spallation of carbon ($^{12}C + \mu \rightarrow ^{10}C + 2n + \mu$) inside the Xe-LS. The expected number of background events is expected to be few counts per year. A dedicated electronic has been developped in order to detect the light produced by the crossing muon followed by detection of the delayed 2.2~MeV $\gamma$-rays produced by the $n-p$ capture of the two neutrons ($T_{1/2}=207 \mu$s). This veto, not yet used in preliminary analysis, allows to reject $95\%$ of the delayed $^{10}$C decay. 

\begin{figure}[!h]
\centering
\includegraphics[scale=0.5]{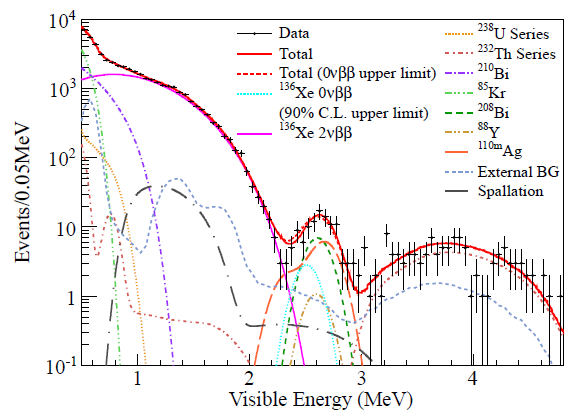}
\caption{Result after 77 days of data collection: energy spectrum of selected $\beta\beta$-decay candidates together with the best-fit backgrounds and $\beta\beta 2\nu$-decays, and the 90\%~C.L. upper limit for $\beta\beta 0\nu$ process.}
\label{fig:kamland-result}
\end{figure}

Figure~\ref{fig:kamland-result} shows the energy spectrum of selected $\beta\beta$ events, with the best fitted background, after the first 77 days of collected data with the effective mass of 129~kg $^{136}$Xe~\cite{kamland-zen}. 
A very nice signal of $\beta\beta 2\nu$-decay (more than 35000 $\beta\beta 2\nu$ events observed in only 77 days !) is well observed, providing an improved measurement of the $^{136}$Xe $\beta\beta 2\nu$-decay half-life: 
$$\mathrm{ T_{1/2}^{\beta\beta 2\nu}(^{136}Xe) = 2.38 \pm 0.02(stat) \pm 0.14(syst) \ 10^{21} \ years}$$
value in agreement with the result obtained by EXO-200. 

But unfortunately an unexpected background has been observed in the $\beta\beta 0\nu$ energy region, with a strong peak significantly above the $Q$-value of the $\beta\beta$ decay.
The collaboration has investigated all the possible contaminants with a peak structure in the 2.4-2.8 MeV energy window. 
It comes that two dominant contaminations can fit well the observed peak: $^{110m}$Ag ($\beta$-decay with associated $\gamma$'s, $T_{1/2}=360$~days, $Q_{\beta}=3.01$~MeV) and $^{208}$Bi (EC-decay, $T_{1/2}=5.31 \ 10^{5}$~years, $Q_{\beta}=2.88$~MeV, $\gamma$ of 2614.5~keV in 100$\%$ of the case).
\begin{itemize}
\item The $^{110m}$Ag isotope is a fission product, which has probably contaminated the  detectors materials by fallout from the Fukushima reactors accident in March 2011. 
Measurements of soil or ocean samples around Fukushima but also of soil samples taken near the inner ballon production facility have confirmed the contamination in $^{110m}$Ag. 
The contamination of the inner ballon by the Fukushima fallout is also consistent with the surface activity ratio of $^{134}$Cs to $^{137}$Cs (0.662 Mev $\gamma$) measured on the surface of the nylon ballon. 
\item The $^{208}$Bi isotope is not a fission product. However it has been conservatively considered as a possible background although its origin is unknown. We mention that its measurement by $\gamma$ spectroscopy is difficult since it decays to the excited $^{208}$Pb state producing in 100$\%$ of the case the same $\gamma$ of 2615~keV as in the $^{208}$Tl decay. 
\end{itemize}
Assuming these two contaminations, a limit to the half-life of the $\beta\beta 0\nu$ decay has beed derived: 
$$\mathrm{ T_{1/2}^{\beta\beta 0\nu}(^{136}Xe) > 5.7 \ 10^{24} \ years~(90\%~C.L.)}$$


KAMLAND-Zen has to be shut down in the next year for security check, allowing some upgrade of the detector. New measurements could start in 2013.

I think that this preliminary result presented by Kamland-Zen collaboration illustrates an important feature: a pure calorimeter detector with a relatively modest energy resolution (FWHM of few percent at $Q_{\beta\beta}$) is a well detector to reach a high sensitivity in case of no background, but cannot distinguish a possible signal to a mimicking $\gamma$ emitter backrgound in case of possible contamination. 
Indeed, looking at the energy spectrum obtained today (Figure~\ref{fig:kamland-result}), it is difficult to distinguish between a possible contamination in $^{208}$Bi, whose origin is unknown and is not understood, or a possible presence of $\beta\beta 0\nu$ signal. Who knows ?...
I would like also to underline that a possible risk of contamination of the xenon itself has to be also considered for next measurements.

\subsection{SNO+ experiment}

SNO$+$~\cite{sno+1}~\cite{sno+2} proposes to fill the Sudbury Neutrino Observatory (SNO)~\cite{sno} with ultrapure Nd-loaded liquid scintillator (LAB), in order to investigate the isotope $^{150}$Nd. 
$^{150}$Nd has the largest phase space factor of all double beta decay isotopes and its $Q_{\beta\beta}$ value of 3.37~MeV places it just above the $Q_{\beta}$ value of $^{214}$Bi (3.27~MeV) and therefore above the Radon background)

The present plan is to dilute 0.1\% in mass of natural neodynium salt in 1~ktons of liquid scintillator, providing a source of 56~kg of enriched $^{150}$Nd (the natural abundance in the $^{150}$Nd isotope is 5.6$\%$). Actually, up to 1\% of Nd can be loaded in liquid scintillator but the light yield becomes too small~\cite{nd-loading}. Optimization studies suggest 0.3\% loading Nd might be a better compromise between light output and statistics.  
Given the liquid scintillator light yield and photocathode coverage of the experiment, an energy resolution performance of about 6.4$\%$ FWHM at $Q_{\beta\beta}$, is expected. 
External backgrounds can be rejected by the external water shielding, self-shielding of the scintillator and with a 50$\%$ fiducial volume selection.

SNO+ requires to modify the support system of the inner acrylic vessel. Indeed, in SNO, the acrylic vessel, which was filled with heavy water, had to be held up. In SNO+, it is filled with LAB and it must be held down. Last data of SNO have shown that the inner surface of the acrylic vessel has been probably contaminated by leaks of mine dust. Therefore the inner surface of vessel must be cleaned. Also, since radiopurity requirements are more stringent for SNO+, the interface regions (like source deployment) must be very tight against leak of external mine air.

A crucial point is the radiopurity of the liquid scintillator. 
An extensive set of scintillator purification systems (distillation and in-situ recirculating purification) are under construction with the aim of reaching the purity levels (10$^{-17}$~g/g of $^{238}$U and $^{232}$Th chain activities) achieved by the Borexino
experiment~\cite{borexino-purif}\cite{sno-purif}. The approach of Borexino is being followed for SNO+. 

Another crucial point is the purification of the neodynium chloride salt. A newly developed ``self-scavenging'' technique has been developped and found to effectively remove thorium and similar containments from the neodymium chloride~\cite{nd-purif}.

\begin{figure}[!h]
\centering
\includegraphics[scale=0.5]{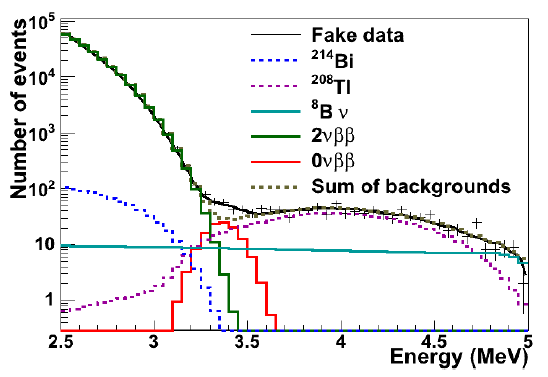}
\caption{Simulated event energy spectrum showing the expected backgrounds and $\beta\beta 0\nu$ signal (assuming $T_{1/2}^{0\nu}=10^{24}$~yrs) for a 3 years run with 0.1\% Nd-loading.}
\label{fig:sno}
\end{figure}

Figure~\ref{fig:sno} shows the expected background, which includes the $\beta\beta 2\nu$ from $^{150}$Nd, $^8$B solar neutrinos, $^{208}$Tl and $^{214}$Bi, for 3 years of data collection. The total expected background is $\approx 40$~counts/(FWHM.year). SNO$+$ compensate this relatively large background (due to a modest low energy resolution) by a large statistic. The expected sensitivity on the $\beta\beta 0\nu$-decay half-life, after 3 years of data collection, is $T_{1/2} > 6 \ 10^{24}$~years~(90\%~C.L.). 
A fiducial volume of 50\% of the acrylic vessel, 80\% detector live time and backgrounds equivalent to those achieved in Borexino, were used in the sensitivity calculations. 

The use of enriched $^{150}$Nd would obviously improve the sensitivity of SNO$+$. 
There are today two possible techniques to enrich Nd: the Atomic Vapour Laser Isotope Separation (AVLIS) and the high-temperature centrifugation.
Both techniques are studied in collaboration with SuperNEMO. 

It is scheduled to start taking data with the acrylic vessel filled with pure water in fall 2012 and to built the
scintillator process systems in parallel. Water will be replaced by purified liquid scintillator in 2013 and data will be collected during few months. First data with $^{nat}$Nd-loaded scintillator are foreseen by the end of 2013 or in 2014.

%
%

\section{Xe TPC experiments}

\subsection{Liquid Xenon TPC: the EXO-200 experiment}

\begin{figure}[!h]
\centering
\includegraphics[scale=0.5]{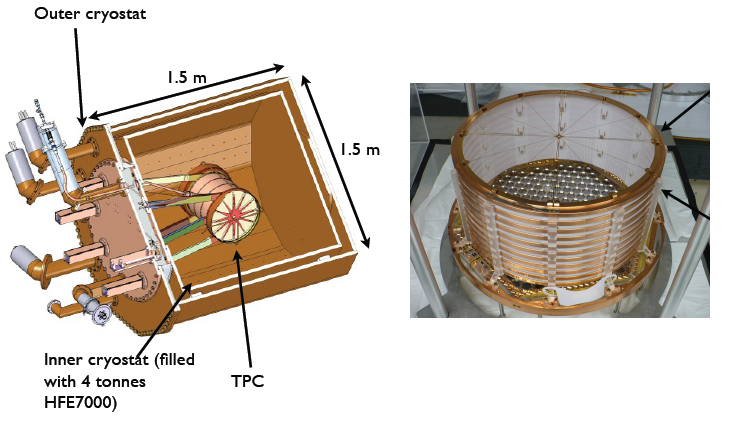}
\caption{(Left) Schematic view of the EXO-200 TPC inside the cryostat; (Right) Picture of the TPC during its assembly with APD holes.}
\label{fig:exo-setup}
\end{figure}

The EXO-200 detector, shown in Figure~\ref{fig:exo-setup}, is a time projection chamber (TPC) using 200~kg of liquid Xenon (enriched at 80\% in $^{136}$Xe), and located in the Wast Isolation Pilot Plant (WIPP, USA, 1600~m.w.e. depth). It started taking data in May 2011.

The main advantage of a liquid TPC is its compact geometry. 
The TPC is a cylinder of 40~cm diameter and 44~cm length with a cathod grid dividing the cylinder into two identical regions. Each end of the TPC contains two wire grids and an array of 250 large-area avalanche photodiodes, that allows for simultaneous readout of ionization and scintillation in liquid xenon. Wire grids provide a 2-dimensional transversal localization and energy information while the third longitudinal coordinate is obtrained from the scintillation light. 
The proved energy resolution is FWHM~$= 3.3 \%$ at $Q_{\beta\beta}$, using the anti-correlation between ionization and scintillation~\cite{exo-energy-resol}

The ability of the TPC to reconstruct energy depositions in space is used to remove interactions at the detector
edges where the background is higher. It reduces to a fiducial volume containing 63~kg of $^{136}$Xe. 
It also provides a discrimination between single-cluster depositions, characteristic of $\beta\beta$ and single $\beta$ decays in the bulk of the Xenon, from multi-cluster ones, generally due to $\gamma$-rays. 

\begin{figure}[!h]
\centering
\includegraphics[scale=0.5]{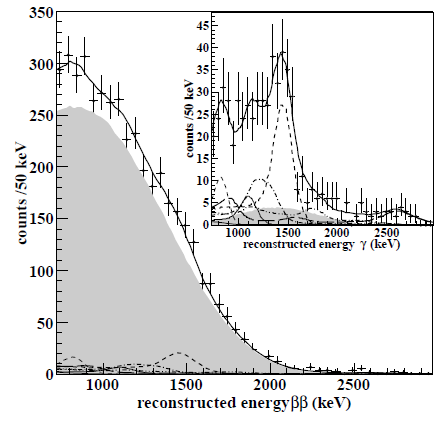}
\caption{Energy distributions from 752.66 hrs of EXO-200 single-cluster events (main panel) and multi-cluster events (inset). The result of the fit (solid line) contains $\beta\beta 2\nu$-decay (shaded region) and prominent background components of the TPC vessel. }
\label{fig:exo-result}
\end{figure}

The first phase of data taking has been devoted for the measurement of the $\beta\beta 2\nu$-decay and a relatively low electric field has been applied in order to provide more stable operation at the expense of the ionization energy resolution. 
The collaboration has recently published the result obtained with first month of data, reporting the first observation of the $\beta\beta 2\nu$-decay of $^{136}$Xe (just few months before Kamland-Zen, which confirmed the result). The measured energy spectrum is presented in Figure~\ref{fig:exo-result} both in single and multi-cluster channels. 
The backgrounds involving $\gamma$-rays are readily identified by their clear multi-cluster signature, while the single-cluster spectrum is dominated by a large structure with a shape consistent with the $\beta\beta 2\nu$-decay of $^{136}$Xe with a half-life of:

$$ \mathrm{T_{1/2}^{2\nu}= 2.11 \pm 0.04(stat) \pm 0.21(syst) \times 10^{21}~yrs } $$

The dominant background in the $\beta \beta 0 \nu$ is expected to come from the radon in the cryostat-lead air-gap and the $^{232}$Th and $^{232}$U in the TPC vessel. 
We mention that a very low level of Radon contamination has been measured inside the Liquid Xenon of 4.5~$\mu$Bq/kg. Only limit has been set for Thoron with an acivity $< 0.04 \mu$Bq/kg. 
Results of the $\beta \beta 0 \nu$ search should be published before summer 2012.

The EXO-200 target is to reach a level of background of $\sim 10^{-3}$~cts/(keV.kg.yr) in the $\beta\beta 0\nu$ energy region. With 200~kg of liquid Xenon, an energy resolution of 3.3\% FWHM at $Q_{\beta\beta}$ and a fiducial volume efficiency of 40\%, it would correspond to a background of about 16~counts/(FWHM.yr). This background is compensated by a large amount of mass (80~kg of $^{136}$Xe) and therefore large statistics. The expected sensitivity is $T_{1/2}^{0\nu} > 6 \ 10^{25}$~yrs~(90\%~C.L.) after 5 years of data collection.

It is clear that an additional background suppression or a higher energy resolution is needed for a ton scale experiment. To reduce the background, the ultimate goal of the EXO collaboration is to develop the so-called {\it barium tagging}~\cite{exo-ba-tag}. It consists of tagging the Ba$^{++}$ ion ($\beta\beta$ decay daughter of $^{136}$Xe) with optical spectroscopy methods by laser fluorescence excitation. Research and development for Barium ion grabbing and tagging is ongoing in parallel with the EXO-200 experiment.

\subsection{Gaseous Xenon TPC}

The use a high pressure gaseous Xenon (HPXe) TPC provides an higher energy resolution and a recontruction of the tracks in order to identify the $\beta\beta$ event topology 

\begin{figure}[!h]
\centering
\includegraphics[scale=0.5]{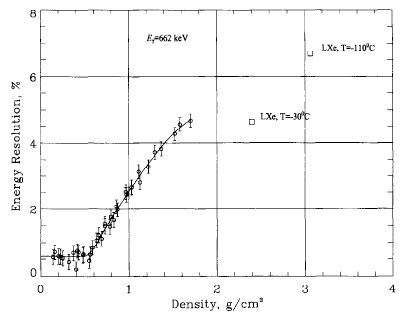}
\caption{The energy resolution (FWHM) for $^{137}$Cs 662~keV $\gamma$-rays, as a function of xenon density, for the ionization signal only~\cite{bolotnikov}.}
\label{fig:next-bolotnikov}
\end{figure}

It has been experimentaly demonstrated~\cite{bolotnikov} that high energy resolution can be obtained in high pressure gaseous xenon chamber, as illustrated in Figure~\ref{fig:next-bolotnikov}
A striking feature in Figure~\ref{fig:next-bolotnikov} is the apparent transition at density $\rho = 0.55$~g/cm~$^3$ (equivalent to about 50~bars). Below this density, the energy resolution is approximately constant. 
Extrapolating this observed resolution as $\sqrt{E}$ to the $^{136}$Xe $Q_{\beta\beta}$ value, one could get an energy resolution of FWHM$\approx 0.3 \%$, close to the ultimate resolution obtained from the Fano factor of gaseous Xenon. And this energy resolution should be constant even at high pressure up to $\approx 50$~bars. 

The second possible advantage of gas relative to liquid is the ability to exploit the topological signal of a $\beta\beta 0\nu$ decay: an ionization track, of about 20 cm length at 15 bar, tortuous because of multiple scattering, and with larger depositions or blobs in both ends.

The Gothard TPC~\cite{gothard} in the 90's represented a pioneering gaseous xenon TPC. Despite a modest energy resolution, it however demonstrated the potential of a gaseous TPC to powerfully utilize the rich topological signature of $\beta\beta 0\nu$ events in the gas, with the characteristic of two blobs topology, to further reduce background. 
Last few years, new projects were proposed using new technologies: electroluminescence, ion TPC and  Micromegas. 

\subsubsection{Electroluminescence and the NEXT-100 experiment}

\begin{figure}[!h]
\centering
\includegraphics[scale=0.4]{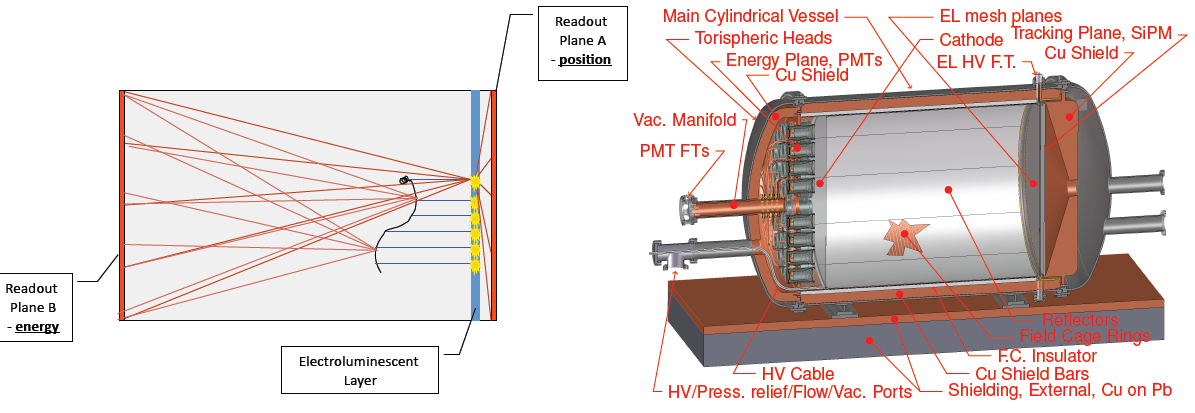}
\caption{(Left) Principle of the electroluminescence TPC; (Right) Design of the NEXT-100 TPC.}
\label{fig:next-setup}
\end{figure}

The NEXT-100 experiment, located in Canfranc Underground Laboratory (Spain), proposes to use the electroluminescence technique for the TPC readout.
The principle of NEXT-100 is as follows (Figure~\ref{fig:next-setup}):
\begin{itemize}
\item The prompt primary scintillation light emission (in VUV) is detected via  photodetectors (60 PMT's) behind a transparent cathode. This faint signal determines the $t_0$ time used for event position along the longitudinal drift.
\item Then ionisation electrons drift toward the opposite anode with a velocity $\approx 1$~mm/$\mu$s in a 0.5~kV/cm electric drift. The diffusion is not negligeable at 10 bar: about 9~mm/$\mathrm{\sqrt{m}}$ transverse, and about 4~mm/$\mathrm{\sqrt{m}}$ longitudinal. 
\item An additional grid in front of the anode creates 0.5~mm thick region of more intense field ($E/p \approx 4$~kV/cm/bar). A secondary scintillation light, named {\it electroluminescence}, is created in between grids by atomic desexcitation, with very linear gain of order 10$^3$ and over a $\approx 2$~$\mu$s interval. Finally, a segmented photodetector plane (7500 SiPM channels), located just behind the anode, performs the ``tracking''
\item The electroluminescence light, emitted isotropically, also reaches the cathode. The same array of PMT's used for $t_0$ measurement is also used for accurate calorimetry. A coating on the inner surface of the vessel improves the light collection.
\end{itemize}

The baseline design of NEXT-100 (Figure~\ref{fig:next-setup}) consists of a Stainless-steal vessel with a drift length of almost 2~m and a diameter of $\simeq$1.3~m (corresponding to a volume of $\simeq$2~m$^3$). The operation pressure ranges from 10 up to 15~bars, corresponding to $\simeq 100-150$~kg of Xenon. 

Two prototypes have been developped:
\begin{itemize}
\item[\textbullet] The NEXT-DBDM prototype (Berkeley) has been developped to demonstrate the capability to obtain a good energy resolution in HPXe using electroluminescence. Only 1 plane of PMT's located in the cathode has been used for calorimetry. An energy resolution of 1\% FWHM at 662~keV ($^{137}$Cs $\gamma$ source) has been obtained at 15~bars. However the prototype is smaller than the final NEXT-100 chamber, although the ratio drift length over diameter is similar. Also a radial dependence (solid angle) of the collected light has been observed and must be corrected carefully. 
\item[\textbullet] The NEXT-DEMO demonstrator (IFIC Valencia) has been developped to demonstrate both the energy resolution and the capability for tracking topology with two planes of photodetectors (PMT and SiPM). The drift region is only 30~cm. First tracks reconstruction have been presented, showing visible blobs at the extremity of the tracks. However the complete analysis has not yet been presented, in order to quantify the background rejection and the $\beta\beta$ efficiency.
\end{itemize}

The NEXT-100 target is to reach an energy resolution better than 1\% FWHM at $Q_{\beta\beta}$,  a level of background of $2 \ 10^{-4}$~cts/(keV.kg.yr) in the $\beta\beta 0 \nu$ energy region, with a $\beta\beta$ efficiency of 30\% due to a $\beta\beta 0\nu$ topological signature for background suppression.
One of the most critical backgrounds comes from the rare 2447~keV $\gamma$-ray emitted by $^{214}$Bi.
When the $\gamma$ is totaly contained, it produces a peak very close to the expected $\beta\beta 0\nu$ signal at $Q_{\beta\beta}=2462$~keV.

The NEXT-100 detector is foreseen to be installed in Canfranc end of 2013, although the construction of the detector is not yet totaly funded. 

\subsubsection{MICROMEGAS and the T-REX project}

Micromegas is a very promising readout technique for a HPXe TPC. 
The Micromegas readouts~\cite{micromegas} make use of a metallic micromesh suspended over a pixellised anode plane by means of insulator pillars, defining an amplification gap of the order of 25 to 150~$\mu$m. Electrons drifting towards the readout, go through the micromesh holes and trigger an avalanche inside the gap, inducing detectable signals both in the anode pixels and in the mesh.

It has been measured recently~\cite{micromegas-radiopurity}  that standard  microbulk  micromegas planes, manufactured out of kapton and copper foils, are very  radiopure with activities $< 30 \ \mu$Bq/cm$^2$ for $^{232}$Th and $^{238}$U chains. Also large surfaces are now available. It makes Micromegas as a very attractive technique for $\beta\beta 0\nu$ researches. 

A R$\&$D program, named T-REX, funded by an european grant, has started few years ago in Zaragoza University and Canfranc (initialy within the NEXT collaboration)~\cite{t-rex}. 

The work performed up to now has been focused in establishing the capability of microbulk readouts to work in high pressure Xe, and more specifically to measure their energy resolution in those conditions. 
For that task two prototypes have been built. 

The first prototype (NEXT-0-MM) is a stainless steel vessel of 2 litres, with a diameter of 14~cm and a drift region of 6~cm, and it is devoted to measurements with small scale readouts, to study gain, operation point, and energy resolution with low energy $\gamma$'s or $\alpha$'s. 
A resolution of 2\% FWHM have been achieved with 5.5~MeV $\alpha$ particles, for pressures up to 5 bar~\cite{next-0-mm}. 
Depending on the ionization quenching of $\alpha$ with respect to electrons, it corresponds to an energy resolution of 1\% to 3\% FWHM at $Q_{\beta\beta}$. 
It is an important result because it shows that microbulk Micromegas work well in pure high pressure Xe without any quencher.

The second prototype (NEXT-1-MM), of much larger size (drift of 35~cm and a readout area of 30~cm diameter), is capable of fully confining a high energy electron track and will therefore probe the detection principle in realistic conditions. 

We mention that another collaboration, EXO-gas, also develops a gas TPC to search for the double beta decay of $^{136}$Xe, and considers Micromegas as an option for its readout~\cite{exo-gas}.

\section{Crystals at room temperature}

\subsection{CANDLES-III}

CANDLES proposes to use natural CaF$_2$ crystals as scintillating detectors for the measurement of the isotope $^{48}$Ca. 
The main advantage of $^{48}$Ca is its high transition energy $Q_{\beta\beta}=4274$~keV, well above the $^{208}$Tl $\gamma$-ray (2.6~MeV), the $^{214}$Bi $\beta$-decay (3.3~MeV end point), and $\alpha$'s from natural radioactivity (max. 2.5~MeV with scintillation quenching factor). Another advantage is its low atomic mass, providing a larger number of nuclei per mass unit. 
However, its natural abundance is very low, only 0.187\% and it is today very difficult to enrich $^{48}$Ca in large quantities.  
I mention that historicaly CaF$_2$ was one of the most sensitive technique used for the search of $\beta\beta 0\nu$~\cite{caf2-1966}.

\begin{figure}[!h]
\centering
\includegraphics{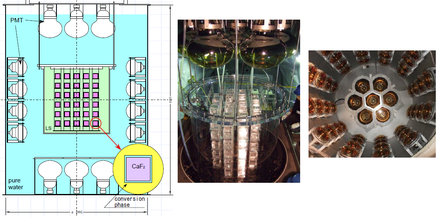}
\caption{Schematic view of the CANDLE-III detector and some pictures.}
\label{fig:candles3}
\end{figure}

CANDLES-III (Figure~\ref{fig:candles3}) is an array of 96 natural pure CaF$_2$ crystals ($10 \times 10 \times 10$~cm$^3$), for a total mass of $\approx 300$~kg, corresponding to about 300~g of $^{48}$Ca. Crystals are immersed in two liquid scintillators:  an internal one (in the vicinity of the crystals) which acts as a wavelength shifter for the UV light emitted by the CaF$_2$ crystals (conversion phase), and an external one which acts as an active veto gainst external background. This setup is installed inside a pure water buffer (3~m diameter and 4~m height) for an additional passive shield. The scintillation light emitted by the CaF$_2$ crystals in the veto liquid scintillator is read out by 48 13'' PMT's and 14 17'' PMT's.
The decay constants of the two involved scintillation processes are different ($\approx 1$~$\mathrm{\mu}$s for CaF$_2$ and $\approx 10$~ns for the liquid scintillator), which provides the capability of a pulse shape discrimination. 
The achieved energy resolution is $\approx 4.5 \%$~FWHM at $Q_{\beta\beta}$. 
The detector has been installed in the Kamiokande Underground Laboratory (Japan) and its commissioning has started in June 2011.

There are two potential backgrounds, coming from $^{238}$U and $^{232}$Th contaminations inside the crystals or in the conversion phase liquid scintillator in the vicinity of the crystals:
\begin{itemize} 
\item[\textbullet] $(\beta + \mathrm{delay} \ \alpha)$ pile-up events from $^{212}$Bi-$^{212}$Po ($\beta$ up to 2.2~MeV and quench $\alpha \approx 2.5$~Mev) and $^{214}$Bi-$^{214}$Po cascades,
\item[\textbullet] $(\beta + \gamma)$ pile-up events from $^{208}$Tl decay ($\beta$ up to 2.6~MeV and $\gamma$ 2.6~MeV).
\end{itemize}
The crystals and the liquid scintillator must be ultra radiopure: the average radiopurities of the crystals are $\approx 36$~$\mathrm{\mu}$Bq/kg and $\approx 28$~$\mathrm{\mu}$Bq/kg in $^{238}$U and $^{232}$Th chains, respectively. 
Residual BiPo $\beta + \mathrm{delay} \alpha$ pile-up events are identified using pulse shape analysis with a rejection efficiency of $\approx 90\%$~\cite{candle}. $\beta + \gamma$ pile-up events are the dominant background. 
Despite a modest energy resolution, the $\beta\beta 2\nu$ background is negligeable, about 0.01~counts per year.
The total expected level of background in the $\beta\beta 0\nu$ energy region (estimated by Monte-Carlo) is about $10^{-3}$~cts/(FWHM.kg.yr), equivalent to about 0.3~counts per year~\cite{candles3-dbd2011}.
It corresponds to an expected sensitivity of $T_{1/2}^{0\nu} > 3.7 \ 10^{24}$~yrs~(90\%~C.L.) with 5~years of collected data.

A R\&D program for $^{48}$Ca enrichment is in progress. Two methods are tested. 
The first one is a chemical process. When a calcium solution migrates through ``crown'' Ether (resin or liquid), Ca$^{2+}$ ions are attracted by negative charged O$^-$ of the CO dipole in Ether. Adsorption of $^{40}$Ca (97\% natural abundance) is slightly more efficient than $^{48}$Ca, providing a chemical enrichment in $^{48}$Ca. This method, first tested with resin crown ether, is now tested in liquid phase (Hiroshima Univ.), in which extraction is expected to be faster.
However this method is not very efficient. The goal is to enrich $^{48}$Ca at a level of $\approx 2\%$ (factor 10). 
The second methode of enrichment is the laser separation by radiation pressure. A bench test is in progress in Fukui Univeristy. 
I also remind that a R\&D program of $^{48}$Ca enrichment using AVLIS technique is in progress in South Corea, within the SuperNEMO collaboration~\cite{ca48-enrichment}. The goal is to provide a significant amount of $^{48}$Ca, about 1~kg within 3 years.

Two other improvements of the CANDLES experiment are investigated by the collaboration: the purification of CaF$_2$ crystals at the level of about 1~$\mu$Bq/kg and the improvement of the energy resolution, by cooling the CaF$_2$ crystals.

A future detector, proposed by the CANDLES Collaboration, would contain 2 tons of CaF$_2$ crystal with 2\%~$^{48}$Ca enrichment, equivalent to about 21~kg of $^{48}$Ca. Assuming an improved energy resolution of 3\%~FWHM at $Q_{\beta\beta}$, and crystals radiopurities of 1~$\mu$Bq/kg, the expected sensitivity would be $T_{1/2}^{0\nu} > 2.5 \ 10^{26}$~yrs~(90\%~C.L.) with 5~years of collected data. The $\beta\beta 2\nu$ background is not anymore negligeable, with about 0.2 expected counts per year in the $\beta\beta 0\nu$ region. 
We note that the current CANDLES-III detector could reach the same sensitivity  if calcium was enriched to 13\% using AVLIS method. 


\subsection{COBRA}

COBRA~\cite{cobra} is a proposed array of $^{116}$Cd-enriched CdZnTe semiconductor detectors at room temperature.
The final aim of the project is to deploy 117~kg of $^{116}$Cd with high granularity. Small scale prototypes
have been realized at LNGS (Italy). The proved energy resolution is 1.9\% FWHM. 
The project is in R\&D phase. 
Recent results on pixellization shows that the COBRA approach may allow an excellent tracking capability, equivalent to a solid state TPC.

\chapter{Summary}
\label{chap:summary}

We are today in a period of intense activities with a large variety of $\beta\beta$ experiments, starting to take data or in the construction phase. 
These projects aim to demonstrate the capability to reduce the background by at least one order of magnitude with at least 100~kg of isotope.

We emphasize that the search of $\beta\beta 0 \nu$-decay requires several experimental techniques and more than one isotope.
This is because there could be unknown background and gamma transitions, and a line observed at the end point in one isotope does not necessarily imply that $\beta\beta 0\nu$ decay was discovered. Nuclear matrix elements are also not very well known. 

We must also always keep in mind that every time one starts running a new $\beta\beta$ detector, a new unexpected background is discovered. The Radon background in the first phase of NEMO-3, the  $\alpha$'s contamination on surface of the materials in TeO$_2$ bolometers, the $^{42}$K contamination in the liquid argon in GERDA, and recently the contaminants from Fukushima fallout (and possibly from other origins) in KAMLAND-Zen are few examples. 
Actually, no $\beta\beta$ experiment has succeeded in obtaining a zero backgroundat the 10~kg scale.
So it is very premature to project experiences at the ton scale.  The first chalenge today is to run a modular pilot experiment with zero background, in the first phase at the 10~kg scale, which can be later enlarged to contain about 100~kg of isotope.

\subsubsection*{Germanium detectors}


The new GERDA experiment (installed in LNGS, Italy) uses Ge detector array, directly immersed inside liquid Argon, which acts both as cryogenic liquid and shield against external $\gamma$-rays. The strings design allows to deploy crystals progressively inside the experiment. 
GERDA started taking data (Phase 1) with refurbished enriched $^{76}$Ge crystals, which were used in the 90's in the previous Heidelberg-Moscow and IGEX experiments. It corresponds to about 20~kg of $^{76}$Ge. 
First data have shown a prominent and unexpected background: the contribution of $^{42}$K, the progeny of $^{42}$Ar, about 20 times higher than expected. This is because of the drift of the $^{42}$K ion towards the germanium crystals before decaying. A copper cylinder surrounding the array of crystals reduced this background to a level of $6 \ 10^{-2}$~cts/(keV.kg.y) in the $\beta\beta 0\nu$ energy region. It is 3 times lower than the level of background measured in Heidelberg-Moscow experiment but still a factor 6 higher than the Phase 1 target ($10^-2$~cts/(keV.kg.y)). 

Two methods are under development in order to reduce the background, with the aim to reach a level of $10^{-3}$~cts/(keV.kg.y). First, new commercially available Broad Energy Germanium (BEGe) detectors will be deployed. They provide a superior pulse shape discrimination performances against multi-compton events, produced by external $\gamma$-rays. The collaboration is also preparing a design for the detection of the scintillation light from argon, as an additional external background veto. 
We underline that the pulse shape discrimination and the LAr scintillation veto are only efficient to reduce external $\gamma$ background. Potential background produced by $\beta$-decay or degraded $\alpha$ from  surface contaminations of the detectors (like $\beta$ from $^{42}$K), or in the vicinity of the germanium crystals could be a limiting background. 

In parallel, the MAJORANA experiment (WIPP, USA) is under construction in the USA. It will measure 20~kg of Ge with similar BEGe detectors, but in an ultra radiopure electroformed cryostat and with a standard lead shield. 

For the next generation of germanium experiments, the most dangereous background is the potential bulk contamination of the germanium crystals in $^{68}$Ge cosmogenic isotope. R\&D on $^{68}$Ge deenrichment will be required for ton-scale projects..

\subsubsection*{Bolometers}

The assembly of CUORE-0 first tower is now completed and data collection will start in 2012 for two years of running, using CUORICINO cryostat (LNGS, Italy). The goal is to demonstrate a background lower than 0.05~cts/(keV.kg.yr). 
The expected sensitivity is $T_{1/2}(\beta\beta 0 \nu) > 6 \ 10^{24}$~y~(90$\%$~C.L.) after 2 years of measurement.
The first data of CUORE in the new large cryostat is expected for 2014. The required background of less than 0.01 cts/(keV.kg.y) should be dominated by degraded $\alpha$'s from the copper surface surrounding the crystal and external $\gamma$'s. With the complete detector (988 crystals, 206~kg of $^{130}$Te), it will correspond to an expected sensitivity of $10^{26}$~y. If SSB detectors are able to reject $\alpha$ background and effectively achieve a background of $10^{-3}$~cts/(keV.kg.y), CUORE could improve its sensitivity up to $7 \ 10^{26}$~y. 

Recent results obtained with scintillating bolometers indicate that it is a very promising calorimetric technique for the search of $\beta\beta0\nu$ decay. First, it provides the possibility to study 4 different isotopes ($^{116}$Cd, $^{82}$Se, $^{100}$Mo and $^{48}$Ca) with a high $\beta \beta 0 \nu$ detection efficiency (almost 100$\%$) and a good energy resolution. The $\alpha$ discrimination allows to reject the dominant background due to degraded $\alpha$ particles and also to identify and reject possible residual bulk contaminations of the crystals in $^{232}$Th and $^{238}$U. Also, the $Q_{\beta\beta}$-values above the 2615~keV $\gamma$-rays from $^{208}$Tl allow to suppress strongly the background due to external $\gamma$'s. The potential residual backgrounds in the $\beta\beta0\nu$ energy region are rare high energy $\gamma$-rays from $^{214}$Bi contamination inside the shield, or coincidences of the 2615~keV and 583~keV $\gamma$-rays emitted by $^{208}$Tl, with the condition that $^{208}$Tl contaminations are sufficiently close to the detectors. 
Assuming the radiocontamination level measured in CUORICINO cryostat and shield, the expected level of background would be in the range 10$^{-3}$ (conservative) to 10$^{-4}$~counts/(keV.kg.yr) in the  $\beta \beta 0 \nu$ energy region.

The LUCIFER project, funded by the european grant ERC, did the choice to use ZnSe crystal because of its large light yield and its ``natural'' radiopurity. Despite an unfavourable quenching factor which should limit the $\beta/\gamma$ to $\alpha$ discrimination, recent preliminar results using a pulse shape analysis of the light signal have shown the possibility to discriminate $\alpha$ particles with a high efficiency. 
Finaly a possible $\beta \beta 0 \nu$ signal discovery could be confirmed or ruled out by the SuperNEMO experiment without any theoretical uncertainty of the nuclear matrix elements, since SuperNEMO will study the same $^{82}$Se isotope. 

ZnMoO$_4$ and CaMoO$_4$ crystals are also excellent candidates because the pulse shape analysis of the heat signal alone, without any need of light readout, is sufficient to discriminate $\beta/\gamma$ to $\alpha$ and therefore to reject the $\alpha$ background. The 7~kg of enriched $^{100}$Mo from NEMO-3 could be reused for this measurement. However development of ZnMoO$_4$ crystal is very recent and only small crystals have been produced so far. It appears also that the pile-up of two successive $\beta \beta 2 \nu$ decays becomes the limiting background, estimated at the level of $2 \ 10^{-3}$~cts/(keV.kg.yr). It is due to the relatively ``low'' $\beta\beta 2\nu$ half-life of the $^{100}$Mo. This background is acceptable for a few tens of kg mass scale experiment. But for larger isotope mass, developments of new NTD, delivering faster phonon signal, are required in order to reduce this unexpected background. ZnMoO$_4$ could be also used in LUCIFER. 
Recently a project of a $\beta \beta$ pilot experiment has been proposed by french, ukrainian and russian collaboration. It is a step-by-step program and the pilot experiment consists of first assembling 4 ZnMoO$_4$ crystals with enriched $^{100}$Mo. The mass of each crystal would be around 400~g, corresponding (with the four crystals) to a total mass of 675~g of $^{100}$Mo and an expected sensitivity on the $\beta\beta 0\nu$ half-life of $10^{24}$~y after 1 year of data collection and an expected zero background. A cryogenic test of a large ZnMoO$_4$ crystal of $\simeq 300$~g is foreseen in the next months in CSNSM Orsay (see level).

For CaMoO$_4$, the $\beta\beta 2\nu$-decay of $^{48}$Ca, although only 0.18\% of natural abundance, will result in a large background in the $\beta\beta 0\nu$ region of $^{100}$Mo.  
The AMORE experiment propose to use $^{40}$Ca$^{100}$MoO$_4$ using the 30~kg of depleted $^{40}$Ca available today in Russia.  

Production of scintillating crystals with enriched isotope is a critical issue because it is necessary to avoid possible contaminations in transition metals, which would deteriorate optical and scintillation quality, and contaminations in $^{238}$U and $^{232}$Th. Recently large enriched $^{116}$CdWO$_4$ crystals have been produced and scintillation performances and radiopurity have been validated.  
Also three  crystals of $^{40}$Ca$^{100}$MoO$_4$ with enriched $^{100}$Mo and depleted $^{40}$Ca have been recently produced by AMORE and should be tested in the Yang Yang Underground Laboratory (South Corea).
Production of very pure enriched Zn$^{82}$Se crystal (with no metallic contamination) is in progress for LUCIFER project. 

Ideally, a large cryostat, like CUORE or EUREKA, would be an ideal facility to accomodate all the four possible crystals, for a unique $\beta\beta$ cryogenic experiment.

\subsubsection*{NEMO-3 and SuperNEMO tracko-calo detectors}

The combination of a tracking detector and a calorimeter provides a direct reconstruction of the tracks of the two emitted electrons from the source foil. It also allows to identify and measure each background component with a high rejection efficiency. However the price is a lower $\beta\beta 0\nu$ efficiency and a lower energy resolution. 
In any case, I think that a direct confirmation by a tracko-calo detector is required if a $\beta\beta 0\nu$ signal is observed by a calorimetric detector. It is thus a complementary approach. 
NEMO could also distinguish the underlying mechanisme (standard process versus V+A right-handed weak current) using the reconstructed angular distribution between the two emitted electrons.

The NEMO-3 detector (Modane, France) has demonstrated the performances of this technique with about 10~kg of isotopes and a sensitivity of $10^{24}$~y with $^{100}$Mo. The residual backgrounds were the tail of $\beta\beta 2\nu$-decay due to the modest energy resolution, the contamination of the $^{100}$Mo source foils in $^{208}$Tl ($\approx 100$~$\mu$Bq/kg), and radon contamination ($\approx 5$~mBq/m$^3$) inside the tracking chamber.

The future SuperNEMO detector is based on an extension and an improvement of the techniques used in NEMO-3.
The goal is to accomodate about 100~kg of enriched $\beta\beta$ isotope in order to reach a sensitivity of 10$^{26}$~y. The design is a planar and modular geometry, composed of 20 modules. The baseline is to measure $^{82}$Se but $^{150}$Nd and $^{48}$Ca are also studied.
The $\beta\beta 2\nu$ background is reduced by the relatively high half-life of  $^{82}$Se (about 14 times higher than $^{100}$Mo), and the improvement of the calorimeter energy resolution ($\approx 4.5\%$ FWHM at $Q_{\beta\beta}$). The radiopurities of the $^{82}$Se foils and the Radon contamination must be improved by a factor $\approx 50$. Developments of new Se purification techniques and foils production are in progress. A dedicated BiPo detector will be installed mid-2012 in Canfranc in order to measure and validate the radiopurities of the $^{82}$Se foils. The design of the SuperNEMO detector is optimized to avoid external Radon diffusion inside the tracking chamber and inner materials are selected to avoid inner Radon emanation.

A first SuperNEMO module, named SuperNEMO demonstrator, will be installed in the Modane Underground Laboratory. 
It will accomodate 7~kg of $^{82}$Se.
The construction of the tracking detector and the $\gamma$ veto has been funded by UK and is under progress, for a delivery in LSM Modane in 2013. 
First data could be taken in 2015.
The sensitivity of the demonstrator will be  $\mathrm{ T_{1/2}^{0 \nu} > 6.6 \ 10^{24} \ years (90\% C.L.)}$, with 2.5~years of collected data. 

Two other isotopes are also studied by the collaboration: $^{150}$Nd and $^{48}$Ca.
A R\&D program of $^{48}$Ca enrichment using the Atomic Vapour Laser Isotope Separation (AVLIS) technique is in progress in South Corea. The goal is to provide a significant amount of $^{48}$Ca, about 1~kg within 3 years.
Enrichment of $^{150}$Nd is also investigated with two possible techniques: the Atomic Vapour Laser Isotope Separation (AVLIS) and the high-temperature centrifugation.

\subsubsection*{Large liquid scintillator detectors}

Recently two large existing ultra radiopure liquid scintillator detectors, Kamland and SNO, initialy developped for neutrino oscillation measurements, have been reused as $\beta\beta$ detectors by adding isotope inside the liquide scintillator. It allows to reach relatively quickly a large amount of isotope ($\approx 100$~kg) but with a limited energy resolution.

KAMLAND-Zen (Kamioka, Japan) proposed to measure $^{136}$Xe isotope. 
This is because $^{136}$Xe is the simplest and least costly $\beta\beta$ isotope to enrich, its high $\beta\beta 2\nu$ half-life ($\approx 2 \ 10^{21}$~y) reduces naturally the $\beta\beta 2\nu$ background, and it is relatively easy to dissolve Xenon gas in liquid scintillator with a mass fraction of few $\%$. 
KAMLAND-Zen started taking data in September 2011, with 300~kg of $^{136}$Xe. 
An energy resolution of $\approx 10\%$ FWHM at $Q_{\beta\beta}$ has been obtained.
A 40\% fiducial volume has been applied in order to reduce external background. 
The $\beta\beta 2\nu$-decay has been measured with a high signal over background ratio and high statistics.
But, unfortunately, unexpected background has been observed in the $\beta\beta 0\nu$ energy region. 
This background is explained by a contamination of some detector materials in $^{110m}$Ag and $^{208}$Bi. $^{110m}$Ag is a known fission product coming from the Fukushima fall-out and its half-life is fortunately relatively small, about 1~year. 
However the origin of $^{208}$Bi contamination (half-life $\approx 5 \ 10^{5}$~years) is still not understood, and it has been conservatively considered as a possible background. A possible risk is a contamination of the xenon itself. This issue has to be considered seriously. New measurements could start in 2013 after a 1~year shut-down.

SNO$+$ (Sudbury, Canada) proposes to fill the SNO detector with ultrapure Nd-loaded liquid scintillator, in order to investigate the isotope $^{150}$Nd. 
$^{150}$Nd has the largest phase space factor of all double beta decay isotopes and its $Q_{\beta\beta}$ value of 3.37~MeV places it just above the $Q_{\beta}$ value of $^{214}$Bi (3.27~MeV) and therefore above the Radon background. However theoretical uncertainties on its nuclear matrix element are relatively large because it is a highly deformed nucleus.
The present plan is to dilute 0.1\% in mass of natural neodynium salt, providing a source of 56~kg of $^{150}$Nd. A larger mass could be loaded in a second phase, up to 150~kg of $^{150}$Nd.
An efficient method has been developped to purify neodinium salt and purification systems of the liquid scintillator are under construction. 
However the expected background is relatively large, $\approx 40$~counts per year in the $\beta\beta 0\nu$ FWHM energy window, due to a modest energy resolution of 6.4$\%$ FWHM at $Q_{\beta\beta}$. 
First data with $^{nat}$Nd-loaded scintillator are foreseen by the end of 2013 with an expected sensitivity of $T_{1/2} > 6 \ 10^{24}$~years~(90\%~C.L.) after 3 years of data collection.
Enrichment of $^{150}$Nd is investigated in collaboration with SuperNEMO.

\subsubsection*{Xenon TPC experiments}

EXO collaboration decided to developp a liquid Xenon TPC with the main advantage to provide a large mass of Xenon with a relatively small TPC chamber.
EXO-200 (WIPP, USA) is a small TPC ($\approx 0.05$~m$^3$) containing 200~kg of liquid Xenon, enriched at 80\% in $^{136}$Xe. An energy resolution of 3.3\% FWHM at $Q_{\beta\beta}$ has been obtained. 
The experiment started taking data in May 2011.
Results obtained with first month of data have been recently published, reporting the first observation of the $\beta\beta 2\nu$-decay of $^{136}$Xe (only few months before Kamland-Zen).
The EXO-200 target is to reach a level of background of $\sim 10^{-3}$~cts/(keV.kg.yr) in the $\beta\beta 0\nu$ energy region. With the fiducial volume efficiency of 40\%, the expected sensitivity is $T_{1/2}^{0\nu} > 6 \ 10^{25}$~y~(90\%~C.L.) after 5 years of data collection.
In order to reduce the background for a ton scale experiment, the ultimate goal of the EXO collaboration  is to develop the so-called barium tagging. It consists of tagging the Ba$^{++}$ ion ($\beta\beta$ decay daughter of $^{136}$Xe) with otpical spectroscopy methods by laser fluorescence excitation. Research and development for Barium ion grabbing and tagging is ongoing in parallel with the EXO-200 experiment.

The use a high pressure gaseous Xenon (HPXe) TPC provides an higher energy resolution and a partial recontruction of the tracks in order to identify the $\beta\beta$ event topology. 
The NEXT collaboration (Canfranc, Spain) proposed to use the electroluminescence technique for the TPC readout.
An energy resolution of 1\% FWHM at  662~keV (corresponding to 0.5\% at $Q_{\beta\beta}$) has been measured with a prototype running at 15~bars.
The NEXT-100 TPC chamber, with a volume of $\approx 2$~m$^3$, is under construction, and will contain 100~kg of enriched Xenon gas at a pressure of 10~bars. 
It is foreseen to be installed in Canfranc end of 2013, although the construction of the detector is not yet totaly funded. 
The target is to reach an energy resolution better than 1\% FWHM at $Q_{\beta\beta}$, with a level of background of $2 \ 10^{-4}$~cts/(keV.kg.yr) in the $\beta\beta 0 \nu$ energy region, and a $\beta\beta$ efficiency of 30\% taking into account the $\beta\beta 0\nu$ topological signature for background suppression.

The T-REX R$\&$D program (Zaragoza-Canfranc, Spain) is studing the MICROMEGAS technique for an alternative TPC readout. Standard  microbulk  micromegas planes, manufactured out of kapton and copper foils, are today available in large surface and are very radiopure.
A first prototype has demonstrated the capability of micromegas readouts to work in high pressure Xe. A second larger prototype is under development in order to measure the energy resolution and the topology signature capability in real conditions.

I emphasize that one of the most critical background for Xenon TPC comes from the rare 2447~keV $\gamma$-ray emitted by $^{214}$Bi.
When the $\gamma$ is totaly contained, it produces a peak very close to the $Q_{\beta\beta}$-value of 2462~keV.

\subsubsection*{Crystals at room temperature}

CANDLES proposes to use natural CaF$_2$ crystals as scintillating detectors for the measurement of the isotope $^{48}$Ca. 
The main advantage of $^{48}$Ca is its high transition energy $Q_{\beta\beta}=4274$~keV. 
Unfortunately, its natural abundance is very low, only 0.187\% and it is today technically very difficult to enrich it.  
CANDLES-III (Kamiokande, Japan) is an array of 96 natural CaF$_2$ crystals, for a total mass of $\approx 300$~kg, corresponding to about 300~g of $^{48}$Ca. Crystals are immersed in a two-phase liquid scintillator which acts as an active veto gainst external background. 
The achieved energy resolution is $\approx 4.5 \%$~FWHM at $Q_{\beta\beta}$. 
The commissioning of the detector has started in June 2011.

Potential $^{238}$U and $^{232}$Th contaminations inside the crystals or in the liquid scintillator in the vicinity of the crystals can produce background by the pile-up of $(\beta + \gamma)$ from $^{208}$Tl decay or by the pile-up of $(\beta + \mathrm{delay} \ \alpha)$ from the $Bi-Po$ cascades. This second background can be identified using a pulse shape analysis with a rejection efficiency of $\approx 90\%$.
Taking into account the radiopurity measurement of the crystals, $\approx 30$~$\mathrm{\mu}$Bq/kg in $^{238}$U and $^{232}$Th chains, the expected level of background in the $\beta\beta 0\nu$ energy region (estimated by Monte-Carlo) is about $10^{-3}$~cts/(FWHM.kg.yr), equivalent to about 0.3~counts per year. It corresponds to an expected sensitivity of $T_{1/2}^{0\nu} > 3.7 \ 10^{24}$~yrs~(90\%~C.L.) with 5~years of collected data.

A R\&D program for $^{48}$Ca enrichment is in progress, with two possible techniques: chemical process by migration of a calcium solution through ``crown'' Ether, and laser separation by radiation pressure. 
I  remind that $^{48}$Ca enrichment using AVLIS technique is also in progress in South Corea, with the aim to enrich 1~kg within 3~years.
Two other improvements of the CANDLES experiment are also investigated: the purification of CaF$_2$ crystals at the level of about 1~$\mu$Bq/kg and the improvement of the energy resolution, by cooling the CaF$_2$ crystals.

We also mention the COBRA project, in R\&D phase. It consists of an array of $^{116}$Cd-enriched CdZnTe semiconductor detectors at room temperature.
Small scale prototypes have been realized at LNGS (Italy). The proved energy resolution is 1.9\% FWHM. 
But level of background is still high.
Recent results on pixellization shows that the COBRA approach may allow an excellent tracking capability, equivalent to a solid state TPC.

\subsubsection*{Summary of the expected sensitivities}

\begin{figure}[!h]
\centering
\includegraphics[scale=0.5]{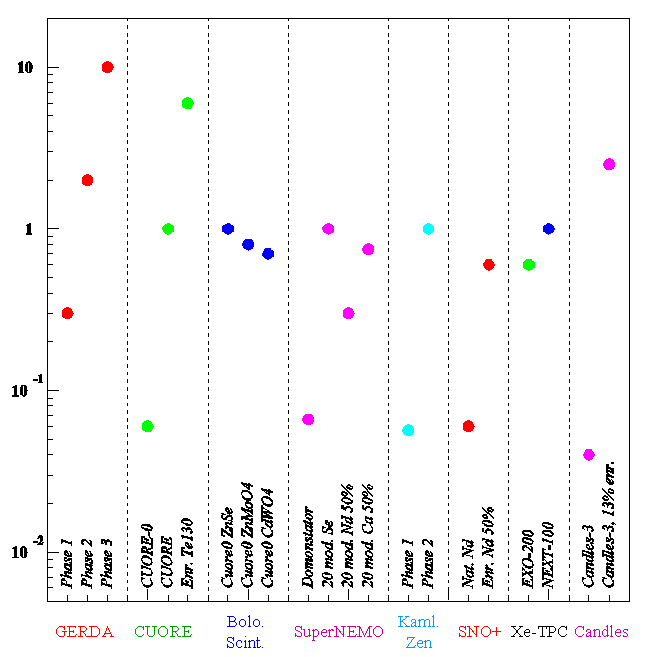}
\caption{Expected sensitivity on $T_{1/2}^{0\nu}$ (90\%~C.L.) in 10$^{26}$~years, for current or possible future experiments. See text for discussion.}
\label{fig:fig1}
\end{figure}

\begin{figure}[!h]
\centering
\includegraphics[scale=0.5]{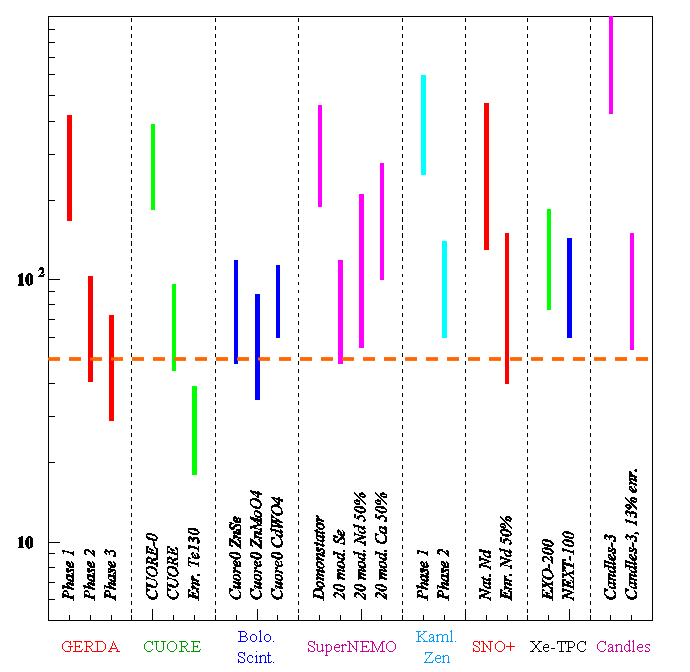}
\caption{Expected sensitivity on the effective Majorana neutrino mass $\langle m_{ee} \rangle$ (in meV), for current or future possible experiments (see text for discussion). The horizontal dashed line corresponds to the upper limit of the inverted hierarchy region.}
\label{fig:fig2}
\end{figure}

I have tried to summarize in Table~\ref{tab:projects-summary} and Figures~\ref{fig:fig1} and \ref{fig:fig2} the expected sensitivities of the new $\beta\beta$ projects, on the limit half-life $T_{1/2}(\beta\beta 0 \nu)$  and on the corresponding effective Majorana neutrino mass $\langle m_{ee} \rangle$. The minimum and maximum $\langle m_{ee} \rangle$ values correspond to the uncertainty range of the nuclear matrix elements given in Tables~\ref{tab:g0nu-nme} and \ref{tab:current-limits}. 

Expected sensitivities have been calculated with 5 years of data collection, except GERDA Phase 1 (1~year), SuperNEMO demonstrator (2.5~years), CUORE-0 (2~years) and SNO$+$ Phase 1 (3~years), and using the Feldman-Cousins statistical method~\cite{feldman-cousins}. 

For the scintillating bolometers, I have assumed a CUORICINO-like setup equipped with enriched scintillating bolometers, assembled in 13~layers of 4~crystals, each one with a dimension $5\times5\times5$~cm$^3$ (like in CUORICINO). It corresponds to a total volume of crystal of 6500~cm$^3$. The densities used for the calculation are: 7.9~g/cm$^3$ for $^{116}$CdWO$_4$, 4.3~g/cm$^3$ for Zn$^{100}$MoO$_4$, 4.7~g/cm$^3$ for $^{40}$Ca$^{100}$MoO$_4$ and 5.26~g/cm$^3$ for Zn$^{82}$Se. I have also assumed a level of background of 10$^{-3}$~counts/(keV.kg.yr), corresponding to the conservative value estimated by Monte-Carlo.

I have assumed a fiducial volume of 50\% for Kamland-Zen and 40\% for EXO-200, and an efficiency of 30\% for NEXT-100, due to topology signature. The efficiencies of the other calorimeter experiments are assumed to be 100\%, although it might be slightly lower ($\approx 90\%$) due to additional background rejection.

I have intentionally not specified projects at the ``ton scale'', because I think that it does note make sense today. 
Each project is separated is different phases. Some are rather realistic, some are optimistic. I let the reader be judge. No... Actually, only real data will be judge !

\begin{landscape}


\begin{table}[!h]
{\footnotesize
\centering
\begin{tabular}{ccccccccccccc}
Project & Phase & Isotope & Mass    & {\scriptsize Energy resol.}   & Efficiency &  {\scriptsize Bkg at $Q_{\beta\beta}$} & {\scriptsize Bkg at $Q_{\beta\beta}$} &  $T_{1/2}(\beta\beta 0 \nu)$ & \multicolumn{2}{c}{$\langle m_{ee} \rangle$ (meV)} & start  \\
        &       &         & isotope & {\scriptsize (FWHM at $Q_{\beta\beta}$)} & $\beta\beta 0 \nu$ & {\scriptsize cts/(keV.kg.yr)}    & {\scriptsize cts/(FWHM.yr)}   & {\scriptsize (90$\%$ C.L.)}  & $m_{min}$ & $m_{max}$ & data \\
\hline
\hline
GERDA    & Phase 1 & $^{76}$Ge & 20 kg   & 4 keV  & $\sim$ 1 & 0.06$^*$ & 5$^*$    & $2 \ 10^{25}$ y & 210  & 530 & 2011 \\
         &         &           &         &        & $\sim$ 1 & $10^{-2}$ & 0.8  & $3 \ 10^{25}$ y    & 167  & 421 &     \\
         & Phase 2 & $^{76}$Ge & 50 kg   & 3 keV  & $\sim$ 1 & $10^{-3}$ & 0.15 & $2 \ 10^{26}$ y    & 41   & 103 & 2013\\
         & Phase 3 & $^{76}$Ge & 200 kg  & 3 keV  & $\sim$ 1 & $10^{-3}$ & 0.3   & $10^{27}$ y     & 29   & 73  &     \\
\hline
SuperNEMO & 1 module & $^{82}$Se & 7 kg & $\approx 200$ keV & $\sim$ 0.2 & $\approx 5 \ 10^{-5}$ & 0.07   &  $6.6 \ 10^{24}$ y & 190 & 460 & End 2014 \\
          & 20 modules & $^{82}$Se & 100 kg & $\approx 200$ keV & $\sim$ 0.2 & $\approx 5 \ 10^{-5}$ & 1  &  $10^{26}$ y       & 48  & 118 & ... \\
          & 20 modules & $^{150}$Nd(50\%) & 50 kg & $\approx 200$ keV & $\sim$ 0.2 & $\approx 5 \ 10^{-5}$ & 1 &  $0.3 \ 10^{26}$ y   & 55  & 210 & ... \\
          & 20 modules & $^{48}$Ca(50\%) & 50 kg & $\approx 200$ keV & $\sim$ 0.2 & $\approx 5 \ 10^{-5}$ & 1  &  $0.75 \ 10^{26}$ y & 100  & 275 & ... \\
\hline
CUORE & CUORE-0 & $^{130}$Te & 11 kg & 5 keV & $\sim$ 1 & 0.05 & 10 &  $6 \ 10^{24}$ y & 184 & 390 & 2012 \\
      & 19 towers & $^{130}$Te & 200 kg & 5 keV & $\sim$ 1 & 0.01 & 37 &  $10^{26}$ y  & 45  & 95  & 2014 \\
      & 19 tow. + ssb & $^{130}$Te & 200 kg & 5 keV & $\sim$ 1 & 0.001 & 3.7 &  $6 \ 10^{26}$ y  & 18  & 39  & ... \\
\hline
ZnSe   & 1 tower & $^{82}$Se  & 19.1 kg & 5 keV & $\sim$ 1 & $10^{-3}$ & 0.17 &  $10^{26}$ y & 48 & 118       & ... \\
ZnMoO4 & 1 tower & $^{100}$Mo & 12.3 kg & 5 keV & $\sim$ 1 & $10^{-3}$ & 0.14 &  $0.7 \ 10^{26}$ y & 37 & 94  & ... \\
CaMoO4 & 1 tower & $^{100}$Mo & 15.3 kg & 5 keV & $\sim$ 1 & $10^{-3}$ & 0.15 &  $0.8 \ 10^{26}$ y & 35 & 88  & ... \\
CdWO4  & 1 tower & $^{116}$Cd & 16.4 kg & 5 keV & $\sim$ 1 & $10^{-3}$ & 0.25 &  $0.7 \ 10^{26}$ y & 60 & 113 & ... \\
\hline
K-Zen & Phase 1 & $^{136}$Xe & 300 kg & 100 keV & $\sim$ 0.5 & ... & $\approx 150^*$ & $5.7 \ 10^{24}$ y$^*$ & 250 & 600 & 2011 \\
      & Phase 2 & $^{136}$Xe & 800 kg & 80 keV  & $\sim$ 0.5 & ... & $1-2$            & $\approx 10^{26}$ y   & 60  & 140 & 2013 \\
\hline
SNO+    & $^{nat}$Nd & $^{150}$Nd & 56 kg & 210 keV & $\sim$ 1 & ... & 40 &  $6 \ 10^{24}$ y & 130 & 470 & 2014 \\
        & $^{enr}$Nd (50\%) & $^{150}$Nd & 500 kg & 210 keV & $\sim$ 1 & ... & 40 &  $6 \ 10^{25}$ y & 40 & 150 & ... \\
\hline
EXO  & EXO-200  & $^{136}$Xe & 160 kg & 80 keV & $\sim$ 0.4 & $10^{-3}$     & 16  & $6 \ 10^{25}$ y & 77 & 185 & 2012\\
NEXT & NEXT-100 & $^{136}$Xe & 90 kg  & 25 keV & $\sim$ 0.3 & $2 \ 10^{-4}$ & 0.5 & $10^{26}$ y & 60 & 143 & 2014 \\
\hline
CANDLES & Candles-III & $^{48}$Ca & 0.3 kg  & 210 keV & $\sim$ 1 & $10^{-3}$ & 0.3 & $4 \ 10^{24}$ y & 430 & 1200 & 2011 \\
        & 13\% enr. & $^{48}$Ca & 20 kg  & 135 keV & $\sim$ 1 & $10^{-3}$ & 0.3 & $2.5 \ 10^{26}$ y & 54 & 150 & 2011 \\
\end{tabular}
\caption{Summary of the characteristics and expected sensitivities of current and possible future experiments. $^*$ refers to already obtained values. Otherwise, it corresponds to expected values. Limits on the $\beta\beta 0\nu$ half-life $T_{1/2}(\beta\beta 0 \nu)$ have been calculated with 5 years of collected data, except GERDA Phase 1 (1 year), SuperNEMO demonstrator (2.5 years) and CUORE-0 (2 years), and using the Feldman-Cousins method~\cite{feldman-cousins}. The corresponding limits on the effective Majorana neutrino mass $\langle m_{ee} \rangle$ have been calculated using the nuclear matrix elements given in Tables~\ref{tab:g0nu-nme} and \ref{tab:current-limits}.}
\label{tab:projects-summary}
}
\end{table}

\end{landscape}

\renewcommand{\baselinestretch}{1}
\normalsize
\addcontentsline{toc}{chapter}{Bibliographie}

\clearoddpage

\renewcommand{\baselinestretch}{\mystretch}

\clearoddpage

\end{document}